\documentclass{article}

\usepackage[main, final]{neurips_2026}

\usepackage[utf8]{inputenc}
\usepackage[T1]{fontenc}
\usepackage{algorithm}
\usepackage{algorithmic}
\usepackage{hyperref}
\usepackage{url}
\usepackage{booktabs}
\usepackage{amsfonts}
\usepackage{nicefrac}
\usepackage{microtype}
\usepackage[table]{xcolor}
\usepackage{amsmath}
\usepackage{amssymb}
\usepackage{graphicx}
\usepackage{capt-of}
\usepackage{wrapfig}
\usepackage{pifont}
\usepackage{longtable}
\usepackage{adjustbox}
\usepackage{placeins}
\usepackage{needspace}
\usepackage{multirow}
\usepackage{enumitem}
\newcommand{\method}{\texorpdfstring{\textsc{CRANE}}{CRANE}}

\title{%
  \raisebox{-0.4\height}{\includegraphics[height=1.3cm]{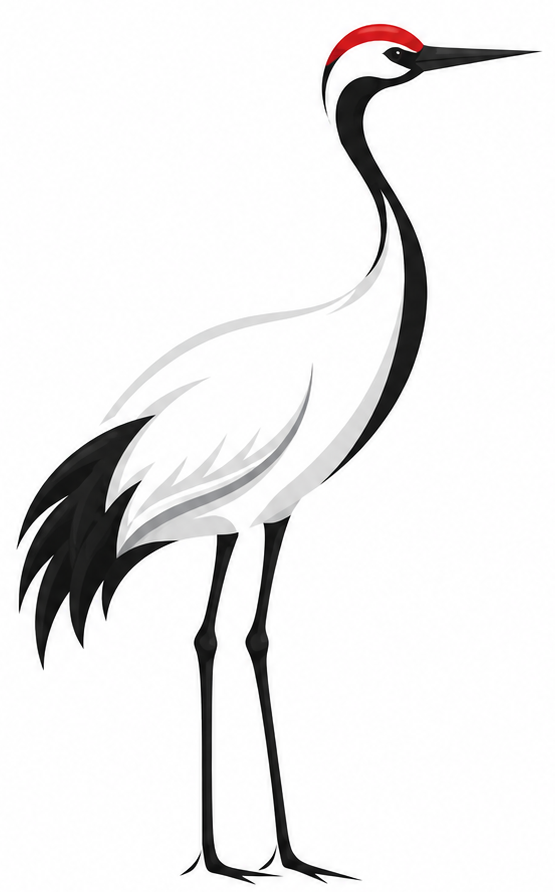}}%
  \hspace{0.5em}%
  \parbox[c]{0.85\textwidth}{\centering
    \method: Constrained Reasoning Injection for\\
    Code Agents via Nullspace Editing
  }%
}

\author{%
    Mingzhi Zhu \\
  Rensselaer Polytechnic Institute\\
  Troy, NY 12180\\
  \texttt{zhum8@rpi.edu} \\
  \And
  Michele Merler \\
  IBM Research \\
  Yorktown Heights, NY 10598  \\
  \texttt{mimerler@us.ibm.com} \\
  \And
  Raju Pavuluri \\
  IBM Research \\
  Yorktown Heights, NY 10598  \\
  \texttt{pavuluri@us.ibm.com} \\
  \And
  Stacy Patterson \\
  Rensselaer Polytechnic Institute \\
  Troy, NY 12180 \\
  \texttt{sep@cs.rpi.edu} \\
}

\begin{document}

\maketitle

\begin{abstract}
    Code agents must both reason over long-horizon repository state and obey strict tool-use protocols. In paired Instruct/Thinking checkpoints, these capabilities are complementary but misaligned. The Instruct model is concise and tool-disciplined, whereas the Thinking model offers stronger planning and recovery behavior but often over-deliberates and degrades agent performance. We present CRANE (Constrained Reasoning Injection for Code Agents via Nullspace Editing), a training-free parameter-editing method that treats the Thinking–Instruct delta as a directional pool of candidate reasoning edits for the Instruct backbone. CRANE combines magnitude thresholding to denoise the delta, a Conservative Taylor Gate to retain edits that are jointly beneficial for reasoning transfer and tool-use preservation, and Graduated Sigmoidal Projection to suppress format-critical update directions. 
    By merging paired Instruct and Thinking checkpoints, CRANE delivers strong gains over either individual model while preserving Instruct-level efficiency: on Roo-Eval it achieves pass@1 of 66.2\% (+19.5\%) for Qwen3-30B-A3B and 81.5\% (+8.7\%) for Qwen3-Next-80B-A3B; on SWE-bench-Verified it resolves up to 14 additional instances at both scales (122/500 and 180/500); and on Terminal-Bench v2 it improves pass@1/pass@5 by up to 2.2\%/7.9\%, reaching 7.6\%/18.0\% and 14.8\%/30.3\%, respectively, consistently outperforming alternative merging strategies across all three benchmarks. Code is available at \url{https://github.com/rpi-nsl/CRANE}.

\end{abstract}

\section{Introduction}
\label{sec:intro}
\begin{figure}[!t]
  \centering  \includegraphics[width=\linewidth]{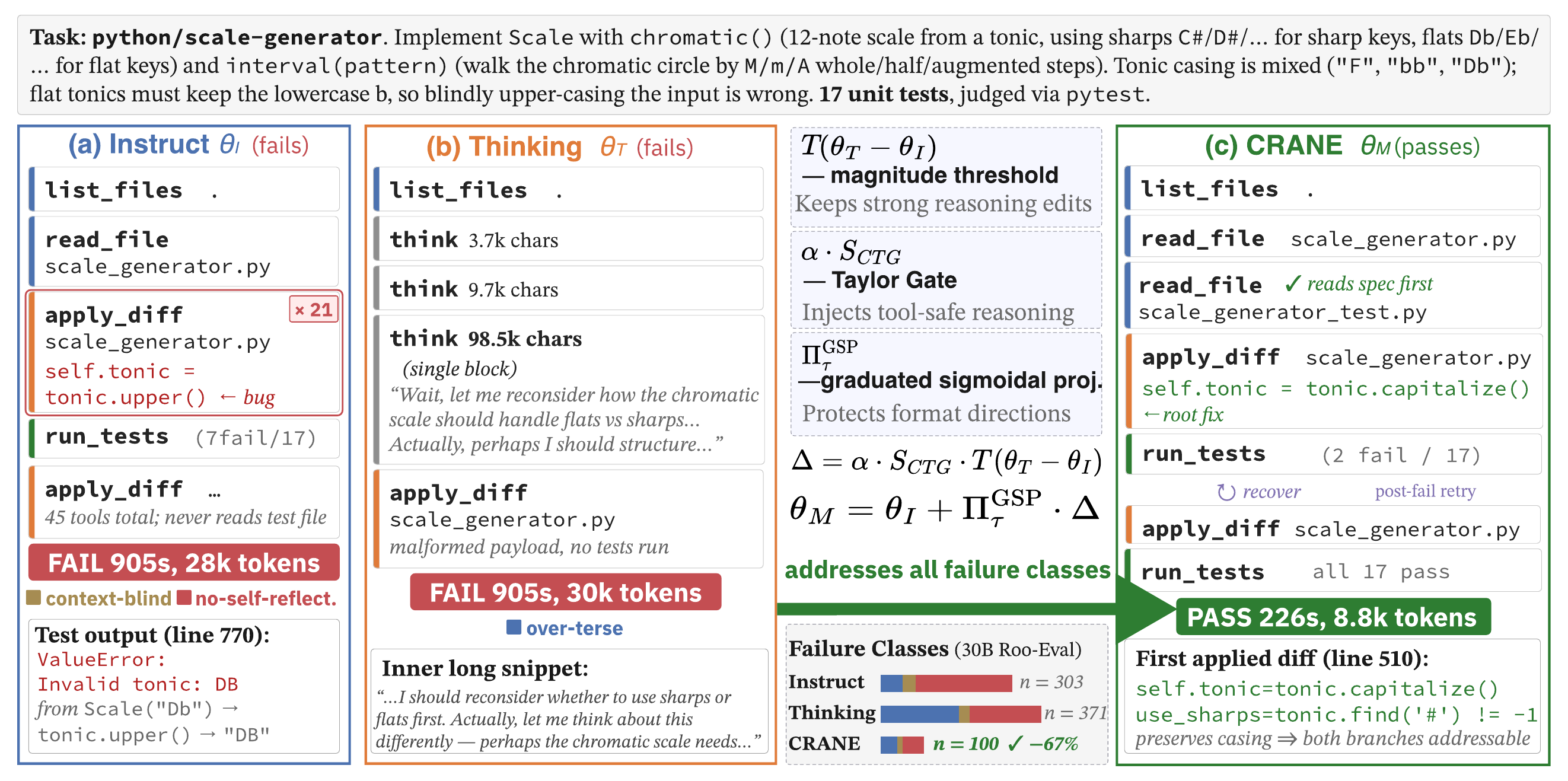}
  \vspace{-0.5cm}
  \caption{Qualitative Roo-Eval trace illustrating the endpoint trade-off that motivates selective injection. On \texttt{python-scale-generator} task, the Instruct endpoint acts quickly but edits before reading the relevant test and then loops on failed tool calls, while the Thinking endpoint shows stronger deliberation but still fails through overlong reasoning without re-testing. \method{} preserves the tool workflow while importing useful planning behavior: it reads the specification first, applies a fix, recovers after a partial failure, and passes all tests. The inset summarizes failure classes over failed Qwen3-30B-A3B Roo-Eval trajectories (Appendix~\ref{app:failure-analysis}); two more trace triples are in Appendix~\ref{app:qual-trace-extra}.}
  \label{fig:qual_trace}
\end{figure}

Modern code agents solve software tasks through long, structured interactions with repositories, tools, and execution environments. Systems such as SWE-agent~\citep{yang2024swe} and OpenHands~\citep{wang2024openhands} make this setting explicit: the model must inspect files, issue edits, execute tests, and react to tool outputs under a constrained agent–computer interface, so success depends on both reasoning quality and protocol fidelity. Yet recent work shows that large reasoning models can sometimes overthink at substantial token cost while actually reducing  performance~\citep{liu2024mindyourstep,li2025thinkbench,zhou2026morethinking}. We confirm this on Roo-Eval~\citep{roocodeevals}, where Thinking checkpoints underperform their Instruct counterparts at two scales, achieving 34.9\% versus 46.7\% pass@1 at 30B (Qwen3-30B-A3B) and 35.4\% versus 72.8\% at 80B (Qwen3-Next-80B-A3B), while generating about 2.7$\times$ more output tokens. Figure~\ref{fig:qual_trace} illustrates this trade-off on a single Roo-Eval task. Based on these observations, this paper studies how to selectively inject the richer planning, context integration, and recovery behavior of Thinking checkpoints into Instruct backbones while strictly preserving the deployed agent protocol: concise tool timing, schema fidelity, and compact outputs.

Prior model-merging works \citep{ilharco2023ta,yu2024dare} and reverse-direction methods such as RAIN-Merging~\citep{huang2026rainmerging} have shown that weight-space editing and task-vector composition can combine capabilities across fine-tuned models without retraining. However, these methods are not designed for the asymmetric code-agent setting, where it is paramount to preserve an Instruct model's tool protocol while importing only those Thinking-side directions that improve agentic reasoning. The challenge is not generic fusion but behavior-conditioned directional editing.

We address this with CRANE (Constrained Reasoning Injection for Code Agents via Nullspace Editing), a training-free parameter-editing method (no parameter is optimized) that treats the Thinking–Instruct difference vector ($\delta = \theta_{\text{think}} - \theta_{\text{inst}}$) as a pool of candidate reasoning edits for the Instruct backbone. CRANE has three stages: (1) a magnitude-thresholding operator $T$ that sparsifies the raw delta and removes low-confidence coordinates; (2) a Conservative Taylor Gate (CTG) that uses gradients of masked calibration losses to give each coordinate of $\delta$ a positive score only when moving it from its Instruct value toward its Thinking value is first-order helpful for both reasoning transfer and tool-use preservation, and then sets each block's injection strength from the sum of these scores; and (3) a Graduated Sigmoidal Projection (GSP) that uses format-critical Instruct activations to suppress update components that would perturb format-control tokens, tool delimiters, or JSON/schema structure. In short, CRANE denoises the candidate delta, scales each block by how much of its edit is both reasoning-helpful and tool-safe, and attenuates edits in the protected format subspace.

We demonstrate empirically that CRANE yields consistent gains across three agentic coding benchmarks (Roo-Eval~\citep{roocodeevals}, SWE-bench-Verified (SWE-V)~\citep{jimenez2024swebench,openai2024swebenchverified}, Terminal-Bench v2 (TB-V2)~\citep{terminalbench}) and two model scales (Qwen3-30B-A3B and Qwen3-Next-80B-A3B). On Roo-Eval, CRANE raises pass@1 to 66.2\% at 30B scale, well above the Instruct (46.7\%) and Thinking (34.9\%) endpoints and the best alternative merge (47.2\%), and to 81.5\% at 80B. On SWE-V it resolves the most instances of any merging baseline at both scales (122/500 and 180/500, respectively), and on TB-V2 it achieves the strongest pass@1/pass@5 results (7.6\%/18.0\% at 30B; 14.8\%/30.3\% at 80B). These gains come with practical efficiency: on Roo-Eval and SWE-V, CRANE has the lowest or near-lowest total token count (TTC) of all methods,
and on TB-V2 the shortest or near-shortest wall time. 
Ablations confirm that each component ($T(\delta)$, CTG, and GSP) contributes meaningfully to the success–cost frontier.

\textbf{Contributions.}
\begin{itemize}[leftmargin=*]
    \item A directional formulation of model merging for paired Instruct/Thinking models, where the Thinking–Instruct delta is treated as a candidate edit pool rather than a symmetric target;
    \item CRANE, a training-free three-stage merge recipe that combines sparse delta extraction, tool-use-aware Conservative Taylor Gating, and format-preserving Graduated Sigmoidal Projection;
    \item A six-setting empirical evaluation across Roo-Eval, SWE-bench-Verified, and Terminal-Bench v2 showing more consistent gains than endpoint substitution or standard global merge baselines;
    \item Ablations and sensitivity analyses that characterize which modules matter and how the performance–efficiency trade-off behaves around the selected merge scale and projection threshold.
\end{itemize}

\section{Related Work}\label{sec:related}

\textbf{Model merging and sparse delta editing.} A broad class of weight-space methods motivates sparse editing, but most prior work targets symmetric endpoint fusion, compression, or generic interference. Task-vector and merge-interference methods such as Task Arithmetic~\citep{ilharco2023ta}, TIES~\citep{yadav2023ties}, DARE~\citep{yu2024dare}, SLERP~\citep{shoemake1985slerp}, RegMean~\citep{jin2023regmean}, AIM~\citep{aim2024}, LEWIS~\citep{lewis2024}, and Fisher-weighted merging~\citep{matena2022fisher} combine or weight endpoint deltas, while pruning methods such as magnitude pruning~\citep{han2015learning,franklelottery}, Wanda~\citep{sun2024simple}, and SparseGPT~\citep{frantar2023sparsegpt} show that many weights can be suppressed with limited immediate degradation. These methods are natural baselines because they edit the same weight-space object, but they do not condition the edit on code-agent behavior. In contrast, our setting is directional and behavior-conditioned. A coordinate is useful only if moving along the actual Thinking--Instruct delta improves reasoning while remaining compatible with tool-use preservation.

\textbf{Preservation-aware merging.} A closer line of work asks which endpoint behavior should be protected while another capability is imported. RAIN-Merging~\citep{huang2026rainmerging} studies the complementary direction. It injects instruction-following ability into a reasoning model while preserving the reasoning model's thinking format. \method{} reverses both the transfer direction and the protected behavior: we inject Thinking-derived reasoning behavior into an Instruct code agent and protect the agent protocol rather than a public chain-of-thought (CoT) format. Other merge variants control the update family rather than explicitly protecting a code-agent protocol: AdaMerging~\citep{yang2024adamerging} learns per-layer scalars, and LoRA-merging methods~\citep{huang2024lorahub} act on low-rank adapters rather than full deltas. Unlike these methods, our preservation mechanism protects activation subspaces tied to code-agent protocol tokens.

\textbf{Reasoning transfer in code-agent settings.} A separate route to importing reasoning behavior is to retrain or distill the target model, but code-agent deployment is more constrained than standalone CoT imitation. Distillation-from-reasoning approaches~\citep{magister2023teaching,deepseekr1distill} teach instruction models to emit CoT, but they retrain the student and must rebuild tool-use formatting from scratch. Code-agent systems and benchmarks such as Roo-Code/Roo-Eval, SWE-bench, SWE-agent, Terminal-Bench, and OpenHands instantiate long-context interactions over repository state, tool observations, and structured tool calls~\citep{roocode,roocodeevals,jimenez2024swebench,yang2024swe,terminalbench,wang2024openhands}. In this setting, useful standalone reasoning can still shift the interaction policy away from tool use, schema fidelity, context-budget discipline, or recovery from tool observations. Our method instead uses Thinking outputs only as calibration targets while the Instruct model supplies the preservation targets. It is complementary to system-level agent acceleration~\citep{lin2026squidagent}.

\section{Method}\label{sec:method}

\begin{figure}[t]
  \centering
  \includegraphics[width=\linewidth]{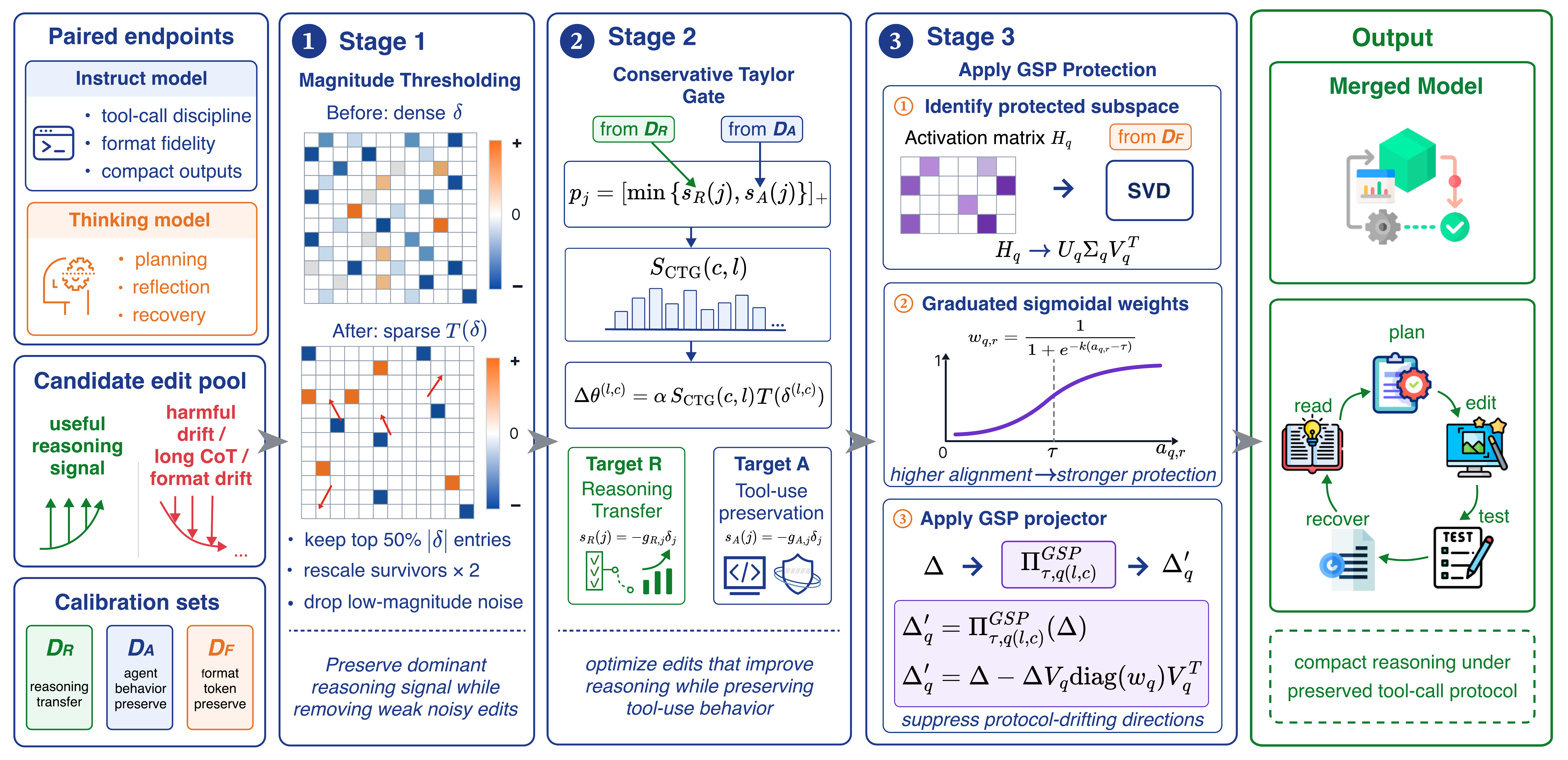}
  \vspace{-0.5cm}
  \caption{\method{} implementation pipeline with three stages: (1) Magnitude thresholding to sparsify $\delta$ and discard low-confidence coordinates; (2) Conservative Taylor Gate that sets per-block injection strength according to how much of the block's edit is first-order beneficial to both reasoning and tool use; (3) Graduated Sigmoidal Projection that attenuates updates along format-critical subspaces.}
  \label{fig:pipeline}
\end{figure}

Starting from a base model with weights $\theta_{\text{base}} \in \mathbb{R}^D$, let $\theta_{\text{inst}} \in \mathbb{R}^D$ denote an instruction-tuned code-agent checkpoint and $\theta_{\text{think}} \in \mathbb{R}^D$ a paired reasoning-tuned checkpoint. We write $\delta=\theta_{\text{think}}-\theta_{\text{inst}}$ for the Thinking--Instruct delta and use $\theta_{\text{merged}}$ for the edited model. The desired endpoint is not a symmetric average. It is an Instruct-style agent that preserves the deployed tool interface of $\theta_{\text{inst}}$ while selectively importing the problem-solving ability exposed by $\theta_{\text{think}}$.

This asymmetric goal leads to three objectives. \textbf{Reasoning transfer} ($R$) uses Thinking-generated continuations conditioned on code-reasoning prompts, capturing planning, context integration, and recovery behavior that we want to inject. \textbf{Format preservation} ($F$) uses Instruct-generated continuations on format-critical prompts, focusing on chat-template tokens, tool-call delimiters, JSON/schema syntax, and other local protocol markers. \textbf{Agent-behavior preservation} ($A$) also uses Instruct-generated continuations, but keeps broader action spans that encode when to call tools, when to read context, and when to stop.
The objectives are complementary because large components of $\delta$ can carry Thinking-side reasoning behavior while overlapping with Instruct-side directions needed for format control and tool-use behavior. A naive linear merge $\theta_{\text{inst}}+\alpha\delta$ may improve reasoning transfer, but it can also damage the Instruct-side agent interface. 

We instead define a three-stage approach that addresses all three objectives (see Figure~\ref{fig:pipeline}):
\begin{equation}
\label{eq:hldecomp}
\theta_{\text{merged}}^{(l,c)} \;=\; \theta_{\text{inst}}^{(l,c)} \;+\; \underbrace{\Pi_{\tau,q(l,c)}^{\text{GSP}}}_{\text{stage 3}}\!\Big(\alpha \cdot \underbrace{S_{\text{CTG}}(c, l)}_{\text{stage 2}} \cdot \underbrace{T(\delta^{(l,c)})}_{\text{stage 1}}\Big)
\end{equation}
where $l \in \{0,\ldots,L-1\}$ indexes the transformer layer and $c \in C$ indexes the parameter component, such as Q/K/V/O attention projections, expert gate/up/down projections, layer norms, and routers. 
Stage 1  removes low-confidence coordinates from $\delta$ via  a conservative sparsifier $T$.
Stage 2 then addresses objectives $R$ and $A$ by scoring, for each block, how much of its $\delta$ is first-order helpful for both reasoning and tool use. We develop a Conservative Taylor Gate (CTG), denoted $S_\text{CTG}(c,l)$, to determine the scaling coefficient for each block.   
Finally, in Stage 3, we address objective $F$ using a  Graduated Sigmoidal Projection (GSP), denoted $\Pi_{\tau,q(l,c)}^{\text{GSP}}$, to project out format-critical directions in the input-side activation space indexed by $q(l,c)$.

 We instantiate the three objectives through model evaluation on three small calibration sets. The sets  $\mathcal{D}_R$ and  $\mathcal{D}_A$ are used to define masked losses for the CTG; $\mathcal{D}_F$ is a set of format traces used to collect the activations protected by GSP. Appendix~\ref{app:calibration-signal} gives construction details.

\subsection{Stage 1: Denoising the Delta by Magnitude Thresholding}
\label{sec:stage1}

Since $\theta_{\mathrm{merged}}$ is obtained by adding an edited delta to $\theta_{\mathrm{inst}}$, each active coordinate moves an Instruct parameter toward its Thinking counterpart.
Small delta entries are less likely to contribute meaningfully to reasoning transfer and may perturb the agent interface. 
We therefore use a conservative sparsification rule that edits only large-magnitude delta coordinates. Following prior sparse-delta merging methods~\citep{yadav2023ties,yu2024dare}, we construct a sparse approximation of $\delta$ using a deterministic median-magnitude threshold with rescaling:
\begin{equation}
\label{eq:tdelta}
T(\delta)_j \;=\; 2\,\delta_j \cdot m_j(\delta), ~~~~~ m_j(\delta)=\mathbf{1}\!\left\{|\delta_j| > \mathrm{median}(|\delta|)\right\}.
\end{equation}
Because the sparsification is deterministic rather than randomized, the factor of two serves only to approximately preserve the overall update scale.
For mixture-of-experts (MoE) layers, $T$ is applied independently to each expert tensor.

\subsection{Stage 2: Tool-Use-Aware Conservative Taylor Gate}
\label{sec:taylor}

Stage 1 reduces element-level noise but still applies a uniform scale to every component and layer. However, reasoning gains and tool-use risks are unevenly distributed across layer-component blocks. A single scale can over-inject fragile blocks while under-utilizing blocks that carry useful reasoning behavior. Stage 2, therefore, determines block-wise importance coefficients for more fine-grained edit scaling. 

We first define the losses for objectives $R$ and $A$.
For $K\in\{R,A\}$, let $\mathcal{D}_K$ contain triples $(x_i^K,y_i^K,m_i^K)$, where $x_i^K$ is the prompt, $y_i^K$ is the endpoint-generated target continuation, and $m_i^K\in\{0,1\}^{S_i^K}$ selects the target tokens that contribute to the loss, where $S_i^K$ is the token length of $z_i^K=[x_i^K;y_i^K]$. With $M_K=\sum_i\sum_s m_{i,s}^K$, define
\begin{equation}
\label{eq:masked_nll}
\mathcal{L}_K(\theta)
=
-\frac{1}{M_K}
\sum_i
\sum_s
m_{i,s}^K
\log p_\theta\!\left(z_{i,s}^K \mid z_{i,<s}^K\right),
\qquad K\in\{R,A\}.
\end{equation}
The implementation value $m_{i,s}^K=0$ corresponds to an ignored label, so prompt tokens and irrelevant continuation positions do not contribute to the loss gradient.

\textbf{Local first-order expansion.} 
Let $g_K = \nabla_\theta \mathcal{L}_K(\theta_{\text{inst}})$ denote the gradient of (\ref{eq:masked_nll}) for $K\in\{R,A\}$. Consider a small coordinate-wise update along the Thinking--Instruct merge direction,
\begin{equation}
\theta_{\text{inst}}+\eta \delta_j e_j,
\end{equation}
where $e_j$ is the unit coordinate vector for the $j$-th entry of the flattened parameter vector. 
Taylor expansion gives
\begin{equation}
\mathcal{L}_K(\theta_{\text{inst}}+\eta \delta_j e_j)=\mathcal{L}_K(\theta_{\text{inst}})+\eta g_{K,j}\delta_j+O(\eta^2\delta_j^2).
\end{equation}
Thus, the first-order change in loss is proportional to $g_{K,j}\delta_j$.
We define the coordinate-wise score
\begin{equation}
s_K(j) = -g_{K,j}\delta_j
\end{equation}
so that $s_K(j)>0$ indicates that moving along the merge direction decreases $\mathcal{L}_K$ to first order. Unlike Fisher-style importance measures~\citep{matena2022fisher}, $s_K(j)$ is signed and direction-aware.

\textbf{Conservative Taylor Gate.} Reasoning transfer and tool-use preservation are not redundant signals. 
We therefore assign positive weight only to coordinates where the same infinitesimal edit is first-order beneficial for both losses. 
CTG uses the positive part of the minimum directional improvement score:
\begin{equation}
\label{eq:ctg}
p_j
=
[\min\left\{s_R(j),s_A(j)\right\}]_+,
\qquad
[u]_+ = \max\{u,0\}.
\end{equation}
Thus, $p_j>0$ only when the Thinking delta is, to first order, a common descent direction for the reasoning loss and the tool-use preservation loss at coordinate $j$.
A coordinate with large reasoning gain but negative tool-use effect receives zero score.

\textbf{Aggregation by component and layer.} Let $\mathcal{B}_{c,l}\subseteq\{1,\ldots,D\}$ be the index set for component $c$ in layer $l$. We aggregate the coordinate scores and define the relative block coefficient directly:
\begin{equation}
\label{eq:ctg_relative}
S_{\mathrm{CTG}}(c,l)
=
\frac{\sum_{j\in\mathcal{B}_{c,l}} p_j
}{\sum_{j\in\mathcal{B}_{b,l}} p_j}
\cdot
\frac{\left\|\theta_{\mathrm{inst}}^{(l,b)}\right\|_F}{\left\|\theta_{\mathrm{inst}}^{(l,c)}\right\|_F}
\end{equation}
where $b$ is the per-layer FFN/expert component, $b\in C$. $\mathcal{B}_{b,l}$ is the union of the gate, up, and down projection indices for dense FFN layers or the union of gate/up/down indices across all experts for MoE layers. 
We normalize each component by the layer FFN/expert block $b$, a common reference scale across components. Because both numerator and denominator aggregate coordinate scores, the coefficient is insensitive to the absolute scale of the losses. Summing rather than averaging coordinate scores also preserves the cumulative CTG-positive contribution of larger blocks. For the hybrid Qwen3-Next-80B-A3B, whose residual stack has three linear-attention layers per full-attention layer, we additionally divide $S_{\mathrm{CTG}}$ of linear-attention components by 3 (Appendix~\ref{app:archnorm}).

The pre-projection block update is
$\Delta\theta^{(l,c)} = \alpha\,S_{\mathrm{CTG}}(c,l)\,T(\delta^{(l,c)})$
where $\alpha$ is a global merge-scale hyperparameter shared by all edited tensors. It controls the overall amount of Thinking--Instruct delta injected after median denoising and CTG component scaling.

Appendix~\ref{app:calibration-robustness} reports a robustness analysis for calibration-set choice. Different calibration subsets preserve the same component ordering and maintain Spearman correlation above 0.990.

\subsection{Stage 3: Format-Preserving Graduated Sigmoidal Projection}
\label{sec:gsp}

Even an importance-weighted delta can violate the Instruct-side protocol if it changes the local computation at tokens that control chat templates, tool-call delimiters, JSON/schema syntax, braces, or schema-critical keys. Stage 3 addresses objective $F$ to preserve these format-critical aspects.

Let $W$ represent the weights of a tensor in $\theta_{\text{inst}}$, and let $h$ represent the input activation vector corresponding to a token we seek to protect.  If the merge proposes an edit $\Delta$, the local output becomes $(W + \Delta) h = Wh + \Delta h$.
Preserving the Instruct computation at format positions asks for $\Delta h\approx0$ on the protected format activations. 

We achieve this by applying a GSP to the proposed tensor edits.
Our formulation borrows from activation-null-space methods used in factual-association editing~\citep{meng2022locating,mengmass} and continual-learning gradient projection~\citep{sahagradient} but replaces hard subspace truncation with a smooth sigmoid mask in singular-value space.
Let $\mathcal{I}_F$ denote the support of the format mask and $\mathcal{N}_\rho(\mathcal{I}_F)$ its local token neighborhood; Appendix~\ref{app:gsp-token-neighborhood} gives both definitions. We let $q$ index the input-side activation space of an edited tensor. Let
$h_q(z_i^F,t;\theta_{\text{inst}})\in\mathbb{R}^{d_q}$
be the input activation vector for token $t$ in space $q$.
The masked activation matrix and its singular value decomposition are
\begin{equation}
\label{eq:gsp_matrix}
H_q
=
\left[
h_q(z_i^F,s;\theta_{\text{inst}})
\right]_{(i,s)\in\mathcal{N}_\rho(\mathcal{I}_F)}
\in
\mathbb{R}^{N_q\times d_q},
\qquad
H_q
=
U_q\Sigma_q V_q^\top .
\end{equation}
Write $V_q=[v_{q,1},\ldots,v_{q,r_q}]$ for the right singular vectors and $\sigma_{q,1}\geq\cdots\geq\sigma_{q,r_q}$ for the corresponding singular values. We then have
\begin{equation}
\label{eq:gsp_format_energy}
\left\|H_q\Delta_q^\top\right\|_F^2
=
\left\|\Sigma_q V_q^\top\Delta_q^\top\right\|_F^2
=
\sum_{r=1}^{r_q}
\sigma_{q,r}^2
\left\|\Delta_q v_{q,r}\right\|_2^2 .
\end{equation}
Directions with large $\sigma_{q,r}$ are the input directions along which an edit most changes the Instruct computation at format-critical positions. Attenuating $\Delta_q v_{q,r}$ for these directions keeps the outputs close to the Instruct endpoint on the masked format traces. The neighborhood $\mathcal{N}_\rho(\mathcal{I}_F)$ extends this protection from literal delimiter tokens to nearby hidden states that condition on those tokens. 

Define the normalized singular amplitude
$
a_{q,r}=\sigma_{q,r}/\sigma_{q,1}
$
and a smooth protection coefficient
\begin{equation}
\label{eq:gsp_weight}
w_{q,r}
=
\frac{1}{1+\exp\!\left(-k(a_{q,r}-\tau)\right)} .
\end{equation}
The slope $k$ controls the width of the transition around the threshold $\tau$; Appendix~\ref{app:gsp-details} gives the exact parameterization.
For a merge delta tensor $\Delta_q\in\mathbb{R}^{d_{\mathrm{out}}\times d_q}$, GSP applies the soft spectral projector
\begin{equation}
\label{eq:gsp_proj}
\Pi_{\tau,q}^{\mathrm{GSP}}(\Delta_q)
=
\Delta_q-\Delta_q V_q\mathrm{diag}(\mathbf{w}_q)V_q^\top .
\end{equation}
After projection, the component of the edit along $v_{q,r}$ is scaled by $1-w_{q,r}$. The sigmoid mask avoids a hard null-space cutoff. High-amplitude format directions are removed almost completely, low-amplitude directions are largely left unchanged, and boundary directions receive partial attenuation that varies continuously with $\tau$. This soft attenuation is better matched to long-context agentic traces. 
For tensors without a matching activation matrix, $\Pi_{\tau,q(l,c)}^{\text{GSP}}$ is the identity.
Appendix~\ref{app:gsp-details} gives the tensor-layout details and the full merge algorithm.

\textbf{Computing the projector.} GSP estimates $V_q$ and $\sigma_{q,r}$ once from the format-trace calibration set $\mathcal{D}_F$, which contains Instruct-generated traces whose masks mark format-critical tokens such as chat-template markers, tool-call delimiters, and JSON/schema syntax (Appendix~\ref{app:calibration-signal}). We run the Instruct model on $\mathcal{D}_F$, record the input activations of each edited tensor at the positions in $\mathcal{N}_\rho(\mathcal{I}_F)$, stack them into $H_q$ (Eq.~\ref{eq:gsp_matrix}), and take its SVD.

\section{Experiments}
\label{sec:experiments}
This section organizes the experiments around three research questions:
\textbf{RQ1:} Does \method{} improve code-agent task success over the Instruct endpoint, standard merge baselines, and lightweight fine-tuning across IDE, repository, and terminal workflows? (\textbf{Tables}~\ref{tab:roo_eval_main},~\ref{tab:swe_results},~\ref{tab:terminalbench_results},~\ref{tab:qlora_main}); \textbf{RQ2:} Do the success gains preserve a compact, Instruct-like rollout footprint, rather than relying on higher aggregate token cost, longer wall time, or Thinking-style output growth?  (\textbf{Tables}~\ref{tab:roo_eval_main},~\ref{tab:swe_results},~\ref{tab:terminalbench_results}, \textbf{Figure}~\ref{fig:token_balance}); and
\textbf{RQ3:} What is the contribution of each component of \method{}, namely $T(\delta)$, CTG, and GSP, to the final performance--cost trade-off? (\textbf{Table}~\ref{tab:ablation_modules}, \textbf{Figure}~\ref{fig:ablation_curves}).

\vspace{-.3cm}
\subsection{Setup}
\label{sec:setup}

\textbf{Models and benchmarks.} We evaluate three tool-using code-agent settings: Roo-Eval,
a five-language in-IDE suite; SWE-bench-Verified (SWE-V),
a repository-level issue-resolution benchmark; and Terminal-Bench v2 (TB-V2),
a long-horizon shell-workflow benchmark. SWE-V and TB-V2 use the OpenHands scaffold~\citep{wang2024openhands}; harness details are in Appendices~\ref{app:swe-bench-details} and~\ref{app:terminalbench-setup}.

For all three datasets, we evaluate paired Instruct/Thinking checkpoints on two different architectures within the same family, at two scales: Qwen3-30B-A3B-Instruct/Thinking-2507~\citep{qwen3} and Qwen3-Next-80B-A3B-Instruct/Thinking~\citep{qwen3next}. \method{} uses $\tau=0.03$ at both scales, with $\alpha=0.25$ at 30B and $\alpha=0.15$ at 80B.

\textbf{Baselines and efficiency metrics.} We compare the original checkpoints with Task Arithmetic,
TIES,
SLERP,
AIM,
LEWIS,
and RAIN-Merging;
hyperparameters and AIM details are in Appendices~\ref{app:baseline-hparams} and~\ref{app:aim-baseline-details}. All models are served locally with vLLM~\citep{kwon2023vllm}. We report $\text{TTC}=N_i+0.1N_c+5N_o$ as an aggregate rollout-footprint proxy, using output tokens and TB-V2 wall time to distinguish compact gains from inflated traces; accounting details are in Appendix~\ref{app:swe-bench-details}.
\vspace{-.4cm}
\subsection{Benchmark Results}
\label{sec:3-ben-results}

\begin{table}[t]
\centering
\scriptsize
\caption{Roo-Eval pass rates and token usage over five languages; details in Appendix~\ref{app:roo-aggregate-results}.}
\label{tab:roo_eval_main}
\begin{adjustbox}{max width=\linewidth}
\begin{tabular}{lccccrrr}
\toprule
Method & pass@1 & pass@3 & pass\_all & TTC & Input tok. & Output tok. & Cached input \\
\midrule
\multicolumn{8}{l}{\textbf{Qwen3-30B-A3B}} \\
\midrule
Instruct (ref) & 91/195 (46.7) & 125/195 (64.1) & 63/195 (32.3) & 181.1M & 43,548,016 & 8,372,134 & 957,076,451 \\
Thinking (ref) & 68/195 (34.9) & 103/195 (52.8) & 35/195 (17.9) & 146.9M & 21,057,008 & 22,786,455 & 119,597,157 \\
\midrule
Task Arithmetic & 92/195 (47.2) & 119/195 (61.0) & 65/195 (33.3) & 208.1M & 50,345,389 & 8,011,542 & 1,177,364,978 \\
TIES & 92/195 (47.2) & 129/195 (66.2) & 57/195 (29.2) & 208.9M & 49,128,311 & 7,644,147 & 1,215,445,711 \\
SLERP & 85/195 (43.6) & 114/195 (58.5) & 58/195 (29.7) & 214.6M & 51,323,145 & 8,418,811 & 1,211,975,312 \\
AIM-TA & 91/195 (46.7) & 126/195 (64.6) & 57/195 (29.2) & 212.6M & 51,338,605 & 7,914,166 & 1,216,900,832 \\
AIM-TIES & 88/195 (45.1) & 120/195 (61.5) & 57/195 (29.2) & 211.3M & 50,606,755 & 8,090,525 & 1,202,205,511 \\
LEWIS & 87/195 (44.6) & 123/195 (63.1) & 54/195 (27.7) & 194.3M & 48,090,553 & 7,657,204 & 1,079,258,386 \\
RAIN & 77/195 (39.5) & 106/195 (54.4) & 42/195 (21.5) & 140.2M & 20,409,513 & 21,681,930 & 113,698,415 \\
\midrule
\textbf{\method{}}  & \textbf{129/195 (66.2)} & \textbf{162/195 (83.1)} & \textbf{86/195 (44.1)} & \textbf{120.9M} & 34,678,861 & 8,759,443 & 424,474,281 \\
\midrule
\multicolumn{8}{l}{\textbf{Qwen3-Next-80B-A3B}} \\
\midrule
Instruct (ref) & 142/195 (72.8) & 170/195 (87.2) & 104/195 (53.3) & 89.6M & 27,444,388 & 6,128,842 & 314,987,867 \\
Thinking (ref) & 69/195 (35.4) & 97/195 (49.7) & 44/195 (22.6) & 109.5M & 18,152,937 & 16,630,299 & 81,763,409 \\
\midrule
Task Arithmetic & 153/195 (78.5) & 173/195 (88.7) & 132/195 (67.7) & 93.1M & 27,492,207 & 6,284,994 & 341,909,682 \\
TIES & 154/195 (79.0) & 172/195 (88.2) & 121/195 (62.1) & \textbf{89.0M} & 26,783,953 & 6,346,889 & 305,139,154 \\
SLERP & 143/195 (73.3) & 169/195 (86.7) & 118/195 (60.5) & 97.6M & 28,915,441 & 6,283,713 & 372,314,291 \\
AIM-TA & 157/195 (80.5) & 171/195 (87.7) & 129/195 (66.2) & 100.0M & 28,687,721 & 6,703,140 & 377,874,779 \\
AIM-TIES & 149/195 (76.4) & \textbf{177/195 (90.8)} & 119/195 (61.0) & 96.0M & 28,855,031 & 6,689,030 & 337,415,124 \\
LEWIS & 155/195 (79.5) & 176/195 (90.3) & 121/195 (62.1) & 95.9M & 28,113,529 & 6,631,916 & 345,905,209 \\
RAIN & 90/195 (46.2) & 114/195 (58.5) & 50/195 (25.6) & 113.2M & 17,933,387 & 17,375,213 & 83,718,010 \\
\midrule
\textbf{\method{}}  & \textbf{159/195 (81.5)} & 176/195 (90.3) & \textbf{139/195 (71.3)} & 89.2M & 26,567,238 & 6,072,681 & 322,364,655 \\
\bottomrule
\end{tabular}
\end{adjustbox}

\end{table}

\begin{figure}[t]
  \centering
  \includegraphics[width=\linewidth]{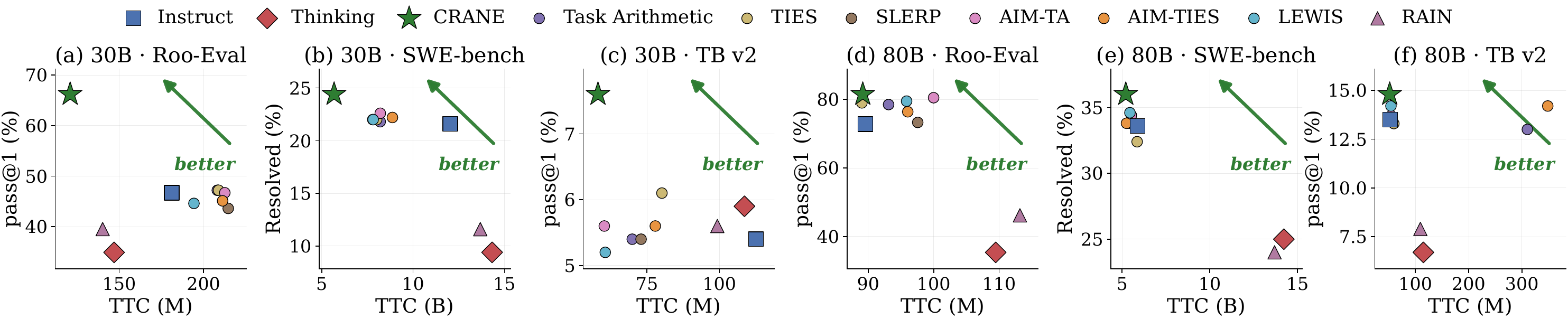}
  \vspace{-0.5cm}
  \caption{TTC vs.\ pass-rate, three benchmarks $\times$ two scales. \textbf{(a--c)} Qwen3-30B-A3B on Roo-Eval, SWE-bench-Verified, Terminal-Bench v2; \textbf{(d--f)} Qwen3-Next-80B-A3B on the same three.}
  \label{fig:token_balance}
\end{figure}

\textbf{Roo-Eval Results.} For RQ1, \method{} improves over the Instruct endpoint by $+19.5$, $+19.0$, and $+11.8$ percentage points on 30B pass@1, pass@3, and pass\_all, respectively; relative to the strongest non-\method{} row for each metric, the corresponding margins are $+19.0$, $+16.9$, and $+10.8$ points. At 80B, \method{} improves over Instruct by $+8.7$ points on pass@1 and $+18.0$ points on pass\_all, beats the strongest non-\method{} pass@1/pass\_all rows by $+1.0$ and $+3.6$ points, and is within $0.5$ points of the best pass@3 row.
For RQ2, these Roo-Eval gains come with lower or comparable TTC. At 30B, \method{} reduces TTC by 60.2M tokens relative to Instruct and by 19.3M relative to the lowest-TTC non-\method{} row while improving all three success metrics. At 80B, \method{} stays within 0.2M TTC of the lowest-TTC alternative and slightly below the Instruct endpoint, while cutting more than 10M output tokens relative to the Thinking and RAIN rows. Figure~\ref{fig:token_balance} visualizes the same success--TTC trade-off across all three benchmarks.

\textbf{SWE-bench-Verified Results.}
For RQ1, \method{} resolves $14$ more instances than the Instruct reference, $9$ more than the strongest merging baseline, and $75$ more than Thinking at 30B. The corresponding 80B gains are $+12$ over Instruct, $+7$ over the strongest merging baseline, and $+55$ over Thinking.
For RQ2, \method{} reaches those higher resolved counts with lower aggregate token cost. Its TTC is 6.36B lower than Instruct and 2.12B lower than the lowest-TTC baseline at 30B. At 80B, the savings are 0.68B relative to Instruct and 0.04B relative to the lowest-TTC non-\method{} row.

\begin{table}[t]
\centering
\scriptsize
\caption{SWE-bench-Verified results. Resolved cells report \texttt{count\,(resolved\%)}; TTC as in Table~\ref{tab:roo_eval_main}.}
\label{tab:swe_results}
\begin{adjustbox}{max width=\linewidth}
\begin{tabular}{lccccrccccr}
\toprule
& \multicolumn{5}{c}{\textbf{Qwen3-30B-A3B}} & \multicolumn{5}{c}{\textbf{Qwen3-Next-80B-A3B}} \\
\cmidrule(lr){2-6} \cmidrule(lr){7-11}
Method & Resolved & Input tok. & Output tok. & Cached input & TTC & Resolved & Input tok. & Output tok. & Cached input & TTC \\
\midrule
Instruct (ref) & 108 (21.6\%) & 2.16B & 353M  & 81.1B & 12.04B & 168 (33.6\%) & 1.96B & 315M  & 23.6B & 5.90B \\
Thinking (ref) & 47 (9.4\%)   & 479M  & 2.15B & 31.0B & 14.33B & 125 (25.0\%) & 1.21B & 2.10B & 25.1B & 14.22B \\
\midrule
Task Arithmetic & 109 (21.8\%) & 1.59B & 322M & 50.0B & 8.20B & 169 (33.8\%) & 1.82B & 318M & 20.7B & 5.48B \\
TIES            & 110 (22.0\%) & 1.66B & 299M & 48.5B & 8.01B & 162 (32.4\%) & 1.91B & 342M & 22.4B & 5.86B \\
SLERP           & 110 (22.0\%) & 1.49B & 331M & 46.5B & 7.80B & 169 (33.8\%) & 1.79B & 326M & 20.5B & 5.47B \\
AIM-TA          & 113 (22.6\%) & 1.61B & 313M & 50.3B & 8.21B & 172 (34.4\%) & 1.81B & 336M & 20.4B & 5.53B \\
AIM-TIES        & 111 (22.2\%) & 1.66B & 350M & 54.6B & 8.87B & 169 (33.8\%) & 1.80B & 311M & 19.0B & 5.26B \\
LEWIS           & 110 (22.0\%) & 1.64B & 303M & 46.6B & 7.82B & 173 (34.6\%) & 1.90B & 312M & 19.9B & 5.45B \\
RAIN            &  58 (11.6\%) & 0.50B & 2.05B & 29.3B & 13.68B & 120 (24.0\%) & 1.22B & 2.00B & 24.7B & 13.69B \\
\midrule
\textbf{\method{}}  & \textbf{122 (24.4\%)} & 1.41B & 373M & 24.0B & \textbf{5.68B} & \textbf{180 (36.0\%)} & 1.81B & 309M & 18.6B & \textbf{5.22B} \\
\bottomrule
\end{tabular}
\end{adjustbox}
\end{table}

\textbf{Terminal-Bench v2 Results.} Terminal-Bench v2 evaluates shell-tool agents on long-horizon command-line workflows in cloud sandboxes. We run all 89 tasks of the \texttt{tb2-zai} dataset~\citep{tb2zai} at $k=5$ attempts/task to match the public Terminal-Bench leaderboard. 
For RQ1, \method{} improves over the strongest non-\method{} rows by $+1.5$ points on pass@1 and $+3.4$ points on pass@5 at 30B, and by $+0.6$ and $+3.3$ points at 80B.
For RQ2, Terminal-Bench provides the clearest wall-time evidence for a compact rollout footprint. At 30B, \method{} is 1h\,56m faster than Instruct and 24m faster than the fastest non-\method{} row, while reducing output by 1.73M tokens relative to Instruct. At 80B, \method{} is 30m faster than Instruct and only 3m slower than the fastest row, while staying within 0.03M output tokens of the lowest-output row.

\begin{table}[t]
\centering
\scriptsize
\caption{Terminal-Bench v2 main results. Test time is the end-to-end harness wall time. Tokens are in millions and Input counts non-cached prefill tokens. Other details are reported in Appendix~\ref{app:terminalbench-detail}.}
\label{tab:terminalbench_results}
\begin{adjustbox}{max width=\linewidth}
\begin{tabular}{lccccrccccr}
\toprule
& \multicolumn{5}{c}{\textbf{Qwen3-30B-A3B}} & \multicolumn{5}{c}{\textbf{Qwen3-Next-80B-A3B}} \\
\cmidrule(lr){2-6} \cmidrule(lr){7-11}
Method & pass@1 & pass@5 & Test time & Input & Output & pass@1 & pass@5 & Test time & Input & Output \\
\midrule
Instruct (ref)  & 4.8 (5.4\%)  & 9 (10.1\%)  & 4h 14m & 16.96 & 5.43  & 12.0 (13.5\%) & 20 (22.5\%) & 2h 28m & 10.84  & 3.85 \\
Thinking (ref)  & 5.2 (5.8\%)  & 12 (13.5\%) & 4h 37m &  4.34 & 18.41 &  6.0 (6.7\%)  & 12 (13.5\%) & 5h 12m &  4.45  & 20.39 \\
\midrule
Task Arithmetic & 4.8 (5.4\%)  & 13 (14.6\%) & 2h 50m &  8.54 & 3.77  & 11.6 (13.0\%) & 22 (24.7\%) & 2h 10m & 266.39 & 3.65 \\
TIES            & 5.4 (6.1\%)  & 12 (13.5\%) & 2h 53m &  9.97 & 4.40  & 11.8 (13.3\%) & 23 (25.8\%) & 1h 55m & 11.71  & 3.86 \\
SLERP           & 4.8 (5.4\%)  & 13 (14.6\%) & 2h 51m &  7.13 & 3.80  & 12.0 (13.5\%) & 24 (27.0\%) & 2h 08m & 12.96  & 3.55 \\
AIM-TA          & 5.0 (5.6\%)  & 12 (13.5\%) & 2h 44m &  7.18 & 3.85  & 12.2 (13.7\%) & 20 (22.5\%) & 2h 00m & 10.10  & 3.72 \\
AIM-TIES        & 5.0 (5.6\%)  & 12 (13.5\%) & 2h 42m &  9.47 & 4.33  & 12.6 (14.2\%) & 22 (24.7\%) & 2h 14m & 301.41 & 3.62 \\
LEWIS           & 4.6 (5.2\%)  & 10 (11.2\%) & 2h 53m &  7.00 & 3.70 & 12.6 (14.2\%) & 23 (25.8\%) & 2h 11m & 10.59  & 3.74 \\
RAIN            & 5.0 (5.6\%)  &  9 (10.1\%) & 4h 05m & 4.01 & 16.76 &  7.0 (7.9\%)  & 14 (15.7\%) & 4h 57m & 4.36 & 19.35 \\
\midrule
\textbf{\method{}} & \textbf{6.8 (7.6\%)} & \textbf{16 (18.0\%)} & 2h 18m & 7.68 & 3.70 & \textbf{13.2 (14.8\%)} & \textbf{27 (30.3\%)} & 1h 58m & 10.42 & 3.58 \\
\bottomrule
\end{tabular}
\end{adjustbox}
\end{table}

\textbf{Cross-benchmark summary.} Across Tables~\ref{tab:roo_eval_main}--\ref{tab:terminalbench_results}, plain merge baselines sometimes improve over a reference checkpoint, especially at 80B, but the gains are inconsistent and RAIN often retains Thinking-like over-deliberation. \method{} turns the endpoint complementarity into more reliable gains across benchmarks and scales while keeping the rollout footprint compact.

\textbf{Inference-time control.} Capping or skipping the Thinking model's reasoning at inference does not recover this gain: the best control reaches 106/195 at pass@3 on 30B Roo-Eval, against 162/195 for \method{} (Appendix~\ref{app:thinkctl}).

\vspace{-.3cm}
\subsection{Ablations}
\label{sec:ablation}

We use ablations to answer RQ3: which parts of the recipe are needed for the observed performance--cost trade-off? One ablation study disables one module at a time ($T(\delta)$, CTG, or GSP), while another varies the global merge scale $\alpha$ and the GSP threshold $\tau$ within a range.

\textbf{Component-importance ablations.} Table~\ref{tab:ablation_modules} shows that no single component can be removed without changing the trade-off. On Roo-Eval 30B, removing GSP causes the largest success drop: $-14.9$, $-11.3$, and $-12.3$ points on pass@1, pass@3, and pass\_all. Without GSP, protected chat-template and tool-call delimiters also leak into 118/195 Roo-Eval exercises, against 3/195 for the full recipe (Appendix~\ref{app:stage3-evidence}). Removing CTG or the sparsifier is less destructive on pass@1/pass@3 but still costs $8.8/3.6$ and $5.7/3.6$ points, respectively; the sparsifier removal is the only variant that improves pass\_all, by $2.1$ points. On Roo-Eval 80B, the full recipe improves pass@1 over all component removals by $2.5$--$4.1$ points and pass\_all by $5.1$--$11.3$ points, while remaining within $1.5$ points of the best pass@3 variant.
The lower block gives the same module removals on Terminal-Bench v2 and SWE-bench-Verified. On Terminal-Bench v2, the full recipe gains $+4.5$ points in 30B pass@5 over the only variant that ties its pass@1, and improves 80B pass@5 by $5.6$--$9.0$ points over all removals. On SWE-bench-Verified, the full recipe resolves $2$--$28$ more 30B instances and $5$--$18$ more 80B instances than the component-removal variants. Overall, the full recipe is strongest on the primary success metrics; single removals occasionally win on a secondary metric or on cost.

\begin{table}[bt!]
\centering
\footnotesize
\setlength{\tabcolsep}{4pt}
\caption{Component-removal ablations. Each row disables one module of \method{}. The upper block reports Roo-Eval; the lower block reports Terminal-Bench v2 and SWE-bench-Verified. Per-variant token breakdowns are in Appendix~\ref{app:roo-ablation-summary}, Tables~\ref{tab:app_ablation_modules_tb_full}--\ref{tab:app_ablation_modules_swe_full}.}
\label{tab:ablation_modules}
\begin{adjustbox}{max width=\linewidth}
\begin{tabular}{lccccccccc}
\toprule
& \multicolumn{4}{c}{\textbf{Qwen3-30B-A3B}} & & \multicolumn{4}{c}{\textbf{Qwen3-Next-80B-A3B}} \\
\cmidrule(lr){2-5} \cmidrule(lr){7-10}
& \multicolumn{4}{c}{Roo-Eval (solved of 195, \%)} & & \multicolumn{4}{c}{Roo-Eval (solved of 195, \%)} \\
\cmidrule(lr){2-5} \cmidrule(lr){7-10}
Method & pass@1 & pass@3 & pass\_all & TTC & & pass@1 & pass@3 & pass\_all & TTC \\
\midrule
\method{} w/o $T(\delta)$ & 118 (60.5) & 155 (79.5) & \textbf{90 (46.2)} & 142.3M & & 154 (79.0) & 177 (90.8) & 129 (66.2) & 97.8M \\
\method{} w/o CTG         & 112 (57.4) & 155 (79.5) & 68 (34.9) & 145.7M & & 151 (77.4) & \textbf{179 (91.8)} & 123 (63.1) & 106.2M \\
\method{} w/o GSP         & 100 (51.3) & 140 (71.8) & 62 (31.8) & \textbf{100.8M} & & 152 (77.9) & 176 (90.3) & 117 (60.0) & 109.7M \\
\midrule
\textbf{\method{}} (full) & \textbf{129 (66.2)} & \textbf{162 (83.1)} & 86 (44.1) & 120.9M & & \textbf{159 (81.5)} & 176 (90.3) & \textbf{139 (71.3)} & \textbf{89.2M} \\
\midrule
& \multicolumn{3}{c}{Terminal-Bench v2} & SWE-V & & \multicolumn{3}{c}{Terminal-Bench v2} & SWE-V \\
\cmidrule(lr){2-4} \cmidrule(lr){5-5} \cmidrule(lr){7-9} \cmidrule(lr){10-10}
Method & pass@1 & pass@5 & TTC (M) & Resolved \,/\, TTC (B) & & pass@1 & pass@5 & TTC (M) & Resolved \,/\, TTC (B) \\
\midrule
\method{} w/o $T(\delta)$ & 6.8 (7.6\%) & 12 (13.5\%)         &  94.1 & 120 (24.0\%) \,/\, 8.43 & & 12.2 (13.7\%) & 21 (23.6\%)         & 52.8 & 164 (32.8\%) \,/\, 5.51 \\
\method{} w/o CTG         & 5.8 (6.5\%) & 14 (15.7\%) & 85.1 & 106 (21.2\%) \,/\, 7.34 & & 11.6 (13.0\%)         & 22 (24.7\%) & \textbf{50.4} & 162 (32.4\%) \,/\, 5.50 \\
\method{} w/o GSP         & 4.8 (5.4\%) & 11 (12.4\%)         &  \textbf{42.5} &  94 (18.8\%) \,/\,\textbf{ 5.35} & & 11.4 (12.8\%)         & 19 (21.3\%)         & 57.3 & 175 (35.0\%) \,/\, 5.35 \\
\midrule
\textbf{\method{}} (full) & \textbf{6.8 (7.6\%)} & \textbf{16 (18.0\%)} & 58.1 & \textbf{122 (24.4\%)} \,/\, 5.68 & & \textbf{13.2 (14.8\%)} & \textbf{27 (30.3\%)} & 51.8 & \textbf{180 (36.0\%)} \,/\, \textbf{5.22} \\
\bottomrule
\end{tabular}
\end{adjustbox}
\end{table}

\begin{figure}[bt!]
  \centering
  \includegraphics[width=\linewidth]{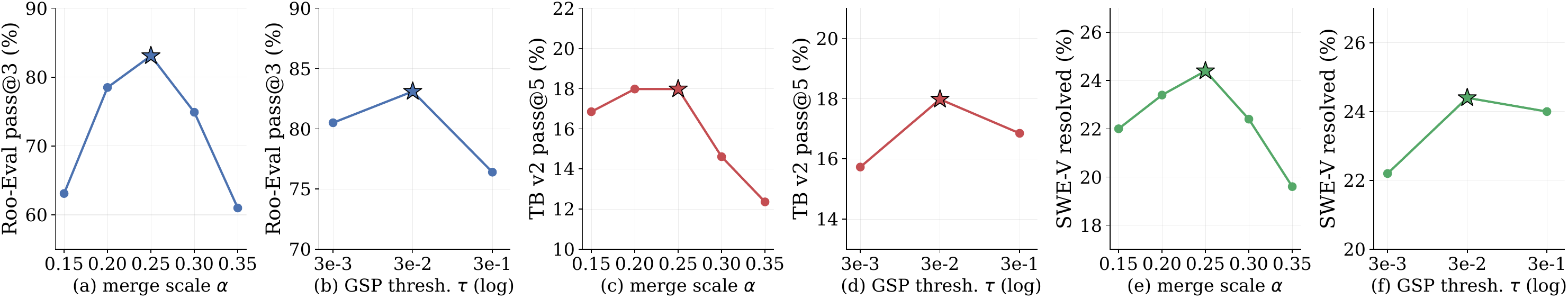}
  \vspace{-0.5cm}
  \caption{Continuous-hyperparameter sensitivity analysis of the \method{} recipe on Qwen3-30B-A3B across three benchmarks, grouped by benchmark. $\alpha$ sweeps fix $\tau{=}0.03$; $\tau$ sweeps fix $\alpha{=}0.25$ and use a log axis. \textbf{(a)--(b)} Roo-Eval pass@3. \textbf{(c)--(d)} TB-V2 pass@5. \textbf{(e)--(f)} SWE-V resolved. Stars mark the reported configuration; Roo-Eval sweep values are tabulated in Appendix~\ref{app:roo-ablation-summary}, Table~\ref{tab:ablation_sweeps}.}
  \label{fig:ablation_curves}
\end{figure}

\textbf{Protocol preservation.} The leakage result replicates on Terminal-Bench v2, where removing GSP raises delimiter leakage from 1.32 to 207.10 per 1k calls at 30B. The full recipe also has a lower JSON-invalid tool-call rate than both endpoints at both scales (Appendix~\ref{app:stage3-evidence}).

\textbf{Hyperparameter sensitivity analysis.} The reported configuration was selected on Roo-Eval only, transfers to TB-V2 and SWE-V without per-benchmark tuning, and remains stable near the chosen point. One step away from it, TB-V2 pass@5 and SWE-V resolved change by at most 3.4 and 2.2 points, whereas Roo-Eval pass@3 drops by 2.6--8.2 points. Raising $\tau$ to 0.3 also weakens protocol preservation: delimiters leak in 26/195 Roo-Eval exercises, against 3/195 at $\tau=0.03$ (Appendix~\ref{app:stage3-evidence}).

\begin{figure}[t]
\centering
\begin{minipage}[t]{0.47\linewidth}
\centering
\vspace{0pt}
\includegraphics[width=\linewidth]{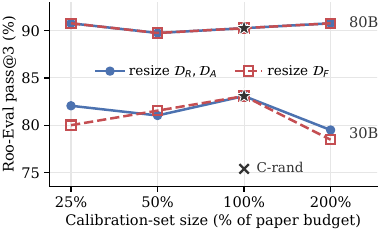}
\caption{Roo-Eval pass@3 with one calibration set resized. Stars: paper checkpoint; C-rand: $S_{\mathrm{CTG}}$ shuffled across layers.}
\label{fig:calib_size_main}
\end{minipage}\hfill
\begin{minipage}[t]{0.49\linewidth}
\centering
\vspace{0pt}
\captionof{table}{Build cost (Appendix~\ref{app:construction-cost}) and Roo-Eval pass@3 (of 195), 30B.}
\label{tab:qlora_main}
\footnotesize
\begin{tabular}{llrr}
\toprule
Model & Teacher & GPU-h & pass@3 \\
\midrule
Instruct & -- & -- & 125 \\
Thinking & -- & -- & 103 \\
\midrule
AIM-TA & -- & 0.3 & 126 \\
AIM-TIES & -- & 0.3 & 120 \\
LEWIS & -- & 0.5 & 123 \\
RAIN & -- & 1.5 & 106 \\
\midrule
QLoRA keepthink & Thinking & 10.1 & 95 \\
QLoRA stripthink & Thinking & 7.7 & 77 \\
QLoRA self-direct & Instruct & 9.8 & 124 \\
\midrule
\method{} & -- & 0.4 & \textbf{162} \\
\bottomrule
\end{tabular}
\end{minipage}
\end{figure}

\textbf{Calibration-set size.} Resizing either calibration set to 25--200\% changes Roo-Eval pass@3 by at most 9 exercises at 30B and 1 at 80B; shuffling the CTG coefficients costs 15 (Figure~\ref{fig:calib_size_main}; Appendix~\ref{app:calibration-size}).

\subsection{Comparison with Lightweight Training}
\label{sec:lightweight-training}

A natural alternative to weight editing is a short fine-tuning run. Three QLoRA~\citep{dettmers2023qlora} students of Qwen3-30B-A3B-Instruct learn from about 800 successful agent turns: Thinking rollouts with the reasoning kept (keepthink) or removed (stripthink), or Instruct's own rollouts (self-direct), which controls for teacher quality (Appendix~\ref{app:qlora}). All students use rank-32 adapters on the attention projections and train for three epochs. None recovers \method{}'s gain (Table~\ref{tab:qlora_main}): the best student only matches Instruct (124 vs.\ 125), while \method{} solves 38 more exercises at 0.4 GPU-hours, in line with the merge baselines (0.3--1.5) and far below the 7.7--10.1 per student. Larger-scale distillation remains a complementary direction.

\section{Limitations}
\label{sec:discussion}

First, \method{} assumes complementary paired endpoints: the Thinking checkpoint must provide useful reasoning behavior, and the Instruct checkpoint must define a useful deployment interface. The two endpoints must also share an architecture and parameterization, typically a common pretrained checkpoint, so that $\delta$ is defined coordinate-wise; transfer across model families would need an extra alignment step and is left to future work. If future Thinking models are already strong in task success, token efficiency, and tool discipline, a simpler endpoint choice or global merge may be competitive. Second, the calibration sets must also cover the deployed tool surface; substantial drift in tools, formatting, or stopping behavior would require re-calibration. Our claims therefore cover structured tool-using agents, where format-critical positions can be identified; unstructured generation tasks are outside this scope. Third, computing the Taylor signal requires backward passes through the Instruct backbone on the calibration sets, the most expensive merge step (about 6\,min at 30B and 27\,min at 80B; Appendix~\ref{app:signal-runtime}); the complete pipeline for Qwen3-235B-A22B takes 133\,min on four H100s. Smaller calibration sets reduce this cost with little loss (Figure~\ref{fig:calib_size_main}). Fourth, on Roo-Eval at 30B, \method{} trails Instruct on Java pass@1 and pass\_all, the language where the Thinking endpoint is weakest, suggesting that the merge can inherit language-specific weaknesses of the Thinking model. A layer-wise analysis of where the CTG coefficients and GSP projection energy concentrate could show where reasoning and protocol behavior reside. We leave it to future work.

\newpage

\begin{ack}
This research is supported in part by the IBM--RPI Future of Computing Research Collaboration. The authors acknowledge the National Artificial Intelligence Research Resource (NAIRR) Pilot, the AI Alliance and resources from IBM, Red Hat and the Mass Open Cloud for contributing to this research result. We gratefully acknowledge use of the research computing resources of the Empire AI Consortium, Inc., with support from Empire State Development of the State of New York, the Simons Foundation, and the Secunda Family Foundation.
\end{ack}

\bibliographystyle{plainnat}
\bibliography{refs}

\appendix
\clearpage

\let\appendixsection\section
\renewcommand{\section}{\FloatBarrier\Needspace{12\baselineskip}\appendixsection}
\let\appendixsubsection\subsection
\renewcommand{\subsection}{\Needspace{8\baselineskip}\appendixsubsection}
\renewcommand{\topfraction}{0.9}
\renewcommand{\bottomfraction}{0.8}
\renewcommand{\textfraction}{0.07}
\renewcommand{\floatpagefraction}{0.8}
\setcounter{topnumber}{4}
\setcounter{bottomnumber}{2}
\setcounter{totalnumber}{6}

\section{Experimental Details}
\label{app:exp-details}

Code for reproducing the merge pipeline and evaluation harness is available at \url{https://github.com/rpi-nsl/CRANE}.

\subsection{Existing Assets and Licenses}
\label{app:asset-licenses}

Table~\ref{tab:app_asset_licenses} lists the external models, benchmarks, datasets, software, and services used in the experiments. Licenses and service terms were checked from official model cards, dataset cards, repositories, or service pages on 2026-05-06. We cite the original creators in the main text and references, use external assets only for evaluation, calibration robustness checks, serving, or sandbox execution, and do not redistribute external model checkpoints or benchmark datasets in the paper package.

\begin{table}[!htbp]
\centering
\scriptsize
\caption{Existing external assets, licenses, terms, and use in this paper.}
\label{tab:app_asset_licenses}
\begin{adjustbox}{max width=\textwidth}
\begin{tabular}{p{0.20\textwidth}p{0.27\textwidth}p{0.20\textwidth}p{0.27\textwidth}}
\toprule
Asset & Official source & License / terms & Use in this paper \\
\midrule
Qwen3 checkpoint pairs~\citep{qwen3,qwen3next} & Hugging Face model cards under \url{https://huggingface.co/Qwen}; exact IDs are listed in Section~\ref{sec:setup} & Apache-2.0 on each listed model card & Source endpoints for merging and evaluation; no redistribution of base checkpoints. \\
\addlinespace
Roo-Code and Roo-Code-Evals~\citep{roocode,roocodeevals} & \url{https://github.com/RooCodeInc/Roo-Code}; \url{https://github.com/RooCodeInc/Roo-Code-Evals} & Apache-2.0 for Roo-Code; MIT for Roo-Code-Evals & Roo-Eval task suite and in-IDE agent environment; no repackaged benchmark release. \\
\addlinespace
SWE-bench and SWE-bench Verified~\citep{jimenez2024swebench,openai2024swebenchverified} & \url{https://github.com/SWE-bench/SWE-bench}; \url{https://huggingface.co/datasets/SWE-bench/SWE-bench_Verified} & MIT for SWE-bench code and data; SWE-bench Verified is the public 500-instance subset released by OpenAI with the SWE-bench authors & Repository-level evaluation and SWE-bench-style calibration prompts; no redistribution of dataset contents. \\
\addlinespace
Terminal-Bench v2 and \texttt{tb2-zai}~\citep{terminalbench,tb2zai} & \url{https://github.com/laude-institute/terminal-bench}; \url{https://huggingface.co/datasets/zai-org/terminal-bench-2-verified} & Apache-2.0 for the Terminal-Bench repository and for the \texttt{tb2-zai} dataset card & Terminal-workflow evaluation on all 89 tasks; no redistribution of task data. \\
\addlinespace
OpenHands~\citep{wang2024openhands} & \url{https://github.com/OpenHands/OpenHands} & MIT for non-enterprise repository content; enterprise directory is separately licensed and not used & Agent scaffold for SWE-bench Verified and Terminal-Bench v2. \\
\addlinespace
vLLM and LiteLLM~\citep{kwon2023vllm,berriai_litellm} & \url{https://github.com/vllm-project/vllm}; \url{https://github.com/BerriAI/litellm} & Apache-2.0 for vLLM; MIT for non-enterprise LiteLLM content, with enterprise directory separately licensed and not used & Local model serving and API transport. \\
\addlinespace
Podman~\citep{podman} & \url{https://github.com/containers/podman} & Apache-2.0 & Rootless container backend for SWE-bench evaluation images. \\
\addlinespace
Public calibration-robustness prompt sources & LiveCodeBench: \url{https://github.com/LiveCodeBench/LiveCodeBench} and \url{https://huggingface.co/datasets/livecodebench/code_generation_lite}; BigCodeBench: \url{https://huggingface.co/datasets/bigcode/bigcodebench}; SWE-bench: \url{https://github.com/SWE-bench/SWE-bench} & LiveCodeBench repository is MIT; its Hugging Face \texttt{code\_generation\_lite} card declares license tag \texttt{cc}; BigCodeBench is Apache-2.0; SWE-bench is MIT & Public-source subsets for the calibration-set robustness check only; the reported paper calibration prompts are original rewrites and the public prompt pools are not redistributed. \\
\addlinespace
Daytona cloud sandboxes & \url{https://www.daytona.io/terms-of-service} & Daytona Terms of Service & Managed sandbox runtime and audit-log cost accounting for Terminal-Bench v2; no Daytona software or service artifact is redistributed. \\
\bottomrule
\end{tabular}
\end{adjustbox}
\end{table}

\subsection{Roo-Eval Evaluation}
\label{app:roo-eval-a2}
Each checkpoint is evaluated on five programming languages with three independent rollouts per exercise. The exercise counts are Python 34, JavaScript 50, Go 36, Java 45, and Rust 30, for 195 exercises and 585 total rollouts per complete sweep.

\begin{table}[!htbp]
\centering
\small
\caption{Roo-Eval serving, judging, and token-accounting protocol used by the result logs.}
\label{tab:app_protocol}
\begin{tabular}{ll}
\toprule
Item & Setting \\
\midrule
Languages & Python, JavaScript, Go, Java, Rust \\
Rollouts & 3 independent iterations per exercise \\
Per-rollout timeout & 900\,s wall-clock \\
Sampling & temperature 0.6, top\_p 0.8, top\_k 20 \\
Context length & $90000$ \\
Eval concurrency & 64 \\
80B serving & vLLM 0.19.0, TP=4, expert parallel enabled, 4$\times$H100 80GB \\
Metrics & pass@1, pass@3, pass\_all, iter pass, TTC \\
\bottomrule
\end{tabular}
\end{table}

\subsection{SWE-bench-Verified Harness}
\label{app:swe-bench-details}

SWE-bench-Verified runs use the OpenHands~\citep{wang2024openhands} agent scaffold over the 500-instance verified subset. All checkpoints are served locally by vLLM~\citep{kwon2023vllm} under the same TP/EP configuration as Roo-Eval; the harness drives OpenHands via litellm~\citep{berriai_litellm}. Table~\ref{tab:app_swe_protocol} records the scaffold and harness settings used for every row of Table~\ref{tab:swe_results}.

\begin{table}[!htbp]
\centering
\small
\caption{SWE-bench-Verified scaffold, container, and harness configuration used for all rows of Table~\ref{tab:swe_results}.}
\label{tab:app_swe_protocol}
\begin{tabular}{ll}
\toprule
Item & Setting \\
\midrule
Subset & SWE-bench-Verified, 500 instances \\
Agent scaffold & OpenHands SDK~\citep{wang2024openhands} \\
Max iterations & 100 per instance \\
Sampling & temperature 0.6, top\_p 0.8, top\_k 20 \\
Serving & vLLM~\citep{kwon2023vllm}, bf16, TP$=4$ \\
GPU & $4\times$H100 80GB \\
Context length & $131072$ \\
Container backend & rootless podman~\citep{podman}  \\
Image registry & Epoch AI ghcr mirror  \\
Per-instance deadline & 60\,min wall-clock; main-thread join cap 61\,min \\
Agent / harness workers & 24 / 24 \\
\bottomrule
\end{tabular}
\end{table}

\paragraph{Sampling.} Without \texttt{top\_k} the Qwen3 checkpoints occasionally drift into long hallucinated continuations that never emit a finish action. We adopt the Qwen3-recommended \texttt{top\_k} $=20$ for every row in Table~\ref{tab:swe_results}, including endpoint references. This setting standardizes decoding across endpoints and reduces stalled-rollout effects in token-usage estimates. The \texttt{litellm} transport timeout is set to 90\,s with 5 retries: the empirical p99 of per-call latency is $\sim$3\,s, so 90\,s gives $\sim$18$\times$ headroom on legitimate calls and bounds unresponsive calls at $\sim$8\,min instead of the OpenHands default of $\sim$30\,min.

\paragraph{Token accounting.} Input tokens are non-cached prefill tokens, computed as \texttt{accumulated\_token\_usage.prompt\_tokens} $-$ \texttt{cache\_read\_tokens}. Cache-read tokens are prompt tokens served by vLLM's prefix cache (requires \texttt{--enable-prompt-tokens-details}). Completion tokens are model outputs. Across SWE-bench-Verified rollouts the agent-loop context is heavily redundant across iterations, and we observe a $\sim$97\% prefix-cache hit rate; cached input is therefore a large term for concise Instruct/merge rows, while output tokens dominate the TTC of over-deliberative Thinking and RAIN rows.
Since major providers price input, cached input, and output tokens differently, we define the Total Token Count (TTC) as a weighted token count:
\begin{equation}
\text{TTC} = w_iN_i + w_cN_c + w_oN_o = N_i + 0.1N_c + 5N_o,
\end{equation}
where $N_i$ is the number of non-cached input tokens, $N_c$ the number of cached input tokens, and $N_o$ the number of output tokens. With $w_i=1$, the weights $w_c$ and $w_o$ are the industry averages of the price ratios in Table~\ref{tab:token_cost}. Because all models are served locally with vLLM, TTC serves as a proxy for the cost of running the same rollouts through a provider. We report TTC on all three benchmarks.

\begin{table}[h!]
\centering
\caption{Token prices of major frontier providers and price ratios relative to input tokens, used to set the TTC weights. Prices are taken from official provider pages on 2026-05-04.}
\resizebox{\textwidth}{!}{
\begin{tabular}{llccccc}
\toprule
\multirow{2}{*}{\textbf{Provider}}& \multirow{2}{*}{\textbf{Model}} & \textbf{Input} & \textbf{Cached Input} & \textbf{Output} & \textbf{Cached\,/\,} & \textbf{Output\,/\,}\\
 & & \textbf{(per 1M tokens)} & \textbf{(per 1M tokens)} & \textbf{(per 1M tokens)} & \textbf{Input} & \textbf{Input} \\
\midrule

\multirow{3}{*}{\textbf{Anthropic}}
  & Claude Opus 4.7     & \$5.00   & \$0.50    & \$25.00  & \multirow{3}{*}{0.10$\times$} & \multirow{3}{*}{5$\times$}\\
  & Claude Sonnet 4.6   & \$3.00   & \$0.30    & \$15.00  & &\\
  & Claude Haiku 4.5    & \$1.00   & \$0.10    & \$5.00   &  &\\

\midrule

\multirow{4}{*}{\textbf{OpenAI}}
  & GPT-5.5             & \$5.00   & \$1.25    & \$30.00  &  \multirow{4}{*}{0.10--0.25$\times$} & \multirow{4}{*}{6$\times$}\\
  & GPT-5.4             & \$2.50   & \$0.25    & \$15.00  &  &\\
  & GPT-5.4 Mini        & \$0.75   & \$0.075   & \$4.50   &  &\\
  & GPT-5.4 Nano        & \$0.20   & \$0.02    & \$1.25   &  &\\

\midrule

\multirow{4}{*}{\textbf{Google}}
  & Gemini 3.1 Pro      & \$2.00   & \$0.20    & \$12.00  &  \multirow{4}{*}{0.10$\times$} & \multirow{4}{*}{6--8$\times$} \\
  & Gemini 3.1 Flash    & \$0.25   & \$0.025    & \$1.50   &  & \\
  & Gemini 2.5 Pro      & \$1.25   & \$0.125   & \$10.00  & & \\
  & Gemini 2.5 Flash    & \$0.30   & \$0.03    & \$2.50   &  & \\

\midrule

\multirow{2}{*}{\textbf{DeepSeek}}
  & V4 Pro    & \$1.74   & \$0.0145  & \$3.48   & \multirow{2}{*}{$\sim$0.01$\times$} & \multirow{2}{*}{2$\times$} \\
  & V4 Flash            & \$0.14   & \$0.0014  & \$0.28   &  &\\

\midrule

\multirow{2}{*}{\textbf{Kimi}}
  & Kimi K2.6           & \$0.74   & \$0.185   & \$3.49   & \multirow{2}{*}{0.25$\times$} & \multirow{2}{*}{4--5$\times$} \\
  & Kimi K2.5           & \$0.60   & \$0.15    & \$2.50   &  \\

\midrule

\textbf{Industry avg.} & & \textbf{1$\times$} & & & $\sim$\textbf{0.1$\times$}  & $\sim$\textbf{5$\times$}  \\

\bottomrule
\end{tabular}
}
\label{tab:token_cost}
\end{table}

\paragraph{Container backend: podman replacing Docker.} Our cluster has no Docker daemon and no \texttt{/etc/subuid} entries for the user, so we run all SWE-bench eval images under rootless podman~\citep{podman}. Two consequences flow from the missing subuid range: (i) podman's namespace is single-UID, so the host UID maps to container UID 0 and nothing else is valid; (ii) the upstream swebench harness's \texttt{copy\_to\_container} tars files with the host UID and calls \texttt{put\_archive}, which podman rejects with \texttt{lchown ... invalid argument}. We patch \texttt{swebench.harness.docker\_utils.copy\_to\_container} to force \texttt{uid=gid=0} in the tarinfo filter; the same patch is applied to every fresh swebench install in the eval venv. The harness reaches podman via \texttt{DOCKER\_HOST=unix:///\ldots/podman.sock} (\texttt{podman system service --time=0}); the OpenHands adapter shells out to \texttt{podman run/exec} directly and does not use the API socket.

\subsection{Terminal-Bench v2 Harness}
\label{app:terminalbench-setup}

Terminal-Bench v2~\citep{terminalbench} evaluates shell-tool agents on long-horizon command-line workflows. We run the official \texttt{openhands} reference agent against the \texttt{tb2-zai} dataset~\citep{tb2zai} on Daytona cloud sandboxes; Table~\ref{tab:app_terminalbench_protocol} records the harness configuration used for every row of Table~\ref{tab:terminalbench_results}.

\begin{table}[!htbp]
\centering
\small
\caption{Terminal-Bench v2 scaffold, sandbox, and reporting configuration used for all rows of Table~\ref{tab:terminalbench_results}.}
\label{tab:app_terminalbench_protocol}
\begin{tabular}{ll}
\toprule
Item & Setting \\
\midrule
Dataset & \texttt{tb2-zai}, all 89 tasks (89-task denominator) \\
Excluded tasks & \texttt{pytorch-model-cli}, \texttt{count-dataset-tokens}, \texttt{mcmc-sampling-stan}, \\
                & \texttt{rstan-to-pystan}, \texttt{reshard-c4-data} \\
Reporting denominator & 89 (matches public Terminal-Bench leaderboard) \\
Agent scaffold & official \texttt{openhands} reference agent, standard online, in-sandbox \\
Attempts per task ($k$) & 5 \\
Sampling & temperature 0.6, top\_p 0.8, top\_k 20 \\
Schedule & longest-first \\
Concurrency & 30B: 20 trials in parallel; 80B: 24 trials in parallel \\
Sandbox runtime & Daytona cloud sandboxes \\
Watchdog & 300\,s sweep interval, 75\,min sandbox age cap \\
Serving & vLLM, TP$=4$, bf16, $4\times$H100 80\,GB, $131{,}072$ ctx, prefix caching on \\
Tool/reasoning parsers & \texttt{--tool-call-parser hermes}; \texttt{--reasoning-parser qwen3}, Thinking only \\
Daytona unit pricing & 1 vCPU \$0.0504/hr; mem \$0.0162/hr/GiB; disk \$0.000108/hr/GiB (5\,GiB free) \\
Default sandbox spec & 1 vCPU / 2\,GiB / 10\,GiB ($\sim$80\% of trials) $\to$ \$0.08334/hr per sandbox \\
Observed spec mix & $\sim$80\% \texttt{1c/2g/10d}; $\sim$16\% \texttt{1c/4g/10d}; $\sim$4\% \texttt{2c/4g/10d} or \texttt{1c/8g/10d} \\
\bottomrule
\end{tabular}
\end{table}

\paragraph{Reporting denominator.} The five excluded tasks fail to launch reliably under our default Daytona sandbox spec budget. Each excluded task is counted as failed for every model, preserving the 89-task denominator. This matches the Terminal-Bench leaderboard convention.

\paragraph{Daytona cost accounting.} Daytona sandboxes are the only billed component of Terminal-Bench v2. We pull per-sandbox lifetimes from the audit-log API (\texttt{/api/audit/organizations/\{orgId\}}), which records every \texttt{create} with its cpu/mem/disk spec and every \texttt{delete} timestamp, and cost each sandbox at the per-spec rate in Table~\ref{tab:app_terminalbench_protocol}. This accounting includes sandbox boot, teardown, and retries, and matches the Daytona dashboard. The 30B sweep contains 3{,}925 billable sandbox creations.

\paragraph{Reasoning-parser configuration on Thinking.} Without \texttt{--reasoning-parser qwen3}, vLLM serves Thinking-checkpoint outputs with \texttt{<think>} blocks landing in the assistant \texttt{content} field, which then accumulates into next-turn prompts and inflates input-token traffic. Every Thinking row in Table~\ref{tab:terminalbench_results} uses the parser-enabled setting.

\paragraph{Token cost.} Terminal-Bench v2 rows report TTC as defined in Appendix~\ref{app:swe-bench-details}. Appendix~\ref{app:tb-headline-full} lists TTC together with the Daytona cost.

\subsection{Baseline Hyperparameters}
\label{app:baseline-hparams}

Baseline rows use the method's paper setting when it fixes the relevant value; otherwise we report the best completed Roo-Eval configuration available for that method at the corresponding scale. Table~\ref{tab:app_baseline_hparams} lists the selected settings used in the main tables.

\begin{table}[!htbp]
\centering
\scriptsize
\caption{Selected baseline hyperparameters for the Roo-Eval results.}
\label{tab:app_baseline_hparams}
\begin{adjustbox}{max width=\textwidth}
\begin{tabular}{llll}
\toprule
Method & 30B setting & 80B setting & Selection note \\
\midrule
Task Arithmetic & $\alpha=0.30$ & $\alpha=0.15$ & Best completed Roo-Eval setting \\
TIES & $\alpha=0.30$, density $=0.50$ & $\alpha=0.15$, density $=0.50$ & Best completed Roo-Eval setting \\
SLERP & $t=0.30$ & $t=0.15$ & Best completed Roo-Eval setting \\
AIM-TA & $\alpha=0.30$, $\omega=0.40$ & $\alpha=0.15$, $\omega=0.40$ & AIM weighting applied to Task Arithmetic \\
AIM-TIES & $\alpha=0.30$, density $=0.50$, $\omega=0.40$ & $\alpha=0.15$, density $=0.50$, $\omega=0.40$ & AIM weighting applied to TIES \\
LEWIS & $\alpha=0.30$, $\gamma=0.30$, $\epsilon=0.80$, density $=0.50$ & $\alpha=0.15$, $\gamma=0.30$, $\epsilon=0.80$, density $=0.50$ & Importance-weighted density schedule \\
RAIN & Reproduction, Thinking proxy base, scaling factor $0.50$ & Reproduction, Thinking proxy base, scaling factor $0.30$ & Reverse-direction diagnostic baseline \\
\bottomrule
\end{tabular}
\end{adjustbox}
\end{table}

\subsubsection{AIM Variants}
\label{app:aim-baseline-details}
AIM is implemented as a channel-wise relaxation on the update produced by another merge rule. For a Linear weight $W_q\in\mathbb{R}^{d_{\mathrm{out}}\times d_{\mathrm{in}}}$, let $m_q\in\mathbb{R}_{\ge 0}^{d_{\mathrm{in}}}$ be the input-channel activation magnitude recorded on the Instruct checkpoint and let
\begin{equation}
s_{q,j}=\frac{m_{q,j}}{\max_{j'}m_{q,j'}},
\qquad
r_{q,j}=1-(1-\omega)s_{q,j},
\qquad \omega=0.40 ,
\end{equation}
when $\max_{j'}m_{q,j'}>0$; otherwise the AIM scaler leaves the update unchanged.
The AIM-adjusted update is applied column-wise,
\begin{equation}
\widetilde{\Delta}_{q,:,j}=r_{q,j}\Delta_{q,:,j}.
\end{equation}
Thus channels that are highly activated by the Instruct model are protected by shrinking the merge update toward an $\omega$ fraction, while low-importance channels keep nearly the full update. AIM-TA sets $\Delta_q=\alpha(\theta_{\mathrm{think},q}-\theta_{\mathrm{inst},q})$. AIM-TIES first computes the usual TIES update after trimming, sign election, and disjoint averaging at density $0.50$, and then applies the same AIM relaxation to the final $\alpha$-scaled update. Biases, embeddings, layer norms, rotary buffers, and Linear weights without a matching AIM importance vector are left unchanged by the AIM post-processing step.

\subsection{Failure-Mode Analysis}
\label{app:failure-analysis}

The failure-mode distribution panel in Figure~\ref{fig:qual_trace} reports rule-based audits of failed Roo-Eval rollouts on 30B for three model variants. The Instruct-side 3-class taxonomy serves as the primary axis; Thinking and \method{} failures are mapped onto it (see the audits below). The audit classifies reasoning failures; rollouts that failed because the harness rejected an \texttt{apply\_diff} payload are excluded.

\paragraph{30B-Instruct audit (303 failed rollouts).} One run per language: Python 52, JavaScript 64, Go 72, Java 57, Rust 58. Each failed rollout is bucketed by parsing its JSONL tool-use stream and applying:
\begin{itemize}\setlength{\itemsep}{2pt}
    \item \textbf{over-terse}: $\le 6$ finalized tool events or $\le 1$ test cycle. The agent converges prematurely without producing an implementation attempt.
    \item \textbf{context-blind}: $\ge 2$ edits with $\le 1$ read, or no read of the test file before editing. The agent fires edits before inspecting the specification scaffold.
    \item \textbf{no-self-reflection}: $\ge 3$ test runs with repeated failure signatures, or $\ge 3$ commands $+\ge 3$ edits. The agent repeats the same approach across multiple failed attempts.
\end{itemize}
Counts: over-terse 50, context-blind 30, no-self-reflection 223.

\paragraph{30B-Thinking audit (371 failed rollouts).} Thinking-native rule labels are mapped to the 3-class taxonomy:
\begin{itemize}\setlength{\itemsep}{2pt}
    \item \textbf{over-terse}: $\le 1$ test cycle (Thinking-native: \emph{premature-end}; budget exhausts at the 900\,s timeout without a productive edit$\to$test cycle).
    \item \textbf{no-self-reflection}: a single \verb|</think>|-bounded inner-monologue block $\ge 20$k chars, OR think text $\ge 50\%$ of total assistant output and total think $\ge 30$k chars (Thinking-native: \emph{monolithic-think}; counts as no-self-reflection because the rollout never alternates between deliberation and tool feedback).
\end{itemize}
Counts: over-terse 180, context-blind 25, no-self-reflection 166.

\paragraph{30B-\method{} audit (100 failed rollouts).} Same 3-class scheme applied. Counts: over-terse 38, context-blind 12, no-self-reflection 50. This is 67\% fewer reasoning failures than Instruct and 73\% fewer than Thinking. The largest reduction is in no-self-reflection; over-terse failures fall well below Thinking and context-blind failures below Instruct, and no \texttt{<think>} blocks appear in any \method{} log.

\paragraph{Over-terse exemplar.} \texttt{python-transpose-iter3-attempt4.log}. 
The agent reads the stub and the test file, then switches to \emph{architect} mode and asks a clarifying question about trailing-space handling rather than implementing the function:

\begin{quote}\small\ttfamily
listFilesRecursive docs $\to$ readFile transpose.py $\to$ \\
readFile transpose\_test.py $\to$ switchMode architect $\to$ \\
ask\_followup\_question("Should the function handle trailing spaces \ldots")
\end{quote}

\noindent The trace contains no edit or test execution. Although the test file specifies the expected behavior, the rollout terminates before implementation.

\paragraph{Context-blind exemplar.} \texttt{javascript-forth-iter1-attempt3.log}. The agent reads only the stub \texttt{forth.js} and never opens \texttt{forth.spec.js}; it then makes three edits guessing the API before running tests for the first time:

\begin{quote}\small\ttfamily
readFile forth.js \\
appliedDiff forth.js (constructor) \\
appliedDiff forth.js (get stack) \\
appliedDiff forth.js (evaluate) \\
execute\_command pnpm test \quad \# forth.spec.js never opened
\end{quote}

\noindent This trace violates the read-before-edit criterion: the specification file defines the API, but the generated implementation is based only on the stub.

\paragraph{No-self-reflection exemplar.} \texttt{python-zipper-iter3-attempt3.log}. After an initial failing test run, the agent applies a near-identical edit to \texttt{zipper.py}'s \texttt{to\_tree} method four consecutive times, each followed by an identical pytest signature:

\begin{quote}\small\ttfamily
EDIT zipper.py (set\_left)             FAIL .....FFFFFF..F \\
EDIT zipper.py (to\_tree, identical)   FAIL .....FF.FFF..F \\
EDIT zipper.py (to\_tree, identical)   FAIL .....FF.FFF..F \\
EDIT zipper.py (to\_tree, identical)   FAIL .....FF.FFF..F \\
\ldots\ \ 12 test cycles, signature unchanged after the first
\end{quote}

\noindent Across 12 test cycles, the failure signature remains unchanged; the trace contains no subsequent test reread, diagnostic instrumentation, or alternative implementation attempt.

\subsection{Additional Qualitative Trace Triples}
\label{app:qual-trace-extra}

Figure~\ref{fig:qual_trace} reports a single triple on \texttt{python-scale-generator}. The two additional triples below were chosen for the same property (Instruct fails, Thinking fails, \method{} succeeds on iter1) and exhibit different but consistent failure modes.

\paragraph{\texttt{javascript-parallel-letter-frequency}.}
\begin{itemize}\setlength{\itemsep}{2pt}
    \item \textbf{Instruct} (\texttt{javascript-parallel-letter-frequency-iter1.log}): 20 tool calls, \emph{zero} edits. The trace contains 14 consecutive \texttt{searchFiles} calls with an empty regex and no edits before the harness emits \texttt{Roo appears to be stuck in a loop}.
    \item \textbf{Thinking} (\texttt{javascript-parallel-letter-frequency-iter1.log}): 12 tool calls but with 47k characters of inner monologue between attempts; four separate \texttt{appliedDiff} revisions on the same Unicode-aware regex regress from 1 failing test to 8 failing tests, then time out.
    \item \textbf{\method{}} (\texttt{javascript-parallel-letter-frequency-iter1.log}): single shot, 7 tools: \texttt{list\_files} $\to$ \texttt{list\_files} $\to$ \texttt{read\_file parallel-letter-frequency.js} $\to$ \texttt{read\_file parallel-letter-frequency.spec.js} $\to$ \texttt{appliedDiff} $\to$ \texttt{pnpm install} $\to$ \texttt{pnpm test} (PASS, all tests). 305\,s, 4k output tokens.
\end{itemize}

\paragraph{\texttt{javascript-tournament}.}
\begin{itemize}\setlength{\itemsep}{2pt}
    \item \textbf{Instruct} (\texttt{javascript-tournament-iter1.log}): 21 tool calls, 6 edits, 4 test runs without convergence; 38k output tokens, 900\,s timeout.
    \item \textbf{Thinking} (\texttt{javascript-tournament-iter1.log}): 8 tool calls dominated by 112k characters of inner monologue, 2 edits, 2 test runs, no recovery; 40k output tokens.
    \item \textbf{\method{}} (\texttt{javascript-tournament-iter1.log}): 9 tool calls in a single attempt; key steps: \texttt{list\_files} $\to$ \texttt{read\_file} stub $\to$ \texttt{read\_file} spec $\to$ short todo $\to$ \texttt{appliedDiff} $\to$ \texttt{pnpm test} (PASS). 79\,s, 2.1k output tokens.
\end{itemize}

\noindent The pattern in both triples mirrors Figure~\ref{fig:qual_trace}: Instruct either edits without reading the specification or repeatedly invokes search tools; Thinking allocates most output tokens to inner monologue; \method{} reads the test/spec file before the first edit and converges in one or two cycles.

\section{Calibration and Signal Computation}
\label{app:calibration-signal}

The reported recipe uses the paper calibration set below. The public-source subsets in Appendix~\ref{app:calibration-robustness} are used for the calibration-set robustness analysis.

\subsection{Calibration Set Construction}
\label{app:calibration-construction}

CTG uses model-generated behavior targets. The Thinking checkpoint supplies reasoning-transfer targets and the Instruct checkpoint supplies agent-behavior preservation targets. Table~\ref{tab:app_calibration} summarizes the calibration inputs: CTG uses $\mathcal{D}_R$ and $\mathcal{D}_A$ as masked-loss sets, and GSP uses the format-trace set $\mathcal{D}_F$ to build activation projectors.

\begin{table}[!htbp]
\centering
\small
\caption{Calibration inputs used by CTG and GSP. CTG uses $\mathcal{D}_R$ and $\mathcal{D}_A$ as masked-loss sets; $\mathcal{D}_F$ provides format traces for GSP.\label{tab:app_calibration}}
\begin{adjustbox}{max width=\textwidth}
\begin{tabular}{lllll}
\toprule
Set & Size & Construction & Target generator & Role \\
\midrule
$\mathcal{D}_R$ & 36 & Original code-agent reasoning prompts: 20 SWE-bench-style, 12 LiveBench-coding-style, 4 LiveCodeBench-style & Thinking & Reasoning-transfer loss \\
$\mathcal{D}_A$ & 16 & Original Roo-style tool-use repair prompts: 14 SWE-bench-style, 2 LiveBench-coding-style & Instruct & Agent-behavior preservation loss \\
$\mathcal{D}_F$ format & 430 & Instruct traces around format-critical tool tokens and local neighborhoods & Instruct & Format activations for GSP; no loss \\
\bottomrule
\end{tabular}
\end{adjustbox}
\end{table}

\paragraph{Reasoning-transfer set $\mathcal{D}_R$.}
The paper calibration set contains 36 $\mathcal{D}_R$ prompts. They are original rewrites in code-agent reasoning styles inspired by SWE-bench~\citep{jimenez2024swebench}, LiveBench coding~\citep{white2024livebench}, and LiveCodeBench~\citep{jain2024livecodebench}. They cover debugging, concurrency, migrations, caching, pagination, parser edge cases, large backfills, rate limiting, pathfinding, and test-design tradeoffs. Each prompt is rendered as a user message; the Thinking checkpoint greedily generates the assistant target. The masked loss is then evaluated at the Instruct endpoint on the generated assistant span.

\paragraph{Agent-behavior set $\mathcal{D}_A$.}
The paper calibration set contains 16 $\mathcal{D}_A$ prompts. They are original Roo-style repository repair instructions. They ask the model to inspect relevant files, patch the smallest correct change, run focused tests, audit scripts or docs, and report intentional non-edits. The Instruct checkpoint generates the preservation target. This set activates the same tool-use and response-format behavior that must be preserved when injecting Thinking-derived deltas.

\paragraph{Format-trace set $\mathcal{D}_F$.}
The 430 format traces are used only by GSP. We locate format-token positions and local neighborhoods in Instruct traces, collect hidden states at the protected sites, and build per-component spectral projectors.

\subsection{Taylor Signal Computation}
\label{app:taylor-signal-computation}

\begin{figure}[!htbp]
  \centering
  \includegraphics[width=0.95\linewidth]{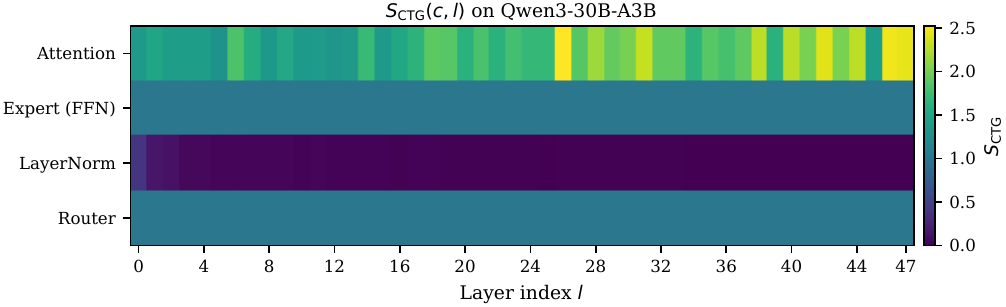}
  \caption{CTG Taylor importance $S_{\mathrm{CTG}}(c,l)$ on Qwen3-30B-A3B, derived automatically from $\mathcal{D}_R$ and $\mathcal{D}_A$ in Table~\ref{tab:app_calibration}. Rows: attention, expert (FFN), layer-norm, and router blocks; columns: layers 0--47. Scores are in expert units, so the expert (FFN) row is the per-layer anchor and equals 1 by construction. Attention receives the largest scale, peaking in late layers; layer norms receive near-zero injection.}
  \label{fig:app_taylor_heatmap}
\end{figure}

For each coordinate $j$, let $\delta_j=\theta_{\mathrm{think},j}-\theta_{\mathrm{inst},j}$. At the Instruct endpoint, we compute gradients of the masked reasoning and agent-behavior losses:
\begin{equation}
g_R=\nabla_\theta \mathcal{L}_R(\theta_{\mathrm{inst}}),
\qquad
g_A=\nabla_\theta \mathcal{L}_A(\theta_{\mathrm{inst}}).
\end{equation}
The equations are written over the full parameter vector, but the implementation computes them shardwise: each shard stores its local entries of $g_R$, $g_A$, and $\delta$, forms local coordinate scores, and contributes the relevant block sums.
The signed first-order improvements along the actual merge direction are
\begin{equation}
s_R(j)=-g_{R,j}\delta_j,
\qquad
s_A(j)=-g_{A,j}\delta_j.
\end{equation}
The Conservative Taylor Gate (CTG) gives a positive score to a coordinate only when the same infinitesimal edit is beneficial for both objectives:
\begin{equation}
p_j=\left[\min\{s_R(j),s_A(j)\}\right]_+ .
\end{equation}
Component/layer scores are obtained by summing $p_j$ within a block, normalizing by the Instruct parameter norm of that block, and then reporting all components in expert units. A block with more CTG-positive coordinates can receive a larger aggregate score after Frobenius normalization. CTG acts at the block level: the final tensor update uses the thresholded delta $T(\delta^{(l,c)})$ scaled by the scalar $S_{\mathrm{CTG}}(c,l)$. The anchor is the per-layer FFN/expert pseudo-component $b$: dense FFN layers use the union of gate/up/down projections, while MoE layers use the union of gate/up/down projections across all expert replicas. The router is not part of this anchor. Figure~\ref{fig:app_taylor_heatmap} shows the resulting Qwen3-30B-A3B table.

\subsection{Robustness to Calibration Set Choice}
\label{app:calibration-robustness}

We assess the robustness of the CTG Taylor scores used by \method{} to calibration-set choice. On Qwen3-30B-A3B, we recompute the full layer-component score table under five independently sampled public calibration subsets, while holding the model pair, target decoding protocol ($T_R{=}4096$, $T_A{=}2048$), layer chunking, and merge equations fixed. The analysis isolates calibration-set variation from the rest of the merge pipeline.

\paragraph{Public mix construction.}
Each \texttt{public\_mix\_seed\{s\}} subset has the same $36{+}16$ prompt budget as the paper calibration set. The frozen reasoning pool has 80 public code-reasoning prompts: 40 from LiveCodeBench code generation~\citep{jain2024livecodebench} and 40 from BigCodeBench~\citep{zhuo2024bigcodebench}. The frozen tool-use pool has 80 SWE-bench issue prompts~\citep{jimenez2024swebench}, excluding SWE-bench Verified instance ids~\citep{openai2024swebenchverified}, wrapped as Roo-style repository repair prompts. For seed $s$, a seeded Python RNG samples 18 LiveCodeBench prompts, 18 BigCodeBench prompts, and 16 SWE-bench prompts without replacement. Items are sorted by source and id before writing the JSONL, making the prompt hash deterministic.

\begin{table}[!htbp]
\centering
\scriptsize
\caption{Robustness to calibration-set choice on Qwen3-30B-A3B. Public mix seeds use 18 LiveCodeBench prompts, 18 BigCodeBench prompts, and 16 SWE-bench issue/tool prompts. Pearson/Spearman are computed over flattened layer-component scores for attention/router/norm against the paper calibration set.\label{tab:app_calibration_robustness}}
\begin{adjustbox}{max width=\textwidth}
\begin{tabular}{lrrrrrrrrrrr}
\toprule
Calibration & $|\mathcal{D}_R|/|\mathcal{D}_A|$ & Attention & Expert & Router & Norm & Pearson & Spearman & Top-10 & Top-20 & Top-30 & Top-48 \\
\midrule
paper calibration & 36/16 & 1.7912 & 1.0000 & 0.3225 & 0.0151 & 1.0000 & 1.0000 & 10/10 & 20/20 & 30/30 & 48/48 \\
public\_mix\_seed0 & 36/16 & 1.7904 & 1.0000 & 0.3378 & 0.0161 & 0.9862 & 0.9917 & 7/10 & 15/20 & 26/30 & 46/48 \\
public\_mix\_seed1 & 36/16 & 1.7706 & 1.0000 & 0.3399 & 0.0161 & 0.9868 & 0.9911 & 6/10 & 15/20 & 25/30 & 46/48 \\
public\_mix\_seed2 & 36/16 & 1.7908 & 1.0000 & 0.3423 & 0.0153 & 0.9856 & 0.9913 & 7/10 & 14/20 & 25/30 & 46/48 \\
public\_mix\_seed3 & 36/16 & 1.7720 & 1.0000 & 0.3447 & 0.0159 & 0.9853 & 0.9906 & 6/10 & 14/20 & 25/30 & 46/48 \\
public\_mix\_seed4 & 36/16 & 1.8066 & 1.0000 & 0.3349 & 0.0164 & 0.9877 & 0.9920 & 7/10 & 15/20 & 25/30 & 46/48 \\
\bottomrule
\end{tabular}
\end{adjustbox}
\end{table}

\begin{table}[!htbp]
\centering
\small
\caption{Dispersion of the five public mix seeds. CV is the coefficient of variation across seeds; drift is relative to the paper 36/16 calibration value.\label{tab:app_calibration_public_dispersion}}
\begin{tabular}{lrrrr}
\toprule
Component & Mean & Std. & CV & Drift vs.\ paper calibration \\
\midrule
attention & 1.7861 & 0.0150 & 0.0084 & -0.28\% \\
expert & 1.0000 & 0.0000 & 0.0000 & +0.00\% \\
router & 0.3399 & 0.0038 & 0.0112 & +5.39\% \\
norm & 0.0160 & 0.0004 & 0.0254 & +5.38\% \\
\bottomrule
\end{tabular}
\end{table}

\paragraph{Findings.}
The five public mix seeds preserve the same component ordering as the paper calibration set, attention $>$ expert baseline $>$ router $\gg$ norm. Their Pearson correlations against the paper calibration set are 0.9853--0.9877 and Spearman correlations are 0.9906--0.9920; the top-48 overlap is 46/48 for every public seed. Per-component variation is small: attention CV is 0.84\%, router CV is 1.12\%, and norm remains near zero. Thus, the layer-component score table used by the merge is insensitive to these calibration-set redraws at the level that determines component ordering and high-score layer selection.

\subsection{Sensitivity to Calibration Set Size}
\label{app:calibration-size}

Calibration only estimates the CTG scaling coefficients and the GSP projectors; no weight is optimized. We test how the size of each calibration set affects the merged model (Table~\ref{tab:app_calibration_size}). One set changes at a time, while the other set and all merge hyperparameters stay at the paper setting. For Stage~2, $|\mathcal{D}_R|/|\mathcal{D}_A|$ is 9/4, 18/8, 36/16, or 72/32 prompts (25\%, 50\%, 100\%, or 200\% of the paper budget). The 25\% and 50\% sets are sampled without replacement from the paper set, and the 200\% set adds prompts from LiveCodeBench~\citep{jain2024livecodebench} and BigCodeBench~\citep{zhuo2024bigcodebench}. For Stage~3, $|\mathcal{D}_F|$ is 107, 215, 430, or 860 traces with the paper's format composition. As a control, C-rand keeps the paper budgets but shuffles the $S_{\mathrm{CTG}}$ coefficients across layers within each component. Each merged model is evaluated with the main Roo-Eval protocol, and Table~\ref{tab:app_calibration_budget_runtime} gives the corresponding construction cost.

At 30B, the 25\% and 50\% Stage~2 sets reach 160 and 158 exercises at pass@3, and the 25\% and 50\% Stage~3 sets reach 156 and 159, against 162 for the paper checkpoint. Doubling either set does not help (155 and 153). C-rand drops to 147, so the estimated block structure is useful, while extra calibration samples add little. At 80B, all six settings reach 175--177 against 176 for the paper checkpoint. These results are consistent with how little each stage estimates from its calibration set: CTG aggregates token-level evidence into about 200 block coefficients, and GSP keeps only the leading directions shared across repeated format patterns. Their sample needs therefore depend on these statistics rather than on the parameter count, in line with the redraw stability in Appendix~\ref{app:calibration-robustness}.

\begin{table}[!htbp]
\centering
\small
\caption{Calibration-set size, Roo-Eval pass@3. One set varies at a time; the other stays at the paper budget (36/16 prompts, 430 traces). C-rand shuffles the $S_{\mathrm{CTG}}$ coefficients across layers within each component. Cells report solved exercises (\%); language columns give the number of exercises.}
\label{tab:app_calibration_size}
\begin{adjustbox}{max width=\linewidth}
\begin{tabular}{lcccccc}
\toprule
$|\mathcal{D}_R|/|\mathcal{D}_A|$, $|\mathcal{D}_F|$ & Python (34) & JavaScript (50) & Go (36) & Java (45) & Rust (30) & \textbf{Overall} (195) \\
\midrule
\multicolumn{7}{l}{\textit{Qwen3-30B-A3B}} \\
36/16, 430 (paper checkpoint) & 31 (91.2) & 42 (84.0) & 30 (83.3) & 37 (82.2) & 22 (73.3) & 162 (83.1) \\
9/4, 430 (25\% $\mathcal{D}_R/\mathcal{D}_A$) & 31 (91.2) & 42 (84.0) & 30 (83.3) & 35 (77.8) & 22 (73.3) & 160 (82.1) \\
18/8, 430 (50\%) & 28 (82.4) & 44 (88.0) & 28 (77.8) & 36 (80.0) & 22 (73.3) & 158 (81.0) \\
72/32, 430 (200\%) & 32 (94.1) & 41 (82.0) & 30 (83.3) & 31 (68.9) & 21 (70.0) & 155 (79.5) \\
36/16, 107 (25\% $\mathcal{D}_F$) & 28 (82.4) & 41 (82.0) & 30 (83.3) & 34 (75.6) & 23 (76.7) & 156 (80.0) \\
36/16, 215 (50\%) & 29 (85.3) & 41 (82.0) & 30 (83.3) & 37 (82.2) & 22 (73.3) & 159 (81.5) \\
36/16, 860 (200\%) & 27 (79.4) & 39 (78.0) & 30 (83.3) & 34 (75.6) & 23 (76.7) & 153 (78.5) \\
36/16, 430 (C-rand) & 28 (82.4) & 40 (80.0) & 25 (69.4) & 33 (73.3) & 21 (70.0) & 147 (75.4) \\
\midrule
\multicolumn{7}{l}{\textit{Qwen3-Next-80B-A3B}} \\
36/16, 430 (paper checkpoint) & 33 (97.1) & 49 (98.0) & 33 (91.7) & 37 (82.2) & 24 (80.0) & 176 (90.3) \\
9/4, 430 (25\% $\mathcal{D}_R/\mathcal{D}_A$) & 31 (91.2) & 48 (96.0) & 34 (94.4) & 36 (80.0) & 28 (93.3) & 177 (90.8) \\
18/8, 430 (50\%) & 32 (94.1) & 48 (96.0) & 33 (91.7) & 36 (80.0) & 26 (86.7) & 175 (89.7) \\
72/32, 430 (200\%) & 34 (100.0) & 49 (98.0) & 35 (97.2) & 34 (75.6) & 25 (83.3) & 177 (90.8) \\
36/16, 107 (25\% $\mathcal{D}_F$) & 31 (91.2) & 48 (96.0) & 33 (91.7) & 36 (80.0) & 29 (96.7) & 177 (90.8) \\
36/16, 215 (50\%) & 30 (88.2) & 48 (96.0) & 34 (94.4) & 36 (80.0) & 27 (90.0) & 175 (89.7) \\
36/16, 860 (200\%) & 32 (94.1) & 47 (94.0) & 33 (91.7) & 39 (86.7) & 26 (86.7) & 177 (90.8) \\
\bottomrule
\end{tabular}
\end{adjustbox}
\end{table}

\subsection{Runtime and Artifacts}
\label{app:signal-runtime}

\begin{table}[!htbp]
\centering
\small
\caption{Measured \method{} signal-computation and merge runtimes. Rows report wall-clock time on the listed hardware. GSP projector construction is a one-time reusable cost.}
\label{tab:app_signal_runtime}
\begin{tabular}{p{0.72\linewidth}r}
\toprule
Stage & Wall time \\
\midrule
30B instruct model load on 2$\times$H100 & $\sim$28 s \\
30B Taylor signal on 2$\times$H100 & $\sim$6 min \\
30B GSP projector build, 96 hidden-state components on 2$\times$H100 & $\sim$2.2 min \\
30B final merge on one H100, 16 shards & $\sim$4 min \\
30B end-to-end signal to merged model, reusing GSP projectors & $\sim$10 min \\
30B end-to-end including GSP projector rebuild & $\sim$13 min \\
\midrule
80B instruct model load on 4$\times$H100 & $\sim$30 s \\
80B Taylor signal on 4$\times$H100  & $\sim$27 min \\
80B GSP projector build, 96 hidden-state components on 4$\times$H100 & $\sim$13 min \\
80B final merge on one H100, 41 shards & 461.7 s ($\sim$7.7 min) \\
80B end-to-end signal to merged model, reusing GSP projectors & $\sim$35 min \\
80B end-to-end including GSP projector rebuild & $\sim$48 min \\
\bottomrule
\end{tabular}
\end{table}

\noindent These are one-time costs. GSP projector construction can be reused across nearby merge-scale sweeps for the same Instruct endpoint and format-trace set, and the Taylor-signal and elementwise-merge steps are naturally shardable.

\paragraph{Cost across calibration budgets.} Table~\ref{tab:app_calibration_budget_runtime} reports how the one-time cost scales with the calibration sets, for the settings of Appendix~\ref{app:calibration-size}. The 30B measurements use one H100, unlike the two H100s in Table~\ref{tab:app_signal_runtime}. At the paper budget, the larger Qwen3-235B-A22B pair (Instruct-2507 and Thinking-2507) takes 77\,min for the Taylor signal, 35\,min for GSP, and 20.8\,min for the merge on 4$\times$H100, or 132.8\,min end to end.

\begin{table}[!htbp]
\centering
\small
\caption{One-time \method{} cost in minutes across calibration budgets. Taylor = Stage~2 signal, GSP = Stage~3 projector build, Merge = final merge, Total = end to end. An asterisk marks a stage whose artifact was reused from the paper-budget build rather than recomputed; its listed time is that build's time.}
\label{tab:app_calibration_budget_runtime}
\begin{adjustbox}{max width=\linewidth}
\begin{tabular}{lrrrrrrrr}
\toprule
& \multicolumn{4}{c}{\textbf{Qwen3-30B-A3B} (1$\times$H100)} & \multicolumn{4}{c}{\textbf{Qwen3-Next-80B-A3B} (4$\times$H100)} \\
\cmidrule(lr){2-5} \cmidrule(lr){6-9}
$|\mathcal{D}_R|/|\mathcal{D}_A|$, $|\mathcal{D}_F|$ & Taylor & GSP & Merge & Total & Taylor & GSP & Merge & Total \\
\midrule
9/4, 430 (25\% $\mathcal{D}_R/\mathcal{D}_A$) & 7 & 2.2$^*$ & 4.1 & 13.3 & 12 & 13$^*$ & 7.7 & 32.7 \\
18/8, 430 (50\%) & 11 & 2.2$^*$ & 4.1 & 17.3 & 19 & 13$^*$ & 7.7 & 39.7 \\
36/16, 430 (paper) & 19 & 2.2$^*$ & 4.1 & 25.3 & 27$^*$ & 13$^*$ & 7.7 & 47.7 \\
72/32, 430 (200\%) & 28 & 2.2$^*$ & 4.1 & 34.3 & 47 & 13$^*$ & 7.7 & 67.7 \\
\midrule
36/16, 107 (25\% $\mathcal{D}_F$) & 19 & 1.0 & 4.1 & 24.1 & 27$^*$ & 5.2 & 7.7 & 39.9 \\
36/16, 215 (50\%) & 19 & 1.4 & 4.1 & 24.5 & 27$^*$ & 7.2 & 7.7 & 41.9 \\
36/16, 860 (200\%) & 19 & 3.7 & 4.1 & 26.8 & 27$^*$ & 19 & 7.7 & 53.7 \\
\bottomrule
\end{tabular}
\end{adjustbox}
\end{table}

\subsection{Construction Cost Compared with Baselines}
\label{app:construction-cost}

Table~\ref{tab:app_construction_cost} compares the one-time cost of producing a deployable checkpoint. \method{} computes one backward pass over the calibration sets but runs no optimization loop: it takes 25.3\,min on one H100 at 30B and 47.7\,min on four H100s at 80B. This lies within the range of the gradient-free merges AIM~\citep{aim2024}, LEWIS~\citep{lewis2024}, and RAIN~\citep{huang2026rainmerging} (18--147\,min) and below Fisher-weighted merging~\citep{matena2022fisher} and AdaMerging~\citep{yang2024adamerging}, while QLoRA distillation~\citep{dettmers2023qlora} takes 7.7--10.1\,h per 30B model.

\begin{table}[!htbp]
\centering
\small
\caption{One-time construction cost, from released checkpoints on disk to a deployable checkpoint on disk, on H100 80GB GPUs. \method{} includes CTG statistics, GSP projectors, and the final merge. QLoRA rows fine-tune Qwen3-30B-A3B-Instruct on about 800 passed agent turns (NF4, rank 32, three epochs) and include the adapter merge; they exclude teacher-rollout generation (3.66 GPU-hours, shared by the three runs).}
\label{tab:app_construction_cost}
\begin{adjustbox}{max width=\linewidth}
\begin{tabular}{llcrrr}
\toprule
Method & Scale & Backward pass & Wall clock (min) & Peak GPU memory (GB) & GPUs \\
\midrule
\method{} & 30B & Yes & 25.3 & 57 & 1 \\
\method{} & 80B & Yes & 47.7 & 161 & 4 \\
\midrule
AIM-TA & 30B & No & 18 & 7 & 1 \\
AIM-TA & 80B & No & 40 & 10 & 1 \\
AIM-TIES & 30B & No & 20.2 & 8 & 1 \\
AIM-TIES & 80B & No & 62 & 12 & 1 \\
LEWIS & 30B & No & 28 & 11 & 1 \\
LEWIS & 80B & No & 69.7 & 23 & 1 \\
RAIN & 30B & No & 90 & 61 & 1 \\
RAIN & 80B & No & 147 & 200 & 4 \\
\midrule
Fisher-weighted merge & 30B & Yes & 32 & 66 & 1 \\
Fisher-weighted merge & 80B & Yes & 161 & 239 & 4 \\
AdaMerging (layer-wise) & 30B & Yes & 120 & 183 & 2 \\
\midrule
QLoRA SFT, keepthink & 30B & Yes & 605 & 81 & 1 \\
QLoRA SFT, stripthink & 30B & Yes & 464 & 81 & 1 \\
QLoRA SFT, self-direct & 30B & Yes & 587 & 81 & 1 \\
\bottomrule
\end{tabular}
\end{adjustbox}
\end{table}

\section{Architecture-Normalized Taylor}
\label{app:archnorm}

This section gives the derivation behind the hybrid-MoE normalization used for the Qwen3-Next-80B-A3B recipe. CTG keeps its layer/component granularity, and architecture families supply only an exposure correction. Let $f=\phi(c)$ map a raw parameter component $c$ to its architecture family, such as full-attention, linear-attention, experts, norms, or routers. The Qwen3-Next-80B-A3B recipe replaces the main coefficient by
\begin{equation}
\label{eq:app_arch_score}
S_{\mathrm{CTG}}^{\mathrm{arch}}(c,l)
=
\frac{1}{\kappa(\phi(c))}
\cdot
\frac{\sum_{j\in\mathcal{B}_{c,l}} p_j
}{\sum_{j\in\mathcal{B}_{b,l}} p_j}
\cdot
\frac{\left\|\theta_{\mathrm{inst}}^{(b,l)}\right\|_F}
{\left\|\theta_{\mathrm{inst}}^{(c,l)}\right\|_F}.
\end{equation}
Here $b$ is the per-layer FFN/expert pseudo-component defined in the main text: the union of gate/up/down projections, across all expert replicas for MoE layers, excluding the router. Q/K/V/O projections, routers, and expert projections keep their own CTG evidence and parameter-norm normalization; the family map only sets the residual-occupation multiplier $\kappa$. With $\kappa(\phi(c))\equiv 1$, Eq.~\ref{eq:app_arch_score} recovers the main-text coefficient.

\subsection{Residual Occupation Measure}

Consider a residual transformer block whose token mixer in layer $l$ has family $f_l$:
\begin{equation}
h_{l+1}
=
h_l + M_{f_l,l}(h_l) + E_l(h_l),
\qquad
f_l\in\{\mathrm{full},\mathrm{linear}\},
\end{equation}
where $M_{f_l,l}$ is the attention or linear-state mixer and $E_l$ denotes the remaining expert/MLP branch. This residual-stack view is consistent with the continuous-depth interpretation of residual networks as ODE discretizations~\citep{chen2018neuralode,ablin2022resnetsode}.

Let a merge induce a small mixer perturbation $\Delta M_{f_l,l}$. If $e_l$ is the hidden-state error between the original and merged networks at layer $l$, then first-order linearization gives
\begin{equation}
\label{eq:app_error_recursion}
e_{l+1}
=
(I+J_l)e_l + \Delta M_{f_l,l}(h_l)
+ O(\|e_l\|^2+\|e_l\|\,\|\Delta M_{f_l,l}\|),
\end{equation}
where $J_l=\partial(M_{f_l,l}+E_l)/\partial h_l$. Dropping higher-order terms and unrolling,
\begin{equation}
\label{eq:app_error_unroll}
e_L
\approx
\sum_l
\mathcal{P}_{L,l+1}\,
\Delta M_{f_l,l}(h_l),
\qquad
\mathcal{P}_{L,l+1}
=
\prod_{m=l+1}^{L-1}(I+J_m).
\end{equation}
Thus the endpoint perturbation contributed by a mixer family is a sum over the layers in which that family appears. If the transported perturbations are bounded by a comparable layerwise scale $a_f$ for family $f$, then
\begin{equation}
\label{eq:app_family_budget}
B(f)
\equiv
\left\|\sum_{l:f_l=f}\mathcal{P}_{L,l+1}\Delta M_{f,l}(h_l)\right\|
\lesssim
\Lambda\,\mu(f)\,a_f,
\qquad
\mu(f)=\sum_l \mathbf{1}\{f_l=f\},
\end{equation}
for a transport bound $\|\mathcal{P}_{L,l+1}\|\le \Lambda$. The linear dependence on $\mu(f)$ models coherent accumulation, which fits a directed model edit such as the Instruct-to-Thinking delta.

\subsection{Full Attention Versus Linear Attention}

A causal full-attention mixer has the form
\begin{equation}
\label{eq:app_full_attn}
M_{\mathrm{full},l}(h)_t
=
W^O_l
\sum_{s\le t}
\mathrm{softmax}
\left(
\frac{q_{l,t}k_{l,s}^{\top}}{\sqrt{d}}
\right)_s
v_{l,s}.
\end{equation}
A Gated DeltaNet-style linear mixer can be abstracted as a recurrent state-space operator,
\begin{align}
\label{eq:app_linear_state}
S_{l,t}
&=
\Gamma_{l,t}S_{l,t-1}
+ U_l(k_{l,t},v_{l,t},S_{l,t-1}),\\
M_{\mathrm{linear},l}(h)_t
&=
W^O_l\!\left(q_{l,t}^{\top}S_{l,t}\right),
\end{align}
with gates, normalization, local convolution, and state-update details absorbed into $\Gamma_{l,t}$ and $U_l$. Equations~\ref{eq:app_full_attn}--\ref{eq:app_linear_state} show that full and linear attention implement different token-mixing operators, and their layerwise output scales are learned.

The factor used in the merge follows from matching family-level residual exposure. Let full attention be the reference family. To keep the integrated first-order update from family $f$ comparable to the reference, Eq.~\ref{eq:app_family_budget} suggests
\begin{equation}
\label{eq:app_budget_match}
\mu(f)a_f
\approx
\mu(\mathrm{full})a_{\mathrm{full}},
\qquad
\frac{a_f}{a_{\mathrm{full}}}
\approx
\frac{\mu(\mathrm{full})}{\mu(f)}.
\end{equation}
Qwen3-Next-80B-A3B has $\mu(\mathrm{linear})=36$ and $\mu(\mathrm{full})=12$, so the architecture coefficient is
\begin{equation}
\label{eq:app_kappa_linear}
\kappa(\mathrm{linear})
=
\frac{\mu(\mathrm{linear})}{\mu(\mathrm{full})}
=
\frac{36}{12}
=
3.
\end{equation}
Since Eq.~\ref{eq:app_arch_score} divides by $\kappa$, each linear-attention layer receives one third of the per-layer merge budget assigned to an otherwise comparable full-attention reference. This is an occupation correction: linear attention appears three times as often in the residual stack, so equal per-layer injection would give the linear family roughly three times the integrated first-order exposure.

\begin{figure}[!htbp]
  \centering
  \includegraphics[width=0.95\linewidth]{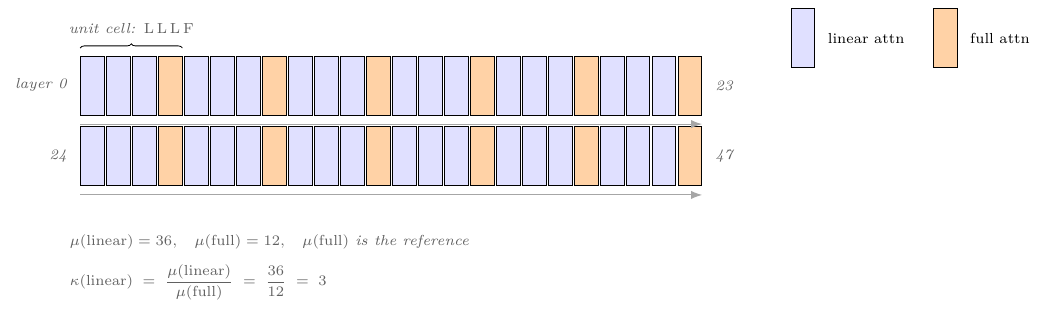}
  \caption{Qwen3-Next-80B-A3B residual stack laid out as 48 mixer slots: linear-attention layers (blue) repeat three times for every full-attention layer (orange), giving $\mu(\mathrm{linear})=36$ and $\mu(\mathrm{full})=12$. The 3:1 occupation is the geometric source of $\kappa(\mathrm{linear})=3$ in Eq.~\ref{eq:app_kappa_linear}.}
  \label{fig:app_residual_occupation}
\end{figure}

If activation-side measurements are available, the architecture-only coefficient can be generalized to
\begin{equation}
\label{eq:app_kappa_meas}
\kappa_{\mathrm{meas}}(f)
=
\frac{\mu(f)a_{\mathrm{meas}}(f)}
{\mu(f_{\mathrm{ref}})a_{\mathrm{meas}}(f_{\mathrm{ref}})},
\qquad
a_{\mathrm{meas}}(f)
=
\mathbb{E}_{l:f_l=f,\;h_l\sim\mathcal{D}_{\mathrm{cal}}}
\left[
\|\Delta M_{f,l}(h_l)\|
\right].
\end{equation}
Here $a_{\mathrm{meas}}(f)$ estimates the absolute layerwise perturbation scale $a_f$ in Eq.~\ref{eq:app_family_budget}, matching the absolute endpoint perturbations controlled by the transport bound. The experiments use the architecture-only version, $a_{\mathrm{meas}}(f)\approx a_{\mathrm{meas}}(f_{\mathrm{ref}})$, so the merge statistics are computed once from masked losses and reused across model shards.

\section{GSP Implementation Details}
\label{app:gsp-details}

This section gives the implementation details of GSP. The format traces define the mask support $\mathcal{I}_F$ for protocol-control positions, and GSP expands $\mathcal{I}_F$ to a local neighborhood before collecting activations.

\subsection{Token Neighborhood}
\label{app:gsp-token-neighborhood}

For the format traces $\mathcal{D}_F$, the format-mask support is
\begin{equation}
\label{eq:gsp_format_support}
\mathcal{I}_F
=
\{(i,s):(x_i^F,y_i^F,m_i^F)\in\mathcal{D}_F,\ m_{i,s}^F=1\}.
\end{equation}
The experiments then use the symmetric token-window expansion
\begin{equation}
\label{eq:gsp_neighborhood}
\mathcal{N}_\rho(\mathcal{I}_F)
=
\{(i,t): \exists (i,s)\in\mathcal{I}_F\ \text{with}\ |t-s|\leq\rho,\ 1\leq t\leq S_i^F\}.
\end{equation}
We set $\rho=2$. The window is applied within each trace before collecting activations, clipped to valid token positions, and deduplicated.

\subsection{SVD Derivation of the GSP Projector}
\label{app:gsp-svd-derivation}

This subsection expands the main-text derivation for Eq.~\ref{eq:gsp_format_energy} and Eq.~\ref{eq:gsp_proj}. Fix an edited tensor and its protected activation space indexed by $q$, meaning the input-side activation space used to construct that tensor's format-preserving projector. Orient the edited tensor as a linear map $W_q\in\mathbb{R}^{d_{\mathrm{out}}\times d_q}$, where $d_q$ is the dimension of the protected input activation. The notation $q(l,c)$ in the main text maps a layer/component tensor to this input-activation space. For a selected format-neighborhood activation $x_n\in\mathbb{R}^{d_q}$, the local output perturbation induced by an additive edit $\Delta_q$ is
\begin{equation}
\label{eq:app_gsp_single_perturb}
(W_q+\Delta_q)x_n-W_qx_n=\Delta_q x_n .
\end{equation}
Stacking all selected activations row-wise gives
\begin{equation}
\label{eq:app_gsp_stack}
H_q=
\begin{bmatrix}
x_1^\top\\
\vdots\\
x_{N_q}^\top
\end{bmatrix}
\in\mathbb{R}^{N_q\times d_q},
\qquad
E_q(\Delta_q)
=
\sum_{n=1}^{N_q}\|\Delta_q x_n\|_2^2
=
\|H_q\Delta_q^\top\|_F^2 .
\end{equation}
Thus GSP uses $E_q(\Delta_q)$ as a local output-preservation surrogate: edits with small $E_q$ leave the immediate module outputs nearly unchanged on the masked format traces.

Let the compact SVD of $H_q$ be
\begin{equation}
\label{eq:app_gsp_svd}
H_q=U_q\Sigma_qV_q^\top,
\qquad
V_q=[v_{q,1},\ldots,v_{q,r_q}],
\qquad
\Sigma_q=\mathrm{diag}(\sigma_{q,1},\ldots,\sigma_{q,r_q}),
\end{equation}
with $\sigma_{q,1}\ge\cdots\ge\sigma_{q,r_q}>0$. By Frobenius-norm invariance under the left-orthogonal factor $U_q$,
\begin{equation}
\label{eq:app_gsp_energy}
E_q(\Delta_q)
=
\|U_q\Sigma_q V_q^\top\Delta_q^\top\|_F^2
=
\|\Sigma_q V_q^\top\Delta_q^\top\|_F^2
=
\sum_{r=1}^{r_q}\sigma_{q,r}^2\|\Delta_q v_{q,r}\|_2^2 .
\end{equation}
The right singular vectors are the relevant directions because the weight edit acts on the input activation dimension: $v_{q,r}$ is an input-space direction, and $\Delta_qv_{q,r}$ is the output change caused by editing along that direction. Large $\sigma_{q,r}$ therefore identifies an input direction that occurs strongly in format-critical traces, so preserving format behavior asks us to suppress the corresponding edit component.

A hard activation-nullspace projection would choose a protected set $P_q$ and remove those components:
\begin{equation}
\label{eq:app_gsp_hard}
\Pi_{P_q}^{\mathrm{hard}}(\Delta_q)
=
\Delta_q\left(I-\sum_{r\in P_q}v_{q,r}v_{q,r}^\top\right).
\end{equation}
\method{} instead uses a smooth mask over singular directions. Define normalized amplitudes
\begin{equation}
\label{eq:app_gsp_amplitude}
a_{q,r}=\frac{\sigma_{q,r}}{\sigma_{q,1}},
\end{equation}
and protection weights $w_{q,r}=\operatorname{sigmoid}(k(a_{q,r}-\tau))\in[0,1]$. The resulting operator is
\begin{equation}
\label{eq:app_gsp_soft}
\Pi_{\tau,q}^{\mathrm{GSP}}(\Delta_q)
=
\Delta_q-\Delta_qV_q\mathrm{diag}(\mathbf{w}_q)V_q^\top
=
\Delta_q\left(I-V_q\mathrm{diag}(\mathbf{w}_q)V_q^\top\right).
\end{equation}
For each retained singular vector,
\begin{equation}
\label{eq:app_gsp_direction_scale}
\Pi_{\tau,q}^{\mathrm{GSP}}(\Delta_q)v_{q,r}
=
(1-w_{q,r})\Delta_qv_{q,r}.
\end{equation}
Therefore high-amplitude format directions are nearly removed, low-amplitude directions are mostly unchanged, and boundary directions are partially attenuated. Substituting Eq.~\ref{eq:app_gsp_direction_scale} into Eq.~\ref{eq:app_gsp_energy} gives the post-projection local surrogate
\begin{equation}
\label{eq:app_gsp_post_energy}
E_q(\Pi_{\tau,q}^{\mathrm{GSP}}(\Delta_q))
=
\sum_{r=1}^{r_q}
\sigma_{q,r}^2(1-w_{q,r})^2\|\Delta_qv_{q,r}\|_2^2 .
\end{equation}
Directions orthogonal to $\mathrm{span}(V_q)$ are unconstrained by the observed activation matrix and pass through unchanged. If no activation matrix with matching input dimension is collected for a tensor, or if the collected matrix is numerically zero, the implementation uses the identity operator for that tensor.

\subsection{Sigmoid Weighting}

\begin{figure}[!htbp]
  \centering
  \includegraphics[width=\linewidth]{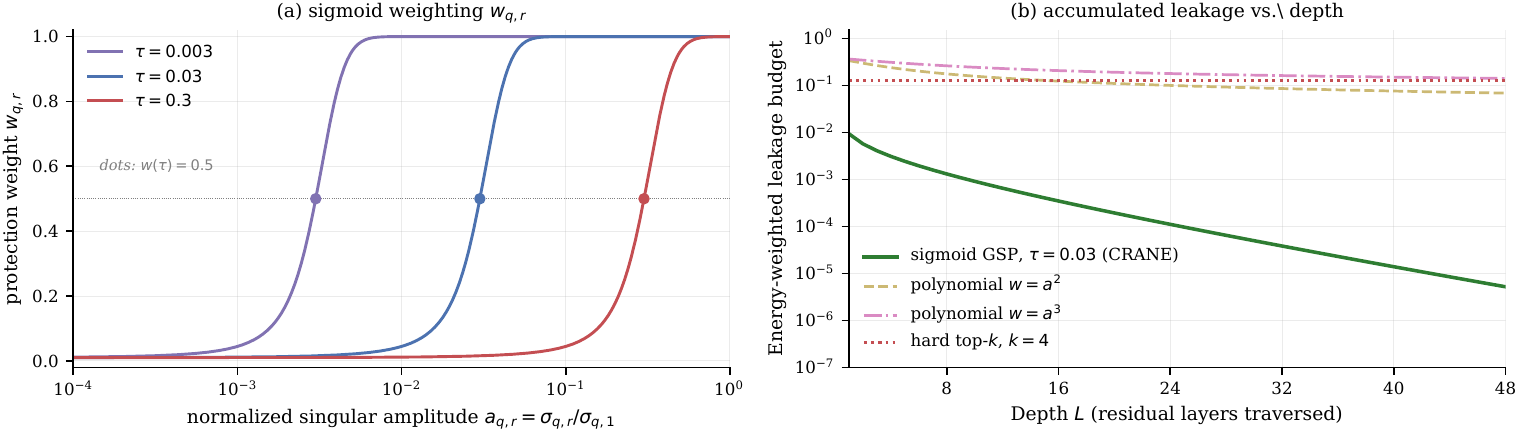}
  \caption{GSP sigmoid-weighting diagnostics. \textbf{(a)} Sigmoid weighting $w_{q,r}=\operatorname{sigmoid}(k(a_{q,r}-\tau))$ with $k=\log(99)/\tau$ for $\tau\in\{0.003, 0.03, 0.3\}$; the dot marks $w(\tau)=0.5$ and the transition band $[0, 2\tau]$ contains all partial attenuation. \textbf{(b)} Energy-weighted residual mask profile along format-protected directions across residual depth for the sigmoid mask, polynomial soft masks ($w=a^{p}$), and a hard top-$k$ mask.}
  \label{fig:app_gsp_properties}
\end{figure}

The experiments use $\tau=0.03$ and set $k=\log(99)/\tau\approx4.6/\tau$ in Eq.~\ref{eq:gsp_weight}; for the default $\tau=0.03$, this gives $k\approx153.2$. The constant $4.6$ is the rounded logit $\log(0.99/0.01)=\log(99)$, chosen so that the sigmoid protection coefficient is approximately $0.01$ at $a_{q,r}=0$, $0.5$ at $a_{q,r}=\tau$, and $0.99$ at $a_{q,r}=2\tau$. The transition from $w\approx0.01$ to $w\approx0.99$ therefore occurs over approximately $[0,2\tau]=[0,0.06]$, so directions near the boundary receive partial attenuation rather than a discontinuous hard projection. Figure~\ref{fig:app_gsp_properties}(a) plots $w_{q,r}$ for several $\tau$ values.

The smooth transition makes $\Pi_{\tau,q}^{\mathrm{GSP}}$ vary continuously with $\tau$, whereas a hard projector can switch a direction from fully removed to fully retained under a small numerical change in $a_{q,r}$. Figure~\ref{fig:app_gsp_properties}(b) visualizes the energy-weighted residual mask profile of the sigmoid mask against polynomial soft masks ($w=a^{2}, a^{3}$) and a hard top-$k$ mask across depth.

\subsection{Tensor Orientation}

Equation~\ref{eq:gsp_proj} is written for tensors whose protected input-activation dimension is on the right, $\Delta_q\in\mathbb{R}^{d_{\mathrm{out}}\times d_q}$. If a stored parameter tensor places that dimension on the left, the implementation applies the same operator after transposing the tensor and then transposes the result back. This changes only the array layout, not the mathematical projection.

\subsection{Protected Activation Map}

The main-text notation $q(l,c)$ maps each layer/component tensor to the input-side activation space used to build its GSP projector. For a linear map whose weight can be oriented as $\Delta_q\in\mathbb{R}^{d_{\mathrm{out}}\times d_q}$, $q(l,c)$ indexes the activation vector multiplied by that weight in the forward pass. GSP is therefore an input-side projector for the edited weight matrix. For Q/K/V, routers, and FFN/expert gate/up projections, the protected input is the residual stream. For output projections and expert down projections, the protected input, when collected, is the attention/mixer or MLP intermediate activation rather than the residual stream. Tensors without a collected activation matrix of matching input dimension, such as scalar biases or unsupported buffers, use the identity projector.

\subsection{Complete Merge Algorithm}
\label{app:algorithm}

\begin{algorithm}[!htbp]
\caption{\method{} merge implementation}
\label{alg:ttg}
\begin{algorithmic}[1]
\REQUIRE $\theta_{\text{inst}}$, $\theta_{\text{think}}$, masked-loss sets $\mathcal{D}_R, \mathcal{D}_A$, format-trace set $\mathcal{D}_F$, GSP projectors $\{V_q,\sigma_q\}_q$, scale $\alpha$, threshold $\tau$
\ENSURE $\theta_{\text{merged}}$
\STATE $\delta \leftarrow \theta_{\text{think}}-\theta_{\text{inst}}$
\STATE compute $g_R=\nabla_\theta \mathcal{L}_R(\theta_{\text{inst}})$ and $g_A=\nabla_\theta \mathcal{L}_A(\theta_{\text{inst}})$
\STATE for each $j$: $s_R(j)\leftarrow-g_{R,j}\delta_j$, $s_A(j)\leftarrow-g_{A,j}\delta_j$, $p_j\leftarrow[\min\{s_R(j),s_A(j)\}]_+$
\STATE aggregate normalized CTG scores into $S_{\mathrm{CTG}}(c,l)$ for each layer/component block
\FOR{each parameter tensor $\theta^{(l,c)}$}
  \STATE $\hat\delta\leftarrow T(\delta^{(l,c)})$
  \STATE $\hat\delta\leftarrow\alpha\,S_{\mathrm{CTG}}(c,l)\,\hat\delta$
  \STATE $\hat\delta\leftarrow\Pi_{\tau,q(l,c)}^{\mathrm{GSP}}(\hat\delta)$
  \STATE $\theta_{\mathrm{merged}}^{(l,c)}\leftarrow\theta_{\mathrm{inst}}^{(l,c)}+\hat\delta$
\ENDFOR
\RETURN $\theta_{\mathrm{merged}}$
\end{algorithmic}
\end{algorithm}

\section{Roo-Eval Detailed Results}
\label{app:roo-aggregate-results}

\begin{figure}[!htbp]
  \centering
  \includegraphics[width=0.95\linewidth]{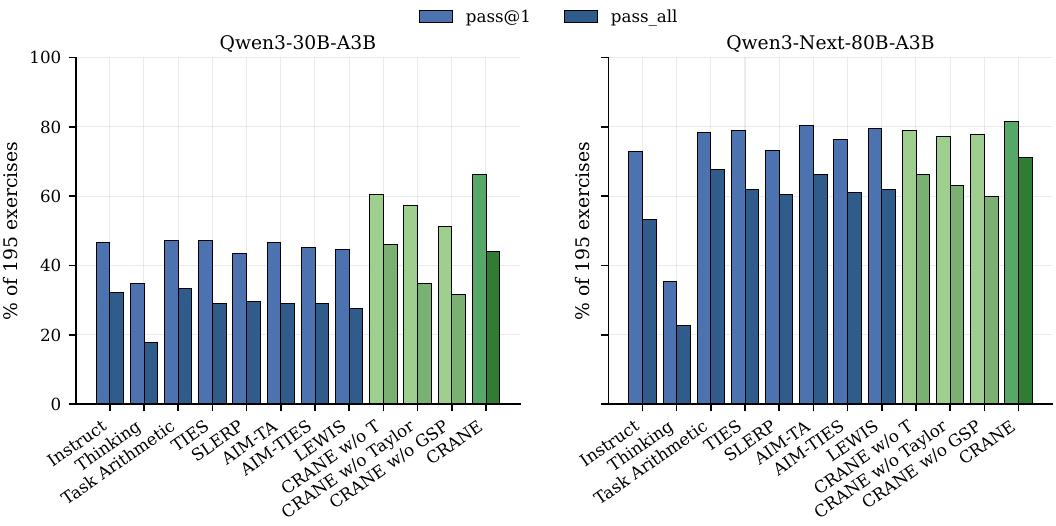}
  \caption{Roo-Eval results across both scales. Per-method pass@1 (light) and pass\_all (dark) on the 195 exercises. Plain merge baselines and \method{} component ablations are reported alongside the full \method{} recipe.}
  \label{fig:app_main_bars}
\end{figure}

This section collects the Roo-Eval results used in the main paper. Figure~\ref{fig:app_main_bars} gives a visual overview of pass@1 and pass\_all across both scales. Appendices~\ref{app:roo-30b-language}--\ref{app:roo-80b-language} report the per-language tables for the main 30B and 80B comparisons. Unlike the headline totals in Table~\ref{tab:roo_eval_main}, these tables retain the full log metrics: pass@1, pass@3, pass\_all, iter pass, TTC, and input/cached/output token counts, where iter pass counts passing rollouts. Appendices~\ref{app:roo-pass1-summary}--\ref{app:roo-passall-summary} give compact pass@1, pass@3, and pass\_all summaries by language. The $\alpha$/$\tau$ sweep tables and component-removal ablations are collected separately in Appendix~\ref{app:roo-ablation-summary}.

\subsection{30B Main Results by Language}
\label{app:roo-30b-language}

\begin{table}[!htbp]
\centering
\scriptsize
\caption{30B Roo-Eval full metrics for Python (34 exercises $\times$ 3 = 102 rollouts).}
\label{tab:app_roo_30b_python_full}
\begin{adjustbox}{max width=\linewidth}
\begin{tabular}{llllllllllll}
\toprule
Model & pass@1 & pass@3 & pass\_all & iter pass & TTC & Input total & Cached total & Output total & Input avg & Cached avg & Output avg \\
\midrule
Instruct (ref) & 15 (44.1\%) & 22 (64.7\%) & 13 (38.2\%) & 50/102 (49.0\%) & 29.3M & 7,622,806 & 146,270,461 & 1,411,369 & 74,733 & 1,434,024 & 13,837 \\
Thinking (ref) & 12 (35.3\%) & 21 (61.8\%) & 7 (20.6\%) & 43/102 (42.2\%) & 23.8M & 3,362,588 & 17,273,159 & 3,745,730 & 32,967 & 169,345 & 36,723 \\
Task Arithmetic & 15 (44.1\%) & 20 (58.8\%) & 12 (35.3\%) & 48/102 (47.1\%) & 32.4M & 8,719,220 & 172,340,721 & 1,296,213 & 85,483 & 1,689,615 & 12,708 \\
TIES & 19 (55.9\%) & 24 (70.6\%) & 13 (38.2\%) & 55/102 (53.9\%) & 31.7M & 7,759,662 & 179,068,578 & 1,211,760 & 76,075 & 1,755,574 & 11,880 \\
SLERP & 17 (50.0\%) & 20 (58.8\%) & 14 (41.2\%) & 52/102 (51.0\%) & 32.2M & 7,880,084 & 179,001,415 & 1,287,865 & 77,256 & 1,754,916 & 12,626 \\
AIM-TA & 17 (50.0\%) & 21 (61.8\%) & 11 (32.4\%) & 50/102 (49.0\%) & 33.6M & 8,551,536 & 185,412,207 & 1,308,627 & 83,839 & 1,817,767 & 12,830 \\
AIM-TIES & 15 (44.1\%) & 21 (61.8\%) & 11 (32.4\%) & 48/102 (47.1\%) & 35.5M & 8,936,722 & 198,989,550 & 1,337,479 & 87,615 & 1,950,878 & 13,113 \\
LEWIS & 18 (52.9\%) & 23 (67.6\%) & 10 (29.4\%) & 51/102 (50.0\%) & 33.4M & 8,508,474 & 180,678,399 & 1,356,050 & 83,416 & 1,771,357 & 13,295 \\
RAIN & 17 (50.0\%) & 21 (61.8\%) & 12 (35.3\%) & 49/102 (48.0\%) & 22.6M & 3,194,566 & 15,928,831 & 3,560,320 & 31,319 & 156,165 & 34,905 \\
\method{} & 27 (79.4\%) & 31 (91.2\%) & 19 (55.9\%) & 74/102 (72.5\%) & 19.4M & 5,605,858 & 63,459,202 & 1,480,496 & 54,959 & 622,149 & 14,514 \\
\bottomrule
\end{tabular}
\end{adjustbox}
\end{table}

\begin{table}[!htbp]
\centering
\scriptsize
\caption{30B Roo-Eval full metrics for JavaScript (50 exercises $\times$ 3 = 150 rollouts).}
\label{tab:app_roo_30b_javascript_full}
\begin{adjustbox}{max width=\linewidth}
\begin{tabular}{llllllllllll}
\toprule
Model & pass@1 & pass@3 & pass\_all & iter pass & TTC & Input total & Cached total & Output total & Input avg & Cached avg & Output avg \\
\midrule
Instruct (ref) & 28 (56.0\%) & 37 (74.0\%) & 20 (40.0\%) & 86/150 (57.3\%) & 45.9M & 11,240,333 & 257,446,200 & 1,786,951 & 74,936 & 1,716,308 & 11,913 \\
Thinking (ref) & 20 (40.0\%) & 27 (54.0\%) & 12 (24.0\%) & 60/150 (40.0\%) & 33.4M & 5,772,708 & 29,434,786 & 4,927,274 & 38,485 & 196,232 & 32,848 \\
Task Arithmetic & 26 (52.0\%) & 35 (70.0\%) & 21 (42.0\%) & 84/150 (56.0\%) & 51.1M & 12,879,073 & 298,843,109 & 1,660,467 & 85,860 & 1,992,287 & 11,070 \\
TIES & 25 (50.0\%) & 33 (66.0\%) & 16 (32.0\%) & 76/150 (50.7\%) & 56.7M & 13,517,910 & 345,238,140 & 1,723,803 & 90,119 & 2,301,588 & 11,492 \\
SLERP & 26 (52.0\%) & 33 (66.0\%) & 16 (32.0\%) & 75/150 (50.0\%) & 54.2M & 13,925,856 & 314,511,653 & 1,755,758 & 92,839 & 2,096,744 & 11,705 \\
AIM-TA & 25 (50.0\%) & 35 (70.0\%) & 17 (34.0\%) & 76/150 (50.7\%) & 55.9M & 13,820,167 & 336,070,898 & 1,699,775 & 92,134 & 2,240,473 & 11,332 \\
AIM-TIES & 28 (56.0\%) & 36 (72.0\%) & 19 (38.0\%) & 84/150 (56.0\%) & 52.0M & 12,355,490 & 314,457,599 & 1,633,606 & 82,370 & 2,096,384 & 10,891 \\
LEWIS & 21 (42.0\%) & 33 (66.0\%) & 17 (34.0\%) & 76/150 (50.7\%) & 51.8M & 12,935,745 & 303,072,729 & 1,713,793 & 86,238 & 2,020,485 & 11,425 \\
RAIN & 26 (52.0\%) & 29 (58.0\%) & 13 (26.0\%) & 68/150 (45.3\%) & 31.9M & 5,752,111 & 28,189,669 & 4,674,807 & 38,347 & 187,931 & 31,165 \\
\method{} & 39 (78.0\%) & 42 (84.0\%) & 30 (60.0\%) & 111/150 (74.0\%) & 26.1M & 8,027,932 & 93,420,273 & 1,753,243 & 53,519 & 622,801 & 11,688 \\
\bottomrule
\end{tabular}
\end{adjustbox}
\end{table}

\begin{table}[!htbp]
\centering
\scriptsize
\caption{30B Roo-Eval full metrics for Go (36 exercises $\times$ 3 = 108 rollouts).}
\label{tab:app_roo_30b_go_full}
\begin{adjustbox}{max width=\linewidth}
\begin{tabular}{llllllllllll}
\toprule
Model & pass@1 & pass@3 & pass\_all & iter pass & TTC & Input total & Cached total & Output total & Input avg & Cached avg & Output avg \\
\midrule
Instruct (ref) & 12 (33.3\%) & 19 (52.8\%) & 6 (16.7\%) & 36/108 (33.3\%) & 35.8M & 8,091,205 & 179,249,487 & 1,955,963 & 74,919 & 1,659,717 & 18,111 \\
Thinking (ref) & 16 (44.4\%) & 23 (63.9\%) & 8 (22.2\%) & 45/108 (41.7\%) & 27.2M & 3,505,650 & 20,108,265 & 4,341,897 & 32,460 & 186,188 & 40,203 \\
Task Arithmetic & 19 (52.8\%) & 22 (61.1\%) & 11 (30.6\%) & 48/108 (44.4\%) & 41.2M & 9,503,998 & 225,774,679 & 1,816,555 & 88,000 & 2,090,506 & 16,820 \\
TIES & 17 (47.2\%) & 26 (72.2\%) & 9 (25.0\%) & 53/108 (49.1\%) & 38.5M & 8,657,592 & 214,344,134 & 1,676,051 & 80,163 & 1,984,668 & 15,519 \\
SLERP & 14 (38.9\%) & 21 (58.3\%) & 10 (27.8\%) & 44/108 (40.7\%) & 41.7M & 9,358,943 & 229,965,935 & 1,866,615 & 86,657 & 2,129,314 & 17,283 \\
AIM-TA & 16 (44.4\%) & 24 (66.7\%) & 10 (27.8\%) & 50/108 (46.3\%) & 39.6M & 9,172,381 & 219,968,657 & 1,694,560 & 84,929 & 2,036,747 & 15,690 \\
AIM-TIES & 13 (36.1\%) & 21 (58.3\%) & 9 (25.0\%) & 44/108 (40.7\%) & 43.4M & 10,082,733 & 238,993,851 & 1,891,978 & 93,359 & 2,212,906 & 17,518 \\
LEWIS & 17 (47.2\%) & 24 (66.7\%) & 8 (22.2\%) & 44/108 (40.7\%) & 35.4M & 8,539,543 & 187,248,082 & 1,625,549 & 79,070 & 1,733,779 & 15,051 \\
RAIN & 14 (38.9\%) & 20 (55.6\%) & 9 (25.0\%) & 47/108 (43.5\%) & 25.9M & 3,443,171 & 19,262,428 & 4,100,136 & 31,881 & 178,355 & 37,964 \\
\method{} & 27 (75.0\%) & 30 (83.3\%) & 18 (50.0\%) & 72/108 (66.7\%) & 21.8M & 6,025,226 & 73,353,048 & 1,684,501 & 55,789 & 679,194 & 15,597 \\
\bottomrule
\end{tabular}
\end{adjustbox}
\end{table}

\begin{table}[!htbp]
\centering
\scriptsize
\caption{30B Roo-Eval full metrics for Java (45 exercises $\times$ 3 = 135 rollouts).}
\label{tab:app_roo_30b_java_full}
\begin{adjustbox}{max width=\linewidth}
\begin{tabular}{llllllllllll}
\toprule
Model & pass@1 & pass@3 & pass\_all & iter pass & TTC & Input total & Cached total & Output total & Input avg & Cached avg & Output avg \\
\midrule
Instruct (ref) & 27 (60.0\%) & 32 (71.1\%) & 19 (42.2\%) & 78/135 (57.8\%) & 40.9M & 9,625,792 & 223,276,324 & 1,792,674 & 71,302 & 1,653,899 & 13,279 \\
Thinking (ref) & 13 (28.9\%) & 21 (46.7\%) & 5 (11.1\%) & 35/135 (25.9\%) & 35.4M & 5,011,844 & 30,749,871 & 5,458,022 & 37,125 & 227,777 & 40,430 \\
Task Arithmetic & 22 (48.9\%) & 27 (60.0\%) & 18 (40.0\%) & 66/135 (48.9\%) & 40.3M & 9,953,576 & 223,712,036 & 1,592,945 & 73,730 & 1,657,126 & 11,800 \\
TIES & 20 (44.4\%) & 29 (64.4\%) & 14 (31.1\%) & 65/135 (48.1\%) & 41.4M & 10,332,428 & 234,942,843 & 1,509,467 & 76,537 & 1,740,317 & 11,181 \\
SLERP & 21 (46.7\%) & 25 (55.6\%) & 14 (31.1\%) & 61/135 (45.2\%) & 42.8M & 10,568,416 & 238,926,759 & 1,670,085 & 78,285 & 1,769,828 & 12,371 \\
AIM-TA & 20 (44.4\%) & 28 (62.2\%) & 14 (31.1\%) & 61/135 (45.2\%) & 43.8M & 10,623,621 & 251,140,388 & 1,608,986 & 78,693 & 1,860,299 & 11,918 \\
AIM-TIES & 21 (46.7\%) & 26 (57.8\%) & 15 (33.3\%) & 63/135 (46.7\%) & 40.9M & 10,494,971 & 224,177,002 & 1,589,514 & 77,741 & 1,660,570 & 11,774 \\
LEWIS & 20 (44.4\%) & 29 (64.4\%) & 13 (28.9\%) & 62/135 (45.9\%) & 36.2M & 9,559,129 & 197,176,539 & 1,382,058 & 70,808 & 1,460,567 & 10,237 \\
RAIN & 12 (26.7\%) & 25 (55.6\%) & 4 (8.9\%) & 36/135 (26.7\%) & 34.8M & 4,844,261 & 30,197,023 & 5,378,656 & 35,883 & 223,681 & 39,841 \\
\method{} & 24 (53.3\%) & 37 (82.2\%) & 10 (22.2\%) & 70/135 (51.9\%) & 32.0M & 9,008,906 & 117,821,938 & 2,247,297 & 66,732 & 872,755 & 16,646 \\
\bottomrule
\end{tabular}
\end{adjustbox}
\end{table}

\begin{table}[!htbp]
\centering
\scriptsize
\caption{30B Roo-Eval full metrics for Rust (30 exercises $\times$ 3 = 90 rollouts).}
\label{tab:app_roo_30b_rust_full}
\begin{adjustbox}{max width=\linewidth}
\begin{tabular}{llllllllllll}
\toprule
Model & pass@1 & pass@3 & pass\_all & iter pass & TTC & Input total & Cached total & Output total & Input avg & Cached avg & Output avg \\
\midrule
Instruct (ref) & 9 (30.0\%) & 15 (50.0\%) & 5 (16.7\%) & 32/90 (35.6\%) & 29.2M & 6,967,880 & 150,833,979 & 1,425,177 & 77,421 & 1,675,933 & 15,835 \\
Thinking (ref) & 7 (23.3\%) & 11 (36.7\%) & 3 (10.0\%) & 18/90 (20.0\%) & 27.2M & 3,404,218 & 22,031,076 & 4,313,532 & 37,825 & 244,790 & 47,928 \\
Task Arithmetic & 10 (33.3\%) & 15 (50.0\%) & 3 (10.0\%) & 28/90 (31.1\%) & 43.2M & 9,289,522 & 256,694,433 & 1,645,362 & 103,217 & 2,852,160 & 18,282 \\
TIES & 11 (36.7\%) & 17 (56.7\%) & 5 (16.7\%) & 33/90 (36.7\%) & 40.7M & 8,860,719 & 241,852,016 & 1,523,066 & 98,452 & 2,687,245 & 16,923 \\
SLERP & 7 (23.3\%) & 15 (50.0\%) & 4 (13.3\%) & 28/90 (31.1\%) & 43.7M & 9,589,846 & 249,569,550 & 1,838,488 & 106,554 & 2,772,995 & 20,428 \\
AIM-TA & 13 (43.3\%) & 18 (60.0\%) & 5 (16.7\%) & 33/90 (36.7\%) & 39.6M & 9,170,900 & 224,308,682 & 1,602,218 & 101,899 & 2,492,319 & 17,802 \\
AIM-TIES & 11 (36.7\%) & 16 (53.3\%) & 3 (10.0\%) & 30/90 (33.3\%) & 39.5M & 8,736,839 & 225,587,509 & 1,637,948 & 97,076 & 2,506,528 & 18,199 \\
LEWIS & 11 (36.7\%) & 14 (46.7\%) & 6 (20.0\%) & 29/90 (32.2\%) & 37.6M & 8,547,662 & 211,082,637 & 1,579,754 & 94,974 & 2,345,363 & 17,553 \\
RAIN & 8 (26.7\%) & 11 (36.7\%) & 4 (13.3\%) & 22/90 (24.4\%) & 25.0M & 3,175,404 & 20,120,464 & 3,968,011 & 35,282 & 223,560 & 44,089 \\
\method{} & 12 (40.0\%) & 22 (73.3\%) & 9 (30.0\%) & 41/90 (45.6\%) & 21.6M & 6,010,939 & 76,419,820 & 1,593,906 & 66,788 & 849,109 & 17,710 \\
\bottomrule
\end{tabular}
\end{adjustbox}
\end{table}

\subsection{80B Main Results by Language}
\label{app:roo-80b-language}

\begin{table}[!htbp]
\centering
\scriptsize
\caption{80B Roo-Eval full metrics for Python (34 exercises $\times$ 3 = 102 rollouts).}
\label{tab:app_roo_80b_python_full}
\begin{adjustbox}{max width=\linewidth}
\begin{tabular}{llllllllllll}
\toprule
Model & pass@1 & pass@3 & pass\_all & iter pass & TTC & Input total & Cached total & Output total & Input avg & Cached avg & Output avg \\
\midrule
Instruct (ref) & 29 (85.3\%) & 31 (91.2\%) & 22 (64.7\%) & 82/102 (80.4\%) & 15.8M & 4,642,554 & 63,011,687 & 971,182 & 45,515 & 617,761 & 9,521 \\
Thinking (ref) & 16 (47.1\%) & 21 (61.8\%) & 11 (32.4\%) & 46/102 (45.1\%) & 17.8M & 2,890,157 & 11,873,010 & 2,735,770 & 28,334 & 116,402 & 26,821 \\
Task Arithmetic & 28 (82.4\%) & 30 (88.2\%) & 24 (70.6\%) & 83/102 (81.4\%) & 16.9M & 4,644,411 & 73,063,904 & 990,290 & 45,533 & 716,312 & 9,708 \\
TIES & 29 (85.3\%) & 30 (88.2\%) & 24 (70.6\%) & 83/102 (81.4\%) & 14.6M & 4,255,004 & 52,634,960 & 1,021,133 & 41,715 & 516,029 & 10,011 \\
SLERP & 28 (82.4\%) & 33 (97.1\%) & 24 (70.6\%) & 86/102 (84.3\%) & 15.5M & 4,251,615 & 65,835,441 & 925,945 & 41,682 & 645,445 & 9,077 \\
AIM-TA & 29 (85.3\%) & 31 (91.2\%) & 26 (76.5\%) & 85/102 (83.3\%) & 17.1M & 4,679,247 & 69,035,610 & 1,106,082 & 45,874 & 676,819 & 10,843 \\
AIM-TIES & 27 (79.4\%) & 31 (91.2\%) & 21 (61.8\%) & 81/102 (79.4\%) & 15.9M & 4,663,119 & 58,860,124 & 1,077,154 & 45,716 & 577,060 & 10,560 \\
LEWIS & 28 (82.4\%) & 31 (91.2\%) & 24 (70.6\%) & 83/102 (81.4\%) & 15.9M & 4,471,974 & 62,461,313 & 1,028,532 & 43,842 & 612,365 & 10,083 \\
\method{} & 30 (88.2\%) & 33 (97.1\%) & 27 (79.4\%) & 90/102 (88.2\%) & 13.1M & 3,807,607 & 46,484,492 & 933,088 & 37,329 & 455,730 & 9,148 \\
\bottomrule
\end{tabular}
\end{adjustbox}
\end{table}

\begin{table}[!htbp]
\centering
\scriptsize
\caption{80B Roo-Eval full metrics for JavaScript (50 exercises $\times$ 3 = 150 rollouts).}
\label{tab:app_roo_80b_javascript_full}
\begin{adjustbox}{max width=\linewidth}
\begin{tabular}{llllllllllll}
\toprule
Model & pass@1 & pass@3 & pass\_all & iter pass & TTC & Input total & Cached total & Output total & Input avg & Cached avg & Output avg \\
\midrule
Instruct (ref) & 42 (84.0\%) & 44 (88.0\%) & 38 (76.0\%) & 124/150 (82.7\%) & 18.7M & 6,100,734 & 62,082,646 & 1,279,387 & 40,671 & 413,884 & 8,529 \\
Thinking (ref) & 18 (36.0\%) & 30 (60.0\%) & 11 (22.0\%) & 60/150 (40.0\%) & 27.5M & 4,812,224 & 19,921,095 & 4,130,946 & 32,081 & 132,807 & 27,539 \\
Task Arithmetic & 44 (88.0\%) & 47 (94.0\%) & 39 (78.0\%) & 132/150 (88.0\%) & 18.2M & 5,775,193 & 64,329,549 & 1,188,490 & 38,501 & 428,863 & 7,923 \\
TIES & 46 (92.0\%) & 49 (98.0\%) & 40 (80.0\%) & 137/150 (91.3\%) & 16.8M & 5,408,355 & 56,427,698 & 1,157,469 & 36,055 & 376,184 & 7,716 \\
SLERP & 45 (90.0\%) & 47 (94.0\%) & 42 (84.0\%) & 134/150 (89.3\%) & 17.8M & 5,732,131 & 60,738,939 & 1,190,274 & 38,214 & 404,926 & 7,935 \\
AIM-TA & 45 (90.0\%) & 46 (92.0\%) & 42 (84.0\%) & 132/150 (88.0\%) & 19.3M & 5,955,332 & 73,104,000 & 1,197,194 & 39,702 & 487,360 & 7,981 \\
AIM-TIES & 44 (88.0\%) & 48 (96.0\%) & 42 (84.0\%) & 135/150 (90.0\%) & 18.4M & 5,941,063 & 64,318,755 & 1,209,895 & 39,607 & 428,791 & 8,065 \\
LEWIS & 46 (92.0\%) & 48 (96.0\%) & 39 (78.0\%) & 132/150 (88.0\%) & 18.6M & 5,901,958 & 64,583,900 & 1,243,850 & 39,346 & 430,559 & 8,292 \\
\method{} & 46 (92.0\%) & 49 (98.0\%) & 42 (84.0\%) & 137/150 (91.3\%) & 17.3M & 5,555,281 & 61,325,457 & 1,130,758 & 37,035 & 408,836 & 7,538 \\
\bottomrule
\end{tabular}
\end{adjustbox}
\end{table}

\begin{table}[!htbp]
\centering
\scriptsize
\caption{80B Roo-Eval full metrics for Go (36 exercises $\times$ 3 = 108 rollouts).}
\label{tab:app_roo_80b_go_full}
\begin{adjustbox}{max width=\linewidth}
\begin{tabular}{llllllllllll}
\toprule
Model & pass@1 & pass@3 & pass\_all & iter pass & TTC & Input total & Cached total & Output total & Input avg & Cached avg & Output avg \\
\midrule
Instruct (ref) & 24 (66.7\%) & 30 (83.3\%) & 17 (47.2\%) & 71/108 (65.7\%) & 12.9M & 4,241,044 & 41,041,332 & 906,858 & 39,268 & 380,012 & 8,396 \\
Thinking (ref) & 19 (52.8\%) & 23 (63.9\%) & 14 (38.9\%) & 56/108 (51.9\%) & 17.9M & 3,009,313 & 13,574,043 & 2,699,233 & 27,864 & 125,685 & 24,992 \\
Task Arithmetic & 32 (88.9\%) & 33 (91.7\%) & 30 (83.3\%) & 95/108 (88.0\%) & 15.1M & 4,410,040 & 50,264,599 & 1,124,352 & 40,833 & 465,412 & 10,410 \\
TIES & 28 (77.8\%) & 33 (91.7\%) & 23 (63.9\%) & 87/108 (80.6\%) & 14.0M & 4,384,786 & 41,693,441 & 1,093,254 & 40,599 & 386,050 & 10,122 \\
SLERP & 26 (72.2\%) & 30 (83.3\%) & 20 (55.6\%) & 78/108 (72.2\%) & 16.5M & 5,075,282 & 62,531,130 & 1,030,326 & 46,993 & 578,991 & 9,540 \\
AIM-TA & 29 (80.6\%) & 31 (86.1\%) & 24 (66.7\%) & 82/108 (75.9\%) & 18.3M & 4,995,851 & 71,376,393 & 1,229,321 & 46,258 & 660,893 & 11,383 \\
AIM-TIES & 28 (77.8\%) & 34 (94.4\%) & 27 (75.0\%) & 92/108 (85.2\%) & 14.0M & 4,296,087 & 43,593,567 & 1,059,110 & 39,778 & 403,644 & 9,806 \\
LEWIS & 31 (86.1\%) & 34 (94.4\%) & 26 (72.2\%) & 89/108 (82.4\%) & 15.7M & 4,452,145 & 55,444,704 & 1,147,981 & 41,223 & 513,376 & 10,629 \\
\method{} & 31 (86.1\%) & 33 (91.7\%) & 29 (80.6\%) & 92/108 (85.2\%) & 16.3M & 4,654,524 & 55,659,080 & 1,209,340 & 43,097 & 515,362 & 11,198 \\
\bottomrule
\end{tabular}
\end{adjustbox}
\end{table}

\begin{table}[!htbp]
\centering
\scriptsize
\caption{80B Roo-Eval full metrics for Java (45 exercises $\times$ 3 = 135 rollouts).}
\label{tab:app_roo_80b_java_full}
\begin{adjustbox}{max width=\linewidth}
\begin{tabular}{llllllllllll}
\toprule
Model & pass@1 & pass@3 & pass\_all & iter pass & TTC & Input total & Cached total & Output total & Input avg & Cached avg & Output avg \\
\midrule
Instruct (ref) & 26 (57.8\%) & 38 (84.4\%) & 12 (26.7\%) & 77/135 (57.0\%) & 23.0M & 7,105,797 & 80,844,477 & 1,566,931 & 52,635 & 598,847 & 11,606 \\
Thinking (ref) & 5 (11.1\%) & 8 (17.8\%) & 1 (2.2\%) & 14/135 (10.4\%) & 28.7M & 4,510,309 & 21,387,607 & 4,410,105 & 33,409 & 158,426 & 32,667 \\
Task Arithmetic & 26 (57.8\%) & 38 (84.4\%) & 18 (40.0\%) & 85/135 (63.0\%) & 25.2M & 7,574,931 & 92,095,924 & 1,682,157 & 56,110 & 682,192 & 12,460 \\
TIES & 28 (62.2\%) & 35 (77.8\%) & 14 (31.1\%) & 76/135 (56.3\%) & 27.0M & 8,077,565 & 96,813,494 & 1,840,404 & 59,833 & 717,136 & 13,632 \\
SLERP & 25 (55.6\%) & 34 (75.6\%) & 17 (37.8\%) & 79/135 (58.5\%) & 27.5M & 8,155,149 & 105,485,058 & 1,761,327 & 60,408 & 781,370 & 13,046 \\
AIM-TA & 32 (71.1\%) & 38 (84.4\%) & 22 (48.9\%) & 91/135 (67.4\%) & 26.2M & 7,786,595 & 96,868,682 & 1,746,001 & 57,678 & 717,545 & 12,933 \\
AIM-TIES & 30 (66.7\%) & 40 (88.9\%) & 15 (33.3\%) & 79/135 (58.5\%) & 28.5M & 8,473,956 & 105,941,200 & 1,877,789 & 62,770 & 784,749 & 13,909 \\
LEWIS & 27 (60.0\%) & 36 (80.0\%) & 15 (33.3\%) & 79/135 (58.5\%) & 27.5M & 8,157,055 & 102,370,544 & 1,826,930 & 60,422 & 758,300 & 13,532 \\
\method{} & 28 (62.2\%) & 37 (82.2\%) & 20 (44.4\%) & 89/135 (65.9\%) & 24.3M & 7,543,322 & 90,934,720 & 1,529,337 & 55,876 & 673,591 & 11,328 \\
\bottomrule
\end{tabular}
\end{adjustbox}
\end{table}

\normalsize

\begin{table}[!htbp]
\centering
\scriptsize
\caption{80B Roo-Eval full metrics for Rust (30 exercises $\times$ 3 = 90 rollouts).}
\label{tab:app_roo_80b_rust_full}
\begin{adjustbox}{max width=\linewidth}
\begin{tabular}{llllllllllll}
\toprule
Model & pass@1 & pass@3 & pass\_all & iter pass & TTC & Input total & Cached total & Output total & Input avg & Cached avg & Output avg \\
\midrule
Instruct (ref) & 21 (70.0\%) & 27 (90.0\%) & 15 (50.0\%) & 62/90 (68.9\%) & 19.2M & 5,354,259 & 68,007,725 & 1,404,484 & 59,491 & 755,641 & 15,605 \\
Thinking (ref) & 11 (36.7\%) & 15 (50.0\%) & 7 (23.3\%) & 32/90 (35.6\%) & 17.7M & 2,930,934 & 15,007,654 & 2,654,245 & 32,565 & 166,751 & 29,491 \\
Task Arithmetic & 23 (76.7\%) & 25 (83.3\%) & 21 (70.0\%) & 69/90 (76.7\%) & 17.8M & 5,087,632 & 62,155,706 & 1,299,705 & 56,529 & 690,618 & 14,441 \\
TIES & 23 (76.7\%) & 25 (83.3\%) & 20 (66.7\%) & 68/90 (75.6\%) & 16.6M & 4,658,243 & 57,569,561 & 1,234,629 & 51,758 & 639,661 & 13,718 \\
SLERP & 19 (63.3\%) & 25 (83.3\%) & 15 (50.0\%) & 62/90 (68.9\%) & 20.4M & 5,701,264 & 77,723,723 & 1,375,841 & 63,347 & 863,596 & 15,287 \\
AIM-TA & 22 (73.3\%) & 25 (83.3\%) & 15 (50.0\%) & 62/90 (68.9\%) & 19.1M & 5,270,696 & 67,490,094 & 1,424,542 & 58,563 & 749,889 & 15,828 \\
AIM-TIES & 20 (66.7\%) & 24 (80.0\%) & 14 (46.7\%) & 59/90 (65.6\%) & 19.3M & 5,480,806 & 64,701,478 & 1,465,082 & 60,897 & 718,905 & 16,278 \\
LEWIS & 23 (76.7\%) & 27 (90.0\%) & 17 (56.7\%) & 66/90 (73.3\%) & 18.2M & 5,130,397 & 61,044,748 & 1,384,623 & 57,004 & 678,274 & 15,384 \\
\method{} & 24 (80.0\%) & 24 (80.0\%) & 21 (70.0\%) & 68/90 (75.6\%) & 18.2M & 5,006,504 & 67,960,906 & 1,270,158 & 55,628 & 755,121 & 14,113 \\
\bottomrule
\end{tabular}
\end{adjustbox}
\end{table}

\subsection{Pass@1 Language Summaries}
\label{app:roo-pass1-summary}

Tables~\ref{tab:app_roo_30b_pass1_summary} and~\ref{tab:app_roo_80b_pass1_summary} summarize Roo-Eval pass@1 by language at the 30B and 80B scales respectively. The Overall column pools all 195 exercises and matches Table~\ref{tab:roo_eval_main}.

\begin{figure}[!htbp]
  \centering
  \includegraphics[width=0.95\linewidth]{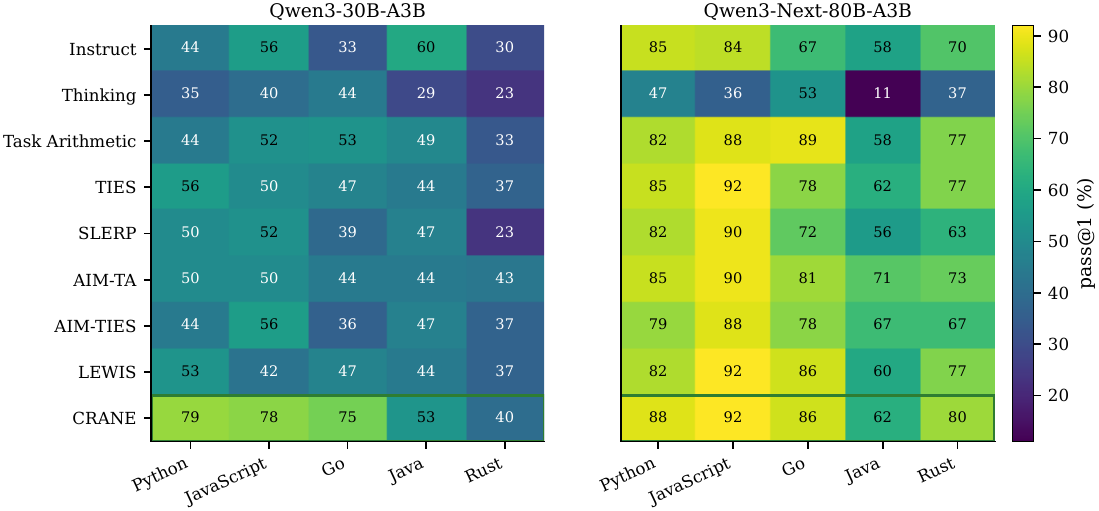}
  \caption{Per-language Roo-Eval pass@1 across methods at both scales. Rows: methods (Instruct, Thinking, plain merges, \method{}); columns: Python, JavaScript, Go, Java, Rust. \method{} achieves the highest pass@1 on Python, JavaScript, and Go at 30B and remains among the top-performing methods at 80B, with the residual Java gap at 30B discussed in Section~\ref{sec:discussion}.}
  \label{fig:app_per_language_heatmap}
\end{figure}

\begin{table}[!htbp]
\centering
\scriptsize
\caption{30B Roo-Eval pass@1 by language.}
\label{tab:app_roo_30b_pass1_summary}
\begin{adjustbox}{max width=\linewidth}
\begin{tabular}{lllllll}
\toprule
Model & Python & JavaScript & Go & Java & Rust & \textbf{Overall} \\
\midrule
Instruct (ref) & 44.1 & 56.0 & 33.3 & 60.0 & 30.0 & 46.7 \\
Thinking (ref) & 35.3 & 40.0 & 44.4 & 28.9 & 23.3 & 34.9 \\
Task Arithmetic & 44.1 & 52.0 & 52.8 & 48.9 & 33.3 & 47.2 \\
TIES & 55.9 & 50.0 & 47.2 & 44.4 & 36.7 & 47.2 \\
SLERP & 50.0 & 52.0 & 38.9 & 46.7 & 23.3 & 43.6 \\
AIM-TA & 50.0 & 50.0 & 44.4 & 44.4 & 43.3 & 46.7 \\
AIM-TIES & 44.1 & 56.0 & 36.1 & 46.7 & 36.7 & 45.1 \\
LEWIS & 52.9 & 42.0 & 47.2 & 44.4 & 36.7 & 44.6 \\
RAIN & 50.0 & 52.0 & 38.9 & 26.7 & 26.7 & 39.5 \\
\method{} & 79.4 & 78.0 & 75.0 & 53.3 & 40.0 & 66.2 \\
\bottomrule
\end{tabular}
\end{adjustbox}
\end{table}

\begin{table}[!htbp]
\centering
\scriptsize
\caption{80B Roo-Eval pass@1 by language.}
\label{tab:app_roo_80b_pass1_summary}
\begin{adjustbox}{max width=\linewidth}
\begin{tabular}{lllllll}
\toprule
Model & Python & JavaScript & Go & Java & Rust & \textbf{Overall} \\
\midrule
Instruct (ref) & 85.3 & 84.0 & 66.7 & 57.8 & 70.0 & 72.8 \\
Thinking (ref) & 47.1 & 36.0 & 52.8 & 11.1 & 36.7 & 35.4 \\
Task Arithmetic & 82.4 & 88.0 & 88.9 & 57.8 & 76.7 & 78.5 \\
TIES & 85.3 & 92.0 & 77.8 & 62.2 & 76.7 & 79.0 \\
SLERP & 82.4 & 90.0 & 72.2 & 55.6 & 63.3 & 73.3 \\
AIM-TA & 85.3 & 90.0 & 80.6 & 71.1 & 73.3 & 80.5 \\
AIM-TIES & 79.4 & 88.0 & 77.8 & 66.7 & 66.7 & 76.4 \\
LEWIS & 82.4 & 92.0 & 86.1 & 60.0 & 76.7 & 79.5 \\
RAIN & 58.8 & 34.0 & 52.8 & 46.7 & 43.3 & 46.2 \\
\method{} & 88.2 & 92.0 & 86.1 & 62.2 & 80.0 & 81.5 \\
\bottomrule
\end{tabular}
\end{adjustbox}
\end{table}

\subsection{Pass@3 Language Summaries}
\label{app:roo-passk-summary}

\begin{table}[!htbp]
\centering
\scriptsize
\caption{30B Roo-Eval pass@3 by language.}
\label{tab:app_roo_30b_passk_summary}
\begin{adjustbox}{max width=\linewidth}
\begin{tabular}{lllllll}
\toprule
Model & Python & JavaScript & Go & Java & Rust & \textbf{Overall} \\
\midrule
\method{} & 91.2 & 84.0 & 83.3 & 82.2 & 73.3 & 83.1 \\
Instruct (ref) & 64.7 & 74.0 & 52.8 & 71.1 & 50.0 & 64.1 \\
Thinking (ref) & 61.8 & 54.0 & 63.9 & 46.7 & 36.7 & 52.8 \\
Task Arithmetic & 58.8 & 70.0 & 61.1 & 60.0 & 50.0 & 61.0 \\
TIES & 70.6 & 66.0 & 72.2 & 64.4 & 56.7 & 66.2 \\
SLERP & 58.8 & 66.0 & 58.3 & 55.6 & 50.0 & 58.5 \\
AIM-TA & 61.8 & 70.0 & 66.7 & 62.2 & 60.0 & 64.6 \\
AIM-TIES & 61.8 & 72.0 & 58.3 & 57.8 & 53.3 & 61.5 \\
LEWIS & 67.6 & 66.0 & 66.7 & 64.4 & 46.7 & 63.1 \\
RAIN & 61.8 & 58.0 & 55.6 & 55.6 & 36.7 & 54.4 \\
\bottomrule
\end{tabular}
\end{adjustbox}
\end{table}

\begin{table}[!htbp]
\centering
\scriptsize
\caption{80B Roo-Eval pass@3 by language.}
\label{tab:app_roo_80b_passk_summary}
\begin{adjustbox}{max width=\linewidth}
\begin{tabular}{lllllll}
\toprule
Model & Python & JavaScript & Go & Java & Rust & \textbf{Overall} \\
\midrule
Instruct (ref) & 91.2 & 88.0 & 83.3 & 84.4 & 90.0 & 87.2 \\
Thinking (ref) & 61.8 & 60.0 & 63.9 & 17.8 & 50.0 & 49.7 \\
Task Arithmetic & 88.2 & 94.0 & 91.7 & 84.4 & 83.3 & 88.7 \\
TIES & 88.2 & 98.0 & 91.7 & 77.8 & 83.3 & 88.2 \\
SLERP & 97.1 & 94.0 & 83.3 & 75.6 & 83.3 & 86.7 \\
AIM-TA & 91.2 & 92.0 & 86.1 & 84.4 & 83.3 & 87.7 \\
AIM-TIES & 91.2 & 96.0 & 94.4 & 88.9 & 80.0 & 90.8 \\
LEWIS & 91.2 & 96.0 & 94.4 & 80.0 & 90.0 & 90.3 \\
RAIN & 61.8 & 58.0 & 63.9 & 57.8 & 50.0 & 58.5 \\
\method{} & 97.1 & 98.0 & 91.7 & 82.2 & 80.0 & 90.3 \\
\bottomrule
\end{tabular}
\end{adjustbox}
\end{table}

\subsection{Pass\_all Language Summaries}
\label{app:roo-passall-summary}

\begin{table}[!htbp]
\centering
\scriptsize
\caption{30B Roo-Eval pass\_all by language, i.e., exercises solved in all three rollouts.}
\label{tab:app_roo_30b_passall_summary}
\begin{adjustbox}{max width=\linewidth}
\begin{tabular}{lllllll}
\toprule
Model & Python & JavaScript & Go & Java & Rust & \textbf{Overall} \\
\midrule
Instruct (ref) & 38.2 & 40.0 & 16.7 & 42.2 & 16.7 & 32.3 \\
Thinking (ref) & 20.6 & 24.0 & 22.2 & 11.1 & 10.0 & 17.9 \\
Task Arithmetic & 35.3 & 42.0 & 30.6 & 40.0 & 10.0 & 33.3 \\
TIES & 38.2 & 32.0 & 25.0 & 31.1 & 16.7 & 29.2 \\
SLERP & 41.2 & 32.0 & 27.8 & 31.1 & 13.3 & 29.7 \\
AIM-TA & 32.4 & 34.0 & 27.8 & 31.1 & 16.7 & 29.2 \\
AIM-TIES & 32.4 & 38.0 & 25.0 & 33.3 & 10.0 & 29.2 \\
LEWIS & 29.4 & 34.0 & 22.2 & 28.9 & 20.0 & 27.7 \\
RAIN & 35.3 & 26.0 & 25.0 & 8.9 & 13.3 & 21.5 \\
\method{} & 55.9 & 60.0 & 50.0 & 22.2 & 30.0 & 44.1 \\
\bottomrule
\end{tabular}
\end{adjustbox}
\end{table}

\begin{table}[!htbp]
\centering
\scriptsize
\caption{80B Roo-Eval pass\_all by language, i.e., exercises solved in all three rollouts.}
\label{tab:app_roo_80b_passall_summary}
\begin{adjustbox}{max width=\linewidth}
\begin{tabular}{lllllll}
\toprule
Model & Python & JavaScript & Go & Java & Rust & \textbf{Overall} \\
\midrule
Instruct (ref) & 64.7 & 76.0 & 47.2 & 26.7 & 50.0 & 53.3 \\
Thinking (ref) & 32.4 & 22.0 & 38.9 & 2.2 & 23.3 & 22.6 \\
Task Arithmetic & 70.6 & 78.0 & 83.3 & 40.0 & 70.0 & 67.7 \\
TIES & 70.6 & 80.0 & 63.9 & 31.1 & 66.7 & 62.1 \\
SLERP & 70.6 & 84.0 & 55.6 & 37.8 & 50.0 & 60.5 \\
AIM-TA & 76.5 & 84.0 & 66.7 & 48.9 & 50.0 & 66.2 \\
AIM-TIES & 61.8 & 84.0 & 75.0 & 33.3 & 46.7 & 61.0 \\
LEWIS & 70.6 & 78.0 & 72.2 & 33.3 & 56.7 & 62.1 \\
RAIN & 38.2 & 20.0 & 44.4 & 13.3 & 16.7 & 25.6 \\
\method{} & 79.4 & 84.0 & 80.6 & 44.4 & 70.0 & 71.3 \\
\bottomrule
\end{tabular}
\end{adjustbox}
\end{table}

\section{Terminal-Bench v2 Detailed Results}
\label{app:terminalbench-detail}

This appendix collects the full Terminal-Bench v2 results. Appendix~\ref{app:tb-headline-full} reports the full per-method table at both scales, including pass@3, pass\_majority, TTC, and Daytona cost. Setup, sandbox specs, Daytona pricing, and parser configuration are documented in Appendix~\ref{app:terminalbench-setup}.

\paragraph{Metric definitions.} pass@1 is the OpenAI-style mean reward $=$ mean(c/5) $\times$ n\_tasks, the expected single-shot pass count. pass@3 is the OpenAI pass@$k$ estimator at $k=3$, $n=5$ attempts: per-task $1 - C(5-c,3)/C(5,3)$, summed over the 89 tasks; this predicts what the same model would have scored with 3 attempts/task instead of 5. pass@5 is best-of-5: a task counts as a pass if any of 5 attempts passed. pass\_majority requires $\geq 3/5$ attempts to pass (per-task rate $\geq 0.60$). pass\_majority differs from pass@3: pass@3 weights by the probability of a 3-shot subsample landing a pass; pass\_majority requires actual $\geq 3$ successes. ``Test time'' is the end-to-end Terminal-Bench harness wall time; tokens are aggregated for the launched attempts, while excluded tasks contribute zero tokens and remain in the 89-task success denominator.

\subsection{Full Per-Method Table}
\label{app:tb-headline-full}

Tables~\ref{tab:app_tb_30b_full} and~\ref{tab:app_tb_80b_full} report the full headline metrics. Bold marks the best value in each pass-rate, test-time, TTC, and Daytona column.

\begin{table}[!htbp]
\centering
\scriptsize
\caption{30B Terminal-Bench v2: full per-method metrics. Tokens and TTC are in millions; Input counts non-cached prefill tokens. Daytona \$ is the sandbox cost from the audit log.}
\label{tab:app_tb_30b_full}
\begin{adjustbox}{max width=\linewidth}
\begin{tabular}{lcccccrrrrr}
\toprule
Method & pass@1 & pass@3 & pass@5 & pass\_maj. & Test time & Input & Cached & Output & TTC & Daytona \$ \\
\midrule
Instruct (ref) & 4.8 (5.4\%) & 7.6 (8.5\%) & 9 (10.1\%) & 4 (4.5\%) & 4h 14m & 16.96 & 685.01 & 5.43 & 112.6 & \$7.34 \\
Thinking (ref) & 5.2 (5.8\%) & 9.4 (10.6\%) & 12 (13.5\%) & 4 (4.5\%) & 4h 37m & 4.34 & 122.24 & 18.41 & 108.6 & \$8.73 \\
Task Arithmetic & 4.8 (5.4\%) & 9.8 (11.0\%) & 13 (14.6\%) & 2 (2.2\%) & 2h 50m & 8.54 & 425.36 & 3.77 & 69.9 & \$4.95 \\
TIES & 5.4 (6.1\%) & 9.6 (10.8\%) & 12 (13.5\%) & 3 (3.4\%) & 2h 53m & 9.97 & 481.93 & 4.40 & 80.2 & \$5.02 \\
SLERP & 4.8 (5.4\%) & 9.9 (11.1\%) & 13 (14.6\%) & 3 (3.4\%) & 2h 51m & 7.13 & 468.41 & 3.80 & 73.0 & \$4.99 \\
AIM-TA & 5.0 (5.6\%) & 9.4 (10.6\%) & 12 (13.5\%) & 4 (4.5\%) & 2h 44m & 7.18 & 338.59 & 3.85 & 60.3 & \$5.00 \\
AIM-TIES & 5.0 (5.6\%) & 9.3 (10.4\%) & 12 (13.5\%) & 3 (3.4\%) & 2h 42m & 9.47 & 467.58 & 4.33 & 77.9 & \$4.67 \\
LEWIS & 4.6 (5.2\%) & 8.2 (9.2\%) & 10 (11.2\%) & 4 (4.5\%) & 2h 53m & 7.00 & 351.21 & 3.70 & 60.6 & \$5.21 \\
RAIN & 5.0 (5.6\%) & 7.9 (8.9\%) & 9 (10.1\%) & 4 (4.5\%) & 4h 05m & 4.01 & 114.61 & 16.76 & 99.3 & \$9.28 \\
\textbf{\method{}} & \textbf{6.8 (7.6\%)} & \textbf{12.4 (13.9\%)} & \textbf{16 (18.0\%)} & \textbf{7 (7.9\%)} & \textbf{2h 18m} & 7.68 & 319.35 & 3.70 & \textbf{58.1} & \textbf{\$4.18} \\
\bottomrule
\end{tabular}
\end{adjustbox}
\end{table}

\begin{table}[!htbp]
\centering
\scriptsize
\caption{80B Terminal-Bench v2: full per-method metrics. Units as in Table~\ref{tab:app_tb_30b_full}.}
\label{tab:app_tb_80b_full}
\begin{adjustbox}{max width=\linewidth}
\begin{tabular}{lcccccrrrrr}
\toprule
Method & pass@1 & pass@3 & pass@5 & pass\_maj. & Test time & Input & Cached & Output & TTC & Daytona \$ \\
\midrule
Instruct (ref) & 12.0 (13.5\%) & 17.4 (19.6\%) & 20 (22.5\%) & 12 (13.5\%) & 2h 28m & 10.84 & 224.62 & 3.85 & 52.6 & \$4.27 \\
Thinking (ref) & 6.0 (6.7\%) & 9.6 (10.8\%) & 12 (13.5\%) & 6 (6.7\%) & 5h 12m & 4.45 & 85.64 & 20.39 & 115.0 & \$12.02 \\
Task Arithmetic & 11.6 (13.0\%) & 19.1 (21.5\%) & 22 (24.7\%) & 11 (12.4\%) & 2h 10m & 266.39 & 255.57 & 3.65 & 310.2 & \$5.01 \\
TIES & 11.8 (13.3\%) & 20.5 (23.0\%) & 23 (25.8\%) & \textbf{13 (14.6\%)} & \textbf{1h 55m} & 11.71 & 285.22 & 3.86 & 59.5 & \textbf{\$4.20} \\
SLERP & 12.0 (13.5\%) & 19.9 (22.4\%) & 24 (27.0\%) & 10 (11.2\%) & 2h 08m & 12.96 & 249.13 & 3.55 & 55.6 & \$4.85 \\
AIM-TA & 12.2 (13.7\%) & 18.0 (20.2\%) & 20 (22.5\%) & 12 (13.5\%) & 2h 00m & 10.10 & 257.56 & 3.72 & 54.5 & \$6.03 \\
AIM-TIES & 12.6 (14.2\%) & 19.1 (21.5\%) & 22 (24.7\%) & 11 (12.4\%) & 2h 14m & 301.41 & 289.77 & 3.62 & 348.5 & \$4.76 \\
LEWIS & 12.6 (14.2\%) & 19.6 (22.0\%) & 23 (25.8\%) & \textbf{13 (14.6\%)} & 2h 11m & 10.59 & 248.36 & 3.74 & 54.1 & \$4.91 \\
RAIN & 7.0 (7.9\%) & 11.5 (12.9\%) & 14 (15.7\%) & 7 (7.9\%) & 4h 57m & 4.36 & 82.32 & 19.35 & 109.3 & \$11.69 \\
\textbf{\method{}} & \textbf{13.2 (14.8\%)} & \textbf{22.1 (24.8\%)} & \textbf{27 (30.3\%)} & 11 (12.4\%) & 1h 58m & 10.42 & 234.57 & 3.58 & \textbf{51.8} & \$4.42 \\
\bottomrule
\end{tabular}
\end{adjustbox}
\end{table}

\section{Additional Baselines}
\label{app:additional-baselines}

\subsection{Inference-Time Reasoning Control}
\label{app:thinkctl}

We test whether simple inference-time controls on the Thinking endpoint recover \method{}'s gain (Table~\ref{tab:app_thinkctl}). Every control runs the unchanged Qwen3-30B-A3B-Thinking-2507 checkpoint behind a proxy that edits only the decoding. C1 closes the think block immediately ($B=0$). C2a and C2b let the model think for $B=960$ or $B=1920$ tokens, then append a fixed exit phrase and close the block. C3 caps each turn at $N=512$ tokens and, if no tool call was emitted, closes the think block and injects a tool-call prefix.

$B=960$ is the budget whose simulated output retention (0.38) matches \method{}'s output-token ratio to Thinking (0.38), $B=1920$ doubles it, and $N=512$ rounds \method{}'s 95th-percentile tokens per step (290) up to a multiple of 256. The harness, decoding settings, and 4$\times$H100 serving follow the main Roo-Eval protocol. The best control, C2b, solves 106/195 exercises at pass@3, three more than uncapped Thinking and 56 fewer than \method{}.

\begin{table}[!htbp]
\centering
\small
\caption{Inference-time reasoning control on the 30B Thinking endpoint, Roo-Eval (195 exercises, three rollouts each). $B$ is the think-token budget; $N$ caps each turn and forces a tool call. Cells report solved exercises (\%); tokens are in millions.}
\label{tab:app_thinkctl}
\begin{adjustbox}{max width=\linewidth}
\begin{tabular}{llccccr}
\toprule
Row & Control & pass@1 & pass@3 & pass\_all & Output tok. & TTC \\
\midrule
Instruct (ref) & -- & 91 (46.7) & 125 (64.1) & 63 (32.3) & 8.37 & 181.1 \\
Thinking (ref) & $B=\infty$ & 68 (34.9) & 103 (52.8) & 35 (17.9) & 22.79 & 146.9 \\
\midrule
C1 & $B=0$ & 73 (37.4) & 101 (51.8) & 31 (15.9) & 22.48 & 145.4 \\
C2a & $B=960$ & 34 (17.4) & 59 (30.3) & 16 (8.2) & 14.80 & \textbf{117.0} \\
C2b & $B=1920$ & 70 (35.9) & 106 (54.4) & 32 (16.4) & 25.20 & 174.0 \\
C3 & $N=512$ & 69 (35.4) & 101 (51.8) & 31 (15.9) & 22.11 & 197.9 \\
\midrule
\method{} & -- & \textbf{129 (66.2)} & \textbf{162 (83.1)} & \textbf{86 (44.1)} & 8.76 & 120.9 \\
\bottomrule
\end{tabular}
\end{adjustbox}
\end{table}

\subsection{Lightweight Distillation with QLoRA}
\label{app:qlora}

We also compare \method{} with lightweight distillation (Table~\ref{tab:app_qlora}). Three QLoRA~\citep{dettmers2023qlora} students start from Qwen3-30B-A3B-Instruct-2507 and train on about 800 per-turn examples from successful agent rollouts (about 760 after a 5\% validation split). The keepthink and stripthink students learn from Thinking rollouts with the reasoning trace kept or removed. The self-direct student learns from Instruct's own direct rollouts, which controls for teacher quality. All three use an NF4 base, rank 32, LoRA scaling 64, dropout 0.05, adapters on the attention projections only (experts and router frozen), three epochs, sequence length 16{,}384, and learning rate $10^{-4}$. Training takes 7.7--10.1 GPU-hours per student on one H100, excluding rollout generation and evaluation (Table~\ref{tab:app_construction_cost}). Each adapter is merged into a bf16 checkpoint and evaluated with the main Roo-Eval protocol.

On Roo-Eval the Thinking teacher is weaker than Instruct, and both Thinking-distilled students fall below Instruct. The self-direct student essentially matches Instruct (124 vs.\ 125 exercises at pass@3) but does not improve on it. \method{} exceeds Instruct and the self-direct student by 19.0 and 19.5 points. At this training budget, lightweight distillation does not recover \method{}'s gain.

\begin{table}[!htbp]
\centering
\small
\caption{QLoRA distillation baselines at 30B, Roo-Eval pass@3. Cells report solved exercises (\%); language columns give the number of exercises.}
\label{tab:app_qlora}
\begin{adjustbox}{max width=\linewidth}
\begin{tabular}{llcccccc}
\toprule
Model & Teacher & Python (34) & JavaScript (50) & Go (36) & Java (45) & Rust (30) & \textbf{Overall} (195) \\
\midrule
Instruct (ref) & -- & 22 (64.7) & 37 (74.0) & 19 (52.8) & 32 (71.1) & 15 (50.0) & 125 (64.1) \\
Thinking (ref) & -- & 21 (61.8) & 27 (54.0) & 23 (63.9) & 21 (46.7) & 11 (36.7) & 103 (52.8) \\
\midrule
QLoRA keepthink & Thinking & 16 (47.1) & 28 (56.0) & 14 (38.9) & 26 (57.8) & 11 (36.7) & 95 (48.7) \\
QLoRA stripthink & Thinking & 16 (47.1) & 24 (48.0) & 11 (30.6) & 20 (44.4) & 6 (20.0) & 77 (39.5) \\
QLoRA self-direct & Instruct & 23 (67.6) & 37 (74.0) & 20 (55.6) & 30 (66.7) & 14 (46.7) & 124 (63.6) \\
\midrule
\method{} & -- & \textbf{31 (91.2)} & \textbf{42 (84.0)} & \textbf{30 (83.3)} & \textbf{37 (82.2)} & \textbf{22 (73.3)} & \textbf{162 (83.1)} \\
\bottomrule
\end{tabular}
\end{adjustbox}
\end{table}

\section{Format Preservation Under Stage 3}
\label{app:stage3-evidence}

We test whether Stage~3 preserves the agent's interaction protocol, using the Qwen3-30B-A3B and Qwen3-Next-80B-A3B runs of the main experiments. Roo-Eval runs its 195 exercises with three rollouts per condition under native OpenAI function calling. Terminal-Bench v2 is an independent scaffold whose archived requests contain complete OpenAI tool specifications with JSON Schemas, so its tool calls can also be checked against the declared schema. A leak is a protected chat-template or tool-call delimiter (Table~\ref{tab:app_stage3_tokens}) that appears in visible assistant text.

The GSP threshold $\tau$ controls this protection on Roo-Eval (Table~\ref{tab:app_stage3_roo}). At 30B, \method{} leaks 0.21 times per 1k turns, in 3/195 exercises. The stronger setting $\tau=0.003$ gives 0.06, while the weaker setting $\tau=0.3$ raises leakage to 7.25 per 1k turns in 26/195 exercises. Removing GSP raises it to 89.10 per 1k turns in 118/195 exercises, mostly through \texttt{<|im\_start|>} and \texttt{<tool\_call>} (Table~\ref{tab:app_stage3_tokens}). At 80B, \method{} leaks 0.15 times per 1k turns in 2/195 exercises, against 3.03 in 17/195 without GSP.

Terminal-Bench v2 replicates the leakage result on a second scaffold (Table~\ref{tab:app_stage3_tb}). \method{} leaks 1.32 and 1.14 times per 1k calls at 30B and 80B, comparable to Instruct at 30B (1.19) and below it at 80B (2.06). Without GSP, the rates rise to 207.10 and 4.92.

\method{} also keeps tool calls well formed on Terminal-Bench v2 (Table~\ref{tab:app_stage3_json}). Its JSON-invalid rate is below both endpoints at both scales. Its schema-invalid rate is 4.12 per 1k tool calls at 30B, close to Thinking (4.04) and below Instruct (10.86), and 2.40 at 80B, below Instruct (2.78). It calls an undeclared tool once at 30B and never at 80B.

\begin{table}[!htbp]
\centering
\small
\caption{Protected-delimiter leakage on Roo-Eval (195 exercises, three rollouts each). Rates are occurrences per 1k model turns. The 30B rows include the GSP threshold sweep; the reported recipe uses $\tau=0.03$.}
\label{tab:app_stage3_roo}
\begin{tabular}{llrrr}
\toprule
Scale & Variant & Turns & Leaks / 1k turns & Exercises with a leak \\
\midrule
30B & Instruct & 44,564 & 0.00 & 0/195 \\
 & Thinking & 26,727 & 0.41 & 2/195 \\
 & \method{}, $\tau=0.003$ & 46,824 & 0.06 & 2/195 \\
 & \method{} & 47,320 & 0.21 & 3/195 \\
 & \method{}, $\tau=0.3$ & 40,551 & 7.25 & 26/195 \\
 & \method{} w/o GSP & 38,180 & 89.10 & 118/195 \\
\midrule
80B & Instruct & 35,226 & 0.03 & 1/195 \\
 & Thinking & 20,576 & 1.94 & 14/195 \\
 & \method{} & 32,928 & 0.15 & 2/195 \\
 & \method{} w/o GSP & 34,686 & 3.03 & 17/195 \\
\bottomrule
\end{tabular}
\end{table}

\begin{table}[!htbp]
\centering
\small
\caption{Leaked protected delimiters by token, 30B Roo-Eval, raw counts over all rollouts.}
\label{tab:app_stage3_tokens}
\begin{tabular}{lrrrr}
\toprule
Variant & \texttt{<|im\_start|>} & \texttt{<|im\_end|>} & \texttt{<tool\_call>} & \texttt{<tool\_response>} \\
\midrule
Instruct & 0 & 0 & 0 & 0 \\
Thinking & 0 & 0 & 11 & 0 \\
\method{}, $\tau=0.003$ & 2 & 1 & 0 & 0 \\
\method{} & 7 & 0 & 3 & 0 \\
\method{}, $\tau=0.3$ & 171 & 25 & 98 & 0 \\
\method{} w/o GSP & 2,308 & 5 & 1,089 & 0 \\
\bottomrule
\end{tabular}
\end{table}

\begin{table}[!htbp]
\centering
\small
\caption{Protected-delimiter leakage on Terminal-Bench v2.}
\label{tab:app_stage3_tb}
\begin{tabular}{llrrr}
\toprule
Scale & Variant & LLM calls & Leaks & Leaks / 1k calls \\
\midrule
30B & Instruct & 16,808 & 20 & 1.19 \\
 & Thinking & 6,543 & 2 & 0.31 \\
 & \method{} & 10,588 & 14 & 1.32 \\
 & \method{} w/o GSP & 11,106 & 2,300 & 207.10 \\
\midrule
80B & Instruct & 9,201 & 19 & 2.06 \\
 & Thinking & 6,039 & 1 & 0.17 \\
 & \method{} & 9,655 & 11 & 1.14 \\
 & \method{} w/o GSP & 9,766 & 48 & 4.92 \\
\bottomrule
\end{tabular}
\end{table}

\begin{table}[!htbp]
\centering
\small
\caption{Tool-call validity on Terminal-Bench v2, checked against the declared JSON Schemas. JSON-invalid is per 1k LLM calls; schema-invalid is per 1k tool calls; unknown tool counts calls to an undeclared tool. Runs as in Table~\ref{tab:app_stage3_tb}.}
\label{tab:app_stage3_json}
\begin{adjustbox}{max width=\linewidth}
\begin{tabular}{llrrrrr}
\toprule
Scale & Variant & LLM calls & Tool calls & JSON-invalid / 1k & Schema-invalid / 1k & Unknown tool \\
\midrule
30B & Instruct & 16,808 & 16,024 & 16.54 & 10.86 & 33 \\
 & Thinking & 6,543 & 6,181 & 16.96 & 4.04 & 20 \\
 & \method{} & 10,588 & 9,707 & 8.50 & 4.12 & 1 \\
\midrule
80B & Instruct & 9,201 & 8,648 & 4.02 & 2.78 & 1 \\
 & Thinking & 6,039 & 4,766 & 7.62 & 0.00 & 0 \\
 & \method{} & 9,655 & 9,151 & 2.90 & 2.40 & 0 \\
\bottomrule
\end{tabular}
\end{adjustbox}
\end{table}

\section{Ablations}
\label{app:roo-ablation-summary}

Table~\ref{tab:ablation_sweeps} reports the Roo-Eval $\alpha$ and $\tau$ sweep values corresponding to the Roo panels in Figure~\ref{fig:ablation_curves} of Section~\ref{sec:ablation}, including the TTC row omitted from the figure. Tables~\ref{tab:app_ablation_modules_tb_full} and~\ref{tab:app_ablation_modules_swe_full} report the full per-variant token breakdowns for the Terminal-Bench v2 and SWE-bench-Verified component-removal ablations summarized in the lower block of Table~\ref{tab:ablation_modules}.

\begin{table}[!htbp]
\centering
\small
\caption{Full per-variant Terminal-Bench v2 component-removal ablations. ``Input'' is non-cached prefill tokens (M); ``Output'' is generated tokens (M); ``TTC'' = $N_i + 0.1 N_c + 5 N_o$ (M).}
\label{tab:app_ablation_modules_tb_full}
\begin{adjustbox}{max width=\linewidth}
\begin{tabular}{lccrrrccrrr}
\toprule
& \multicolumn{5}{c}{\textbf{Qwen3-30B-A3B}} & \multicolumn{5}{c}{\textbf{Qwen3-Next-80B-A3B}} \\
\cmidrule(lr){2-6} \cmidrule(lr){7-11}
Method & pass@1 & pass@5 & Input & Output & TTC & pass@1 & pass@5 & Input & Output & TTC \\
\midrule
\method{} w/o $T(\delta)$ & 6.80 (7.6\%) & 12 (13.5\%)          & 13.47 & 4.92 & 94.1 & 12.20 (13.7\%) & 21 (23.6\%)         & 10.86 & 3.49 & 52.8 \\
\method{} w/o CTG         & 5.80 (6.5\%) & 14 (15.7\%)          & 12.02 & 4.61 & 85.1 & 11.60 (13.0\%) & 22 (24.7\%)          &  9.95 & 3.61 & \textbf{50.4} \\
\method{} w/o GSP         & 4.80 (5.4\%) & 11 (12.4\%)          &  4.78 & 3.56 & \textbf{42.5} & 11.40 (12.8\%) & 19 (21.3\%)          & 11.80 & 3.79 & 57.3 \\
\midrule
\textbf{\method{}} (full) & \textbf{6.80 (7.6\%)} & \textbf{16 (18.0\%)} & 7.68 & 3.70 & 58.1 & \textbf{13.20 (14.8\%)} & \textbf{27 (30.3\%)} & 10.42 & 3.58 & 51.8 \\
\bottomrule
\end{tabular}
\end{adjustbox}
\end{table}

\begin{table}[!htbp]
\centering
\small
\caption{Full per-variant SWE-bench-Verified component-removal ablations. ``Compl.'' counts patches that completed grading; ``Empty'' counts predictions filtered for empty patches before grading; ``Output'' is generated tokens (M); ``TTC'' = $N_i + 0.1 N_c + 5 N_o$ (B). The full-recipe row's ``Empty'' is omitted because the headline run did not log it separately.}
\label{tab:app_ablation_modules_swe_full}
\begin{adjustbox}{max width=\linewidth}
\begin{tabular}{lccrrrcccrr}
\toprule
& \multicolumn{5}{c}{\textbf{Qwen3-30B-A3B}} & \multicolumn{5}{c}{\textbf{Qwen3-Next-80B-A3B}} \\
\cmidrule(lr){2-6} \cmidrule(lr){7-11}
Method & Resolved & Compl. & Empty & Output & TTC & Resolved & Compl. & Empty & Output & TTC \\
\midrule
\method{} w/o $T(\delta)$ & 120 (24.0\%)          & 439 &  60 & 316 & 8.43 & 164 (32.8\%)          & 488 & 10 & 305 & 5.51 \\
\method{} w/o CTG         & 106 (21.2\%)          & 454 &  43 & 308 & 7.34 & 162 (32.4\%)          & 483 & 15 & 313 & 5.50 \\
\method{} w/o GSP         &  94 (18.8\%)          & 374 & 116 & 476 & \textbf{5.35} & 175 (35.0\%)          & 485 & 12 & 334 & 5.35 \\
\midrule
\textbf{\method{}} (full) & \textbf{122 (24.4\%)} & 460 & --- & 373 & 5.68 & \textbf{180 (36.0\%)} & 487 & --- & 309 & \textbf{5.22} \\
\bottomrule
\end{tabular}
\end{adjustbox}
\end{table}

\begin{table}[!htbp]
\centering
\small
\caption{Continuous-hyperparameter sweeps of the \method{} recipe on Qwen3-30B-A3B Roo-Eval. The bold column is the reported configuration ($\alpha=0.25$, $\tau=0.03$); the $\alpha$ sweep varies $\alpha$ at fixed $\tau=0.03$, and the $\tau$ sweep varies $\tau$ at fixed $\alpha=0.25$. pass@1 / pass@3 / pass\_all are exercise-weighted aggregates over the 195 Roo-Eval exercises; per-language splits follow below.}
\label{tab:ablation_sweeps}
\begin{adjustbox}{max width=\linewidth}
\begin{tabular}{lcccccccc}
\toprule
& \textbf{reported} & \multicolumn{4}{c}{$\alpha$ sweep ($\tau=0.03$)} & & \multicolumn{2}{c}{$\tau$ sweep ($\alpha=0.25$)} \\
\cmidrule(lr){3-6} \cmidrule(lr){8-9}
Metric & \textbf{$\alpha=0.25$, $\tau=0.03$} & $\alpha=0.15$ & $\alpha=0.20$ & $\alpha=0.30$ & $\alpha=0.35$ & & $\tau=0.003$ & $\tau=0.3$ \\
\midrule
pass@1 (\%)    & \textbf{66.2}  & 47.2  & 63.1  & 54.4  & 39.5  & & 63.1  & 52.3 \\
pass@3 (\%)    & \textbf{83.1}  & 63.1  & 78.5  & 74.9  & 61.0  & & 80.5  & 76.4 \\
pass\_all (\%) & \textbf{44.1}  & 33.3  & 47.7  & 31.8  & 16.9  & & 43.1  & 29.7 \\
TTC (M)        & \textbf{120.9} & 148.6 & 129.0 & 94.2 & 80.5 & & 120.8 & 104.4 \\
\bottomrule
\end{tabular}
\end{adjustbox}
\end{table}

This section contains two groups of tables. The first group is four pass@1 summary tables: Tables~\ref{tab:app_roo_alpha_summary},~\ref{tab:app_roo_tau_summary}, and~\ref{tab:app_roo_component_summary} report 30B Roo-Eval pass@1 percentages by language for the $\alpha$ sweep, $\tau$ sweep, and component-removal ablations respectively, and Table~\ref{tab:app_roo_component_80b_summary} reports the corresponding 80B \method{} component ablations. The final column of each summary reports the five-language TTC computed from recorded local-vLLM token usage. The second group is four detail tables (Tables~\ref{tab:app_roo_alpha_detail_by_language},~\ref{tab:app_roo_tau_detail_by_language},~\ref{tab:app_roo_component_detail_by_language}, and~\ref{tab:app_roo_component_80b_detail_by_language}) that group each ablation family by programming language and retain pass@1, pass@3, pass\_all, iter pass, TTC, and recorded input/cached/output token totals and averages.

\begin{table}[!htbp]
\centering
\scriptsize
\caption{Global merge-scale $\alpha$ sweep on the 30B \method{} recipe.}
\label{tab:app_roo_alpha_summary}
\begin{adjustbox}{max width=\linewidth}
\begin{tabular}{llllllllr}
\toprule
Variant  & Python & JavaScript & Go & Java & Rust & \textbf{Overall} & TTC \\
\midrule
$\alpha=0.15$ & 70.6 & 72.0 & 52.8 & 4.4 & 36.7 & 47.2 & 148.6M \\
$\alpha=0.20$ & 61.8 & 78.0 & 66.7 & 51.1 & 53.3 & 63.1 & 129.0M \\
$\alpha=0.30$ & 61.8 & 66.0 & 52.8 & 48.9 & 36.7 & 54.4 & 94.2M \\
$\alpha=0.35$ & 50.0 & 46.0 & 38.9 & 31.1 & 30.0 & 39.5 & 80.5M \\
\method{} & 79.4 & 78.0 & 75.0 & 53.3 & 40.0 & 66.2 & 120.9M \\
\bottomrule
\end{tabular}
\end{adjustbox}
\vspace{0.25em}
\end{table}

\begin{table}[!htbp]
\centering
\scriptsize
\caption{GSP threshold sweep on the 30B \method{} recipe.}
\label{tab:app_roo_tau_summary}
\begin{adjustbox}{max width=\linewidth}
\begin{tabular}{lrrrrrrr}
\toprule
Variant & Python & JavaScript & Go & Java & Rust & \textbf{Overall} & TTC \\
\midrule
\method{} ($\tau=0.003$) & 70.6 & 76.0 & 63.9 & 53.3 & 46.7 & 63.1 & 120.8M \\
\method{} ($\tau=0.3$) & 55.9 & 62.0 & 50.0 & 53.3 & 33.3 & 52.3 & 104.4M \\
\method{} ($\tau=0.03$) & 79.4 & 78.0 & 75.0 & 53.3 & 40.0 & 66.2 & 120.9M \\
\bottomrule
\end{tabular}
\end{adjustbox}
\end{table}

\begin{table}[!htbp]
\centering
\scriptsize
\caption{Component-removal ablations for the 30B \method{} recipe.}
\label{tab:app_roo_component_summary}
\begin{adjustbox}{max width=\linewidth}
\begin{tabular}{lrrrrrrr}
\toprule
Variant & Python & JavaScript & Go & Java & Rust & \textbf{Overall} & TTC \\
\midrule
\method{} w/o $T(\delta)$ & 73.5 & 70.0 & 58.3 & 57.8 & 36.7 & 60.5 & 142.3M \\
\method{} w/o CTG & 58.8 & 70.0 & 61.1 & 48.9 & 43.3 & 57.4 & 145.7M \\
\method{} w/o GSP & 58.8 & 58.0 & 30.6 & 51.1 & 56.7 & 51.3 & 100.8M \\
\method{} (full) & 79.4 & 78.0 & 75.0 & 53.3 & 40.0 & 66.2 & 120.9M \\
\bottomrule
\end{tabular}
\end{adjustbox}
\end{table}

\begin{table}[!htbp]
\centering
\scriptsize
\caption{Component-removal ablations for the 80B \method{} recipe. The full recipe uses $\alpha=0.15$, $\tau=0.03$, architecture-normalized CTG, and GSP for attention, linear-attention inner slots, and routers.}
\label{tab:app_roo_component_80b_summary}
\begin{adjustbox}{max width=\linewidth}
\begin{tabular}{lrrrrrrr}
\toprule
Variant & Python & JavaScript & Go & Java & Rust & \textbf{Overall} & TTC \\
\midrule
\method{} w/o $T(\delta)$ & 85.3 & 90.0 & 86.1 & 66.7 & 63.3 & 79.0 & 97.8M \\
\method{} w/o CTG & 88.2 & 92.0 & 83.3 & 51.1 & 73.3 & 77.4 & 106.2M \\
\method{} w/o GSP & 88.2 & 72.0 & 86.1 & 73.3 & 73.3 & 77.9 & 109.7M \\
\method{} (full) & 88.2 & 92.0 & 86.1 & 62.2 & 80.0 & 81.5 & 89.2M \\
\bottomrule
\end{tabular}
\end{adjustbox}
\end{table}

\paragraph{Alpha sweep detailed per-language results.}

Table~\ref{tab:app_roo_alpha_detail_by_language} reports per-language pass metrics, TTC, and recorded local-vLLM token usage for each row in this ablation family.

\begin{table}[!htbp]
\centering
\tiny
\setlength{\tabcolsep}{2pt}
\caption{30B alpha sweep detailed Roo-Eval metrics by language, including token usage.}
\label{tab:app_roo_alpha_detail_by_language}
\begin{adjustbox}{max width=\linewidth}
\begin{tabular}{llllllllllll}
\toprule
Model & pass@1 & pass@3 & pass\_all & iter pass & TTC & Input total & Cached total & Output total & Input avg & Cached avg & Output avg \\
\midrule
\multicolumn{12}{l}{\textbf{Python (34 exercises $\times$ 3 = 102 rollouts)}} \\
\midrule
$\alpha=0.15$ & 24 (70.6\%) & 29 (85.3\%) & 18 (52.9\%) & 70/102 (68.6\%) & 23.5M & 6,310,035 & 96,501,372 & 1,499,610 & 61,863 & 946,091 & 14,702 \\
$\alpha=0.20$ & 21 (61.8\%) & 26 (76.5\%) & 19 (55.9\%) & 67/102 (65.7\%) & 21.1M & 5,976,877 & 68,969,826 & 1,636,553 & 58,596 & 676,174 & 16,044 \\
$\alpha=0.30$ & 21 (61.8\%) & 26 (76.5\%) & 15 (44.1\%) & 63/102 (61.8\%) & 15.1M & 4,810,657 & 45,377,736 & 1,149,974 & 47,163 & 444,879 & 11,274 \\
$\alpha=0.35$ & 17 (50.0\%) & 25 (73.5\%) & 8 (23.5\%) & 50/102 (49.0\%) & 13.5M & 4,612,640 & 37,437,300 & 1,021,834 & 45,221 & 367,032 & 10,017 \\
\method{} ($\alpha=0.25$) & 27 (79.4\%) & 31 (91.2\%) & 19 (55.9\%) & 74/102 (72.5\%) & 19.4M & 5,605,858 & 63,459,202 & 1,480,496 & 54,959 & 622,149 & 14,514 \\
\midrule
\multicolumn{12}{l}{\textbf{JavaScript (50 exercises $\times$ 3 = 150 rollouts)}} \\
\midrule
$\alpha=0.15$ & 36 (72.0\%) & 42 (84.0\%) & 29 (58.0\%) & 109/150 (72.7\%) & 33.9M & 9,533,664 & 146,786,597 & 1,931,367 & 63,557 & 978,577 & 12,875 \\
$\alpha=0.20$ & 39 (78.0\%) & 42 (84.0\%) & 35 (70.0\%) & 115/150 (76.7\%) & 28.9M & 8,366,804 & 107,114,322 & 1,961,024 & 55,778 & 714,095 & 13,073 \\
$\alpha=0.30$ & 33 (66.0\%) & 40 (80.0\%) & 22 (44.0\%) & 97/150 (64.7\%) & 23.3M & 7,714,444 & 76,644,299 & 1,591,815 & 51,429 & 510,961 & 10,612 \\
$\alpha=0.35$ & 23 (46.0\%) & 34 (68.0\%) & 12 (24.0\%) & 68/150 (45.3\%) & 18.6M & 6,894,052 & 55,705,859 & 1,226,625 & 45,960 & 371,372 & 8,177 \\
\method{} ($\alpha=0.25$) & 39 (78.0\%) & 42 (84.0\%) & 30 (60.0\%) & 111/150 (74.0\%) & 26.1M & 8,027,932 & 93,420,273 & 1,753,243 & 53,519 & 622,801 & 11,688 \\
\midrule
\multicolumn{12}{l}{\textbf{Go (36 exercises $\times$ 3 = 108 rollouts)}} \\
\midrule
$\alpha=0.15$ & 19 (52.8\%) & 25 (69.4\%) & 12 (33.3\%) & 57/108 (52.8\%) & 29.1M & 7,730,307 & 114,301,139 & 1,981,231 & 71,576 & 1,058,343 & 18,344 \\
$\alpha=0.20$ & 24 (66.7\%) & 29 (80.6\%) & 19 (52.8\%) & 74/108 (68.5\%) & 23.8M & 6,213,547 & 84,881,875 & 1,809,862 & 57,532 & 785,943 & 16,757 \\
$\alpha=0.30$ & 19 (52.8\%) & 26 (72.2\%) & 13 (36.1\%) & 60/108 (55.6\%) & 16.4M & 5,133,709 & 49,002,330 & 1,274,416 & 47,534 & 453,725 & 11,800 \\
$\alpha=0.35$ & 14 (38.9\%) & 18 (50.0\%) & 4 (11.1\%) & 33/108 (30.6\%) & 15.9M & 5,179,880 & 46,117,197 & 1,214,226 & 47,961 & 427,011 & 11,242 \\
\method{} ($\alpha=0.25$) & 27 (75.0\%) & 30 (83.3\%) & 18 (50.0\%) & 72/108 (66.7\%) & 21.8M & 6,025,226 & 73,353,048 & 1,684,501 & 55,789 & 679,194 & 15,597 \\
\midrule
\multicolumn{12}{l}{\textbf{Java (45 exercises $\times$ 3 = 135 rollouts)}} \\
\midrule
$\alpha=0.15$ & 2 (4.4\%) & 7 (15.6\%) & 0 (0.0\%) & 11/135 (8.1\%) & 34.5M & 10,678,999 & 155,441,947 & 1,663,109 & 79,103 & 1,151,421 & 12,319 \\
$\alpha=0.20$ & 23 (51.1\%) & 34 (75.6\%) & 13 (28.9\%) & 74/135 (54.8\%) & 28.7M & 8,297,995 & 103,038,820 & 2,027,443 & 61,466 & 763,250 & 15,018 \\
$\alpha=0.30$ & 22 (48.9\%) & 33 (73.3\%) & 7 (15.6\%) & 63/135 (46.7\%) & 21.9M & 6,956,536 & 76,181,499 & 1,473,820 & 51,529 & 564,307 & 10,917 \\
$\alpha=0.35$ & 14 (31.1\%) & 26 (57.8\%) & 6 (13.3\%) & 50/135 (37.0\%) & 18.8M & 6,408,116 & 58,432,458 & 1,301,752 & 47,467 & 432,833 & 9,642 \\
\method{} ($\alpha=0.25$) & 24 (53.3\%) & 37 (82.2\%) & 10 (22.2\%) & 70/135 (51.9\%) & 32.0M & 9,008,906 & 117,821,938 & 2,247,297 & 66,732 & 872,755 & 16,646 \\
\midrule
\multicolumn{12}{l}{\textbf{Rust (30 exercises $\times$ 3 = 90 rollouts)}} \\
\midrule
$\alpha=0.15$ & 11 (36.7\%) & 20 (66.7\%) & 6 (20.0\%) & 40/90 (44.4\%) & 27.6M & 7,018,272 & 117,218,617 & 1,777,625 & 77,980 & 1,302,429 & 19,751 \\
$\alpha=0.20$ & 16 (53.3\%) & 22 (73.3\%) & 7 (23.3\%) & 43/90 (47.8\%) & 26.6M & 6,752,178 & 99,484,904 & 1,974,889 & 75,024 & 1,105,387 & 21,943 \\
$\alpha=0.30$ & 11 (36.7\%) & 21 (70.0\%) & 5 (16.7\%) & 37/90 (41.1\%) & 17.5M & 5,310,968 & 55,219,647 & 1,324,501 & 59,010 & 613,551 & 14,716 \\
$\alpha=0.35$ & 9 (30.0\%) & 16 (53.3\%) & 3 (10.0\%) & 31/90 (34.4\%) & 13.8M & 4,542,968 & 41,701,994 & 1,010,861 & 50,477 & 463,355 & 11,231 \\
\method{} ($\alpha=0.25$) & 12 (40.0\%) & 22 (73.3\%) & 9 (30.0\%) & 41/90 (45.6\%) & 21.6M & 6,010,939 & 76,419,820 & 1,593,906 & 66,788 & 849,109 & 17,710 \\
\bottomrule
\end{tabular}
\end{adjustbox}
\end{table}
\paragraph{Tau (GSP threshold) sweep detailed per-language results.}

Table~\ref{tab:app_roo_tau_detail_by_language} reports per-language pass metrics, TTC, and recorded local-vLLM token usage for each row in this ablation family.

\begin{table}[!htbp]
\centering
\tiny
\setlength{\tabcolsep}{2pt}
\caption{30B GSP-threshold $\tau$ sweep detailed Roo-Eval metrics by language, including token usage.}
\label{tab:app_roo_tau_detail_by_language}
\begin{adjustbox}{max width=\linewidth}
\begin{tabular}{llllllllllll}
\toprule
Model & pass@1 & pass@3 & pass\_all & iter pass & TTC & Input total & Cached total & Output total & Input avg & Cached avg & Output avg \\
\midrule
\multicolumn{12}{l}{\textbf{Python (34 exercises $\times$ 3 = 102 rollouts)}} \\
\midrule
\method{} ($\tau=0.003$) & 24 (70.6\%) & 27 (79.4\%) & 19 (55.9\%) & 71/102 (69.6\%) & 19.9M & 5,733,004 & 63,733,531 & 1,566,538 & 56,205 & 624,838 & 15,358 \\
\method{} ($\tau=0.3$) & 19 (55.9\%) & 26 (76.5\%) & 17 (50.0\%) & 64/102 (62.7\%) & 15.3M & 4,760,484 & 46,344,278 & 1,188,137 & 46,671 & 454,355 & 11,648 \\
\midrule
\multicolumn{12}{l}{\textbf{JavaScript (50 exercises $\times$ 3 = 150 rollouts)}} \\
\midrule
\method{} ($\tau=0.003$) & 38 (76.0\%) & 43 (86.0\%) & 34 (68.0\%) & 114/150 (76.0\%) & 25.8M & 7,970,896 & 90,562,897 & 1,753,504 & 53,139 & 603,752 & 11,690 \\
\method{} ($\tau=0.3$) & 31 (62.0\%) & 40 (80.0\%) & 24 (48.0\%) & 97/150 (64.7\%) & 25.0M & 7,887,087 & 85,462,251 & 1,714,622 & 52,580 & 569,748 & 11,430 \\
\midrule
\multicolumn{12}{l}{\textbf{Go (36 exercises $\times$ 3 = 108 rollouts)}} \\
\midrule
\method{} ($\tau=0.003$) & 23 (63.9\%) & 30 (83.3\%) & 15 (41.7\%) & 68/108 (63.0\%) & 22.5M & 6,062,990 & 74,891,532 & 1,782,621 & 56,138 & 693,440 & 16,505 \\
\method{} ($\tau=0.3$) & 18 (50.0\%) & 28 (77.8\%) & 8 (22.2\%) & 53/108 (49.1\%) & 19.1M & 5,541,504 & 60,447,267 & 1,500,727 & 51,310 & 559,696 & 13,895 \\
\midrule
\multicolumn{12}{l}{\textbf{Java (45 exercises $\times$ 3 = 135 rollouts)}} \\
\midrule
\method{} ($\tau=0.003$) & 24 (53.3\%) & 35 (77.8\%) & 10 (22.2\%) & 70/135 (51.9\%) & 28.4M & 8,258,313 & 100,277,698 & 2,014,566 & 61,172 & 742,797 & 14,922 \\
\method{} ($\tau=0.3$) & 24 (53.3\%) & 35 (77.8\%) & 7 (15.6\%) & 66/135 (48.9\%) & 24.5M & 7,497,429 & 82,818,956 & 1,745,685 & 55,536 & 613,473 & 12,931 \\
\midrule
\multicolumn{12}{l}{\textbf{Rust (30 exercises $\times$ 3 = 90 rollouts)}} \\
\midrule
\method{} ($\tau=0.003$) & 14 (46.7\%) & 22 (73.3\%) & 6 (20.0\%) & 45/90 (50.0\%) & 24.2M & 6,419,556 & 88,119,661 & 1,794,838 & 71,328 & 979,107 & 19,942 \\
\method{} ($\tau=0.3$) & 10 (33.3\%) & 20 (66.7\%) & 2 (6.7\%) & 31/90 (34.4\%) & 20.5M & 5,793,208 & 73,444,760 & 1,471,464 & 64,368 & 816,052 & 16,349 \\
\bottomrule
\end{tabular}
\end{adjustbox}
\end{table}
\paragraph{Component-ablation detailed per-language results.}

Table~\ref{tab:app_roo_component_detail_by_language} reports per-language pass metrics, TTC, and recorded local-vLLM token usage for each row in this ablation family.

\begin{table}[!htbp]
\centering
\tiny
\setlength{\tabcolsep}{2pt}
\caption{30B component ablation detailed Roo-Eval metrics by language, including token usage.}
\label{tab:app_roo_component_detail_by_language}
\begin{adjustbox}{max width=\linewidth}
\begin{tabular}{llllllllllll}
\toprule
Model & pass@1 & pass@3 & pass\_all & iter pass & TTC & Input total & Cached total & Output total & Input avg & Cached avg & Output avg \\
\midrule
\multicolumn{12}{l}{\textbf{Python (34 exercises $\times$ 3 = 102 rollouts)}} \\
\midrule
\method{} w/o $T(\delta)$ & 25 (73.5\%) & 28 (82.4\%) & 19 (55.9\%) & 70/102 (68.6\%) & 23.1M & 5,998,159 & 97,013,123 & 1,487,554 & 58,805 & 951,109 & 14,583 \\
\method{} w/o CTG & 20 (58.8\%) & 27 (79.4\%) & 15 (44.1\%) & 62/102 (60.8\%) & 24.9M & 6,698,445 & 104,306,164 & 1,563,142 & 65,671 & 1,022,609 & 15,324 \\
\method{} w/o GSP & 20 (58.8\%) & 25 (73.5\%) & 15 (44.1\%) & 59/102 (57.8\%) & 18.0M & 5,487,685 & 54,728,526 & 1,401,681 & 53,800 & 536,554 & 13,741 \\
\midrule
\multicolumn{12}{l}{\textbf{JavaScript (50 exercises $\times$ 3 = 150 rollouts)}} \\
\midrule
\method{} w/o $T(\delta)$ & 35 (70.0\%) & 43 (86.0\%) & 31 (62.0\%) & 111/150 (74.0\%) & 31.9M & 9,261,800 & 130,686,941 & 1,910,699 & 61,745 & 871,246 & 12,737 \\
\method{} w/o CTG & 35 (70.0\%) & 45 (90.0\%) & 25 (50.0\%) & 105/150 (70.0\%) & 33.6M & 9,674,038 & 145,091,970 & 1,876,068 & 64,493 & 967,279 & 12,507 \\
\method{} w/o GSP & 29 (58.0\%) & 41 (82.0\%) & 19 (38.0\%) & 93/150 (62.0\%) & 23.1M & 7,487,656 & 72,072,493 & 1,687,293 & 49,917 & 480,483 & 11,248 \\
\midrule
\multicolumn{12}{l}{\textbf{Go (36 exercises $\times$ 3 = 108 rollouts)}} \\
\midrule
\method{} w/o $T(\delta)$ & 21 (58.3\%) & 28 (77.8\%) & 17 (47.2\%) & 68/108 (63.0\%) & 26.2M & 6,660,774 & 99,054,445 & 1,923,562 & 61,673 & 917,170 & 17,810 \\
\method{} w/o CTG & 22 (61.1\%) & 30 (83.3\%) & 9 (25.0\%) & 61/108 (56.5\%) & 25.4M & 6,598,698 & 104,522,691 & 1,678,456 & 61,099 & 967,802 & 15,541 \\
\method{} w/o GSP & 11 (30.6\%) & 22 (61.1\%) & 8 (22.2\%) & 43/108 (39.8\%) & 19.8M & 5,751,413 & 60,167,905 & 1,609,440 & 53,253 & 557,110 & 14,902 \\
\midrule
\multicolumn{12}{l}{\textbf{Java (45 exercises $\times$ 3 = 135 rollouts)}} \\
\midrule
\method{} w/o $T(\delta)$ & 26 (57.8\%) & 33 (73.3\%) & 16 (35.6\%) & 76/135 (56.3\%) & 34.3M & 9,172,836 & 142,735,263 & 2,164,033 & 67,946 & 1,057,298 & 16,029 \\
\method{} w/o CTG & 22 (48.9\%) & 33 (73.3\%) & 13 (28.9\%) & 70/135 (51.9\%) & 34.1M & 9,112,509 & 150,035,462 & 1,992,039 & 67,500 & 1,111,373 & 14,755 \\
\method{} w/o GSP & 23 (51.1\%) & 31 (68.9\%) & 13 (28.9\%) & 69/135 (51.1\%) & 23.1M & 7,085,661 & 75,620,397 & 1,695,966 & 52,486 & 560,151 & 12,562 \\
\midrule
\multicolumn{12}{l}{\textbf{Rust (30 exercises $\times$ 3 = 90 rollouts)}} \\
\midrule
\method{} w/o $T(\delta)$ & 11 (36.7\%) & 23 (76.7\%) & 7 (23.3\%) & 43/90 (47.8\%) & 26.8M & 6,916,585 & 108,020,668 & 1,817,406 & 76,850 & 1,200,229 & 20,193 \\
\method{} w/o CTG & 13 (43.3\%) & 20 (66.7\%) & 6 (20.0\%) & 38/90 (42.2\%) & 27.7M & 7,107,117 & 118,415,728 & 1,752,833 & 78,967 & 1,315,730 & 19,475 \\
\method{} w/o GSP & 17 (56.7\%) & 21 (70.0\%) & 7 (23.3\%) & 45/90 (50.0\%) & 16.7M & 4,971,062 & 55,546,738 & 1,244,441 & 55,234 & 617,185 & 13,827 \\
\bottomrule
\end{tabular}
\end{adjustbox}
\end{table}

\paragraph{80B component-ablation detailed per-language results.}

Table~\ref{tab:app_roo_component_80b_detail_by_language} reports the 80B \method{} full recipe and its one-component removals by language. All rows use the same $\alpha=0.15$, $\tau=0.03$, Qwen3-Next-80B-A3B Instruct/Thinking pair, and Roo-Eval serving configuration; each ablation removes exactly one of $T(\delta)$, CTG, or GSP.

\begin{table}[!htbp]
\centering
\tiny
\setlength{\tabcolsep}{2pt}
\caption{80B \method{} component-ablation detailed Roo-Eval metrics by language, including token usage.}
\label{tab:app_roo_component_80b_detail_by_language}
\begin{adjustbox}{max width=\linewidth}
\begin{tabular}{llllllllllll}
\toprule
Model & pass@1 & pass@3 & pass\_all & iter pass & TTC & Input total & Cached total & Output total & Input avg & Cached avg & Output avg \\
\midrule
\multicolumn{12}{l}{\textbf{Python (34 exercises $\times$ 3 = 102 rollouts)}} \\
\midrule
\method{} w/o $T(\delta)$ & 29 (85.3\%) & 33 (97.1\%) & 24 (70.6\%) & 85/102 (83.3\%) & 13.9M & 4,035,791 & 52,833,706 & 912,765 & 39,567 & 517,978 & 8,949 \\
\method{} w/o CTG & 30 (88.2\%) & 33 (97.1\%) & 25 (73.5\%) & 89/102 (87.3\%) & 16.1M & 4,499,559 & 67,386,007 & 977,139 & 44,113 & 660,647 & 9,580 \\
\method{} w/o GSP & 30 (88.2\%) & 32 (94.1\%) & 23 (67.6\%) & 83/102 (81.4\%) & 20.1M & 4,847,925 & 102,658,102 & 1,004,775 & 47,529 & 1,006,452 & 9,851 \\
\method{} (full) & 30 (88.2\%) & 33 (97.1\%) & 27 (79.4\%) & 90/102 (88.2\%) & 13.1M & 3,807,607 & 46,484,492 & 933,088 & 37,329 & 455,730 & 9,148 \\
\midrule
\multicolumn{12}{l}{\textbf{JavaScript (50 exercises $\times$ 3 = 150 rollouts)}} \\
\midrule
\method{} w/o $T(\delta)$ & 45 (90.0\%) & 47 (94.0\%) & 44 (88.0\%) & 137/150 (91.3\%) & 18.6M & 5,968,854 & 69,574,697 & 1,133,874 & 39,792 & 463,831 & 7,559 \\
\method{} w/o CTG & 46 (92.0\%) & 48 (96.0\%) & 42 (84.0\%) & 137/150 (91.3\%) & 19.4M & 5,810,693 & 81,208,201 & 1,099,342 & 38,738 & 541,388 & 7,329 \\
\method{} w/o GSP & 36 (72.0\%) & 46 (92.0\%) & 31 (62.0\%) & 117/150 (78.0\%) & 21.7M & 6,491,278 & 99,738,353 & 1,037,504 & 43,275 & 664,922 & 6,917 \\
\method{} (full) & 46 (92.0\%) & 49 (98.0\%) & 42 (84.0\%) & 137/150 (91.3\%) & 17.3M & 5,555,281 & 61,325,457 & 1,130,758 & 37,035 & 408,836 & 7,538 \\
\midrule
\multicolumn{12}{l}{\textbf{Go (36 exercises $\times$ 3 = 108 rollouts)}} \\
\midrule
\method{} w/o $T(\delta)$ & 31 (86.1\%) & 34 (94.4\%) & 25 (69.4\%) & 91/108 (84.3\%) & 14.2M & 4,075,592 & 48,670,131 & 1,059,954 & 37,737 & 450,649 & 9,814 \\
\method{} w/o CTG & 30 (83.3\%) & 34 (94.4\%) & 25 (69.4\%) & 87/108 (80.6\%) & 22.6M & 6,650,666 & 87,594,557 & 1,432,894 & 61,580 & 811,061 & 13,268 \\
\method{} w/o GSP & 31 (86.1\%) & 32 (88.9\%) & 24 (66.7\%) & 84/108 (77.8\%) & 19.1M & 5,038,572 & 85,575,230 & 1,102,255 & 46,653 & 792,363 & 10,206 \\
\method{} (full) & 31 (86.1\%) & 33 (91.7\%) & 29 (80.6\%) & 92/108 (85.2\%) & 16.3M & 4,654,524 & 55,659,080 & 1,209,340 & 43,097 & 515,362 & 11,198 \\
\midrule
\multicolumn{12}{l}{\textbf{Java (45 exercises $\times$ 3 = 135 rollouts)}} \\
\midrule
\method{} w/o $T(\delta)$ & 30 (66.7\%) & 38 (84.4\%) & 20 (44.4\%) & 91/135 (67.4\%) & 31.6M & 9,168,853 & 122,372,257 & 2,034,221 & 67,917 & 906,461 & 15,068 \\
\method{} w/o CTG & 23 (51.1\%) & 37 (82.2\%) & 13 (28.9\%) & 74/135 (54.8\%) & 28.3M & 8,457,768 & 108,610,839 & 1,805,331 & 62,650 & 804,525 & 13,373 \\
\method{} w/o GSP & 33 (73.3\%) & 39 (86.7\%) & 22 (48.9\%) & 94/135 (69.6\%) & 24.9M & 7,603,925 & 98,546,623 & 1,480,701 & 56,325 & 729,975 & 10,968 \\
\method{} (full) & 28 (62.2\%) & 37 (82.2\%) & 20 (44.4\%) & 89/135 (65.9\%) & 24.3M & 7,543,322 & 90,934,720 & 1,529,337 & 55,876 & 673,591 & 11,328 \\
\midrule
\multicolumn{12}{l}{\textbf{Rust (30 exercises $\times$ 3 = 90 rollouts)}} \\
\midrule
\method{} w/o $T(\delta)$ & 19 (63.3\%) & 25 (83.3\%) & 16 (53.3\%) & 63/90 (70.0\%) & 19.5M & 5,328,776 & 73,042,138 & 1,373,749 & 59,209 & 811,579 & 15,264 \\
\method{} w/o CTG & 22 (73.3\%) & 27 (90.0\%) & 18 (60.0\%) & 69/90 (76.7\%) & 19.7M & 5,372,024 & 73,642,864 & 1,399,325 & 59,689 & 818,254 & 15,548 \\
\method{} w/o GSP & 22 (73.3\%) & 27 (90.0\%) & 17 (56.7\%) & 65/90 (72.2\%) & 24.0M & 6,267,397 & 112,976,962 & 1,281,274 & 69,638 & 1,255,300 & 14,236 \\
\method{} (full) & 24 (80.0\%) & 24 (80.0\%) & 21 (70.0\%) & 68/90 (75.6\%) & 18.2M & 5,006,504 & 67,960,906 & 1,270,158 & 55,628 & 755,121 & 14,113 \\
\bottomrule
\end{tabular}
\end{adjustbox}
\end{table}
\FloatBarrier

\end{document}